\documentclass[journal]{IEEEtran}
\usepackage{cite}
\usepackage{threeparttable}
\usepackage{graphicx}
\usepackage{picinpar}
\usepackage{stfloats}
\usepackage{setspace}
\usepackage[all]{xy}
\usepackage{float}
\usepackage{amsmath}
\usepackage{array}
\usepackage{mdwmath}
\usepackage{mdwtab}
\usepackage{eqparbox}
\usepackage{algorithmic}
\usepackage{algorithm}
\usepackage{exscale}
\usepackage{relsize}
\usepackage{supertabular}
\usepackage{subfigure}
\usepackage{threeparttable}
\usepackage{caption}
\usepackage{booktabs}
\usepackage{multirow}
\usepackage{etoolbox}
\usepackage{makecell}
\usepackage{amssymb}
\usepackage{mathrsfs}
\usepackage{color}
\definecolor{purple}{RGB}{160,32,240}
\definecolor{darkred}{RGB}{255,0,255}
\usepackage{stfloats}
\usepackage{bm}
\usepackage{flushend}
\usepackage[justification=centering]{caption}

\renewcommand{\raggedright}{\leftskip=0pt \rightskip=0pt plus 0cm}

\begin{document}

\newtheorem{lemma}{Lemma}
\newtheorem{corol}{Corollary}
\newtheorem{theorem}{Theorem}
\newtheorem{proposition}{Proposition}
\newtheorem{definition}{Definition}
\newcommand{\e}{\begin{equation}}
\newcommand{\ee}{\end{equation}}
\newcommand{\eqn}{\begin{eqnarray}}
\newcommand{\eeqn}{\end{eqnarray}}
\title{Joint Activity and Blind Information Detection
for UAV-Assisted Massive IoT Access}
%
\author{Li Qiao, Jun Zhang, Zhen Gao, Dezhi Zheng, Md. Jahangir Hossain, Yue Gao, \\
Derrick Wing Kwan 
Ng,~\IEEEmembership{Fellow,~IEEE}, and Marco Di Renzo,~\IEEEmembership{Fellow,~IEEE}

\vspace{-10mm}
\thanks{\color{black}This paper was presented in part at the 2021 IEEE/CIC International Conference on Communications in China (ICCC), 2021\cite{ICCC2021}. The codes and some other materials about this work may be available at https://gaozhen16.github.io}
\thanks{L. Qiao, J. Zhang, Z. Gao, and D. Zheng are with the School of Information and Electronics and the Advanced Research Institute of Multidisciplinary Science, Beijing Institute of Technology, Beijing 100081, China, also with the Yangtze Delta Region Academy of Beijing Institute of Technology, Jiaxing 314001, China (e-mail: qiaoli@bit.edu.cn; buaazhangjun@vip.sina.com; gaozhen16@bit.edu.cn; zhengdezhi@buaa.edu.cn).}
\thanks{M. J. Hossain is with the School of Engineering, The University of British Columbia, Kelowna, BC V1V 1V7, Canada (e-mail: jahangir.hossain@ubc.ca).}
\thanks{Y. Gao is with the Department of Electrical and Electronic Engineering,
University of Surrey, Surrey GU2 7XH, U.K. (e-mail: yue.gao@ieee.org).}
\thanks{D. W. K. Ng is with the School of Electrical Engineering and Telecommunications, University of New South Wales, Sydney,
NSW 2025, Australia (email: w.k.ng@unsw.edu.au).}
\thanks{M. Di Renzo is with Universit\'e Paris-Saclay, CNRS, CentraleSup\'elec, Laboratoire des Signaux et Syst\`emes, 3 Rue Joliot-Curie, 91192 Gif-sur-Yvette, France. (marco.di-renzo@universite-paris-saclay.fr)}.
}

\maketitle

\begin{abstract}
Grant-free non-coherent index-modulation (NC-IM) has been recently considered as an efficient massive access scheme for enabling cost- and energy-limited Internet-of-Things (IoT) devices that transmit small data packets. This paper investigates the grant-free NC-IM scheme combined with orthogonal frequency division multiplexing for applicant to unmanned aerial vehicle (UAV)-based massive IoT access. Specifically, each device is assigned a unique non-orthogonal signature sequence codebook. Each active device transmits one of its signature sequences in the given time-frequency resources, by modulating the information in the index of the transmitted signature sequence. For small-scale multiple-input multiple-output (MIMO) deployed at the UAV-based aerial base station (BS), by jointly exploiting the space-time-frequency domain  device activity, we propose a computationally efficient space-time-frequency joint activity and blind information detection (JABID) algorithm with significantly improved detection performance. Furthermore, for large-scale MIMO deployed at the aerial BS, by leveraging the sparsity of the virtual angular-domain channels, we propose an angular-domain based JABID algorithm for improving the system performance with reduced access latency. In addition, for the case of high mobility IoT devices and/or UAVs, we introduce a time-frequency spread transmission (TFST) strategy for the proposed JABID algorithms to combat doubly-selective fading channels. Finally, extensive simulation results are illustrated to verify the superiority of the proposed algorithms and the TFST strategy over known state-of-the-art algorithms.
\end{abstract}


\IEEEpeerreviewmaketitle
\vspace{-1.5mm}
\section{Introduction}

{\color{black}The most important paradigm shift of the future sixth-generation (6G) networks is global coverage, which enables broadband access anytime and anywhere by developing a space-air-ground-sea integrated communication network \cite{mMTC1, Liu-NGMA}. It can also allow wide-area Internet-of-Things (IoT) access in areas not covered by terrestrial networks, ensuring emergency communications, agriculture monitoring, as well as collection of information on marine buoys, etc. {\color{black}With their inherent properties such as flexibility, mobility, and adaptive altitude, unmanned aerial vehicles (UAVs) have been considered as an indispensable component in future integrated terrestrial and aerial 6G networks to efficiently support
massive access for wide-area IoT \cite{UAV-Zeng-New},{\color{black}\cite{Xiao}}.} Due to the low-cost and low-energy consumption features of typical IoT devices, as well as the complicated propagation condition of UAV-ground channels, disruptive techniques are needed to efficiently enable UAV-enabled wide-area massive IoT access\cite{UAV-6G,yuanwei-2,yuanwei-3,UAV-Ploss1,WFeng2}.}




To accommodate massive IoT devices, existing random access protocols can be mainly divided into two categories: grant-based random access protocols and grant-free random access (GFRA) protocols \cite{JSAC-Editor}. As for grant-based access protocols, the devices are required to first transmit a scheduling request and then to wait for an uplink (UL) grant from the targeted base station (BS). Such a complicated handshaking procedure would generally result in an excessive signaling overhead and long access latency in IoT networks. To reduce the required access latency and signaling overhead, the recently emerging GFRA protocols have drawn much attention, as they allow the IoT devices to send sporadic small data directly to the BS without acquiring any permission, thereby reducing the signaling overhead for massive IoT access \cite{JSAC-Editor}. In particular, due to the sporadic traffic of typical IoT devices, only a fraction of devices are simultaneously active in one transmission block \cite{JSAC-Editor}. Hence, compressed sensing (CS) techniques have been integrated in the  transceiver design to facilitate grant-free massive IoT access, e.g., \cite{yuwei18,shaoxiaodan19,shim19,kemalong20,wangbichai16,duyang18,shim18-2,weichao17,X.Ma-CL,TWC-Qiao,X.Ma-ACCESS,L.Qiao-TVT,NC-IM-Pro1,NC-IM-Tcom,NC-IM-Access}.

\subsection{Related Work}

\begin{figure*}[t]
\vspace{-7mm}
     \centering
     \includegraphics[width = 1.67\columnwidth,keepaspectratio]
     {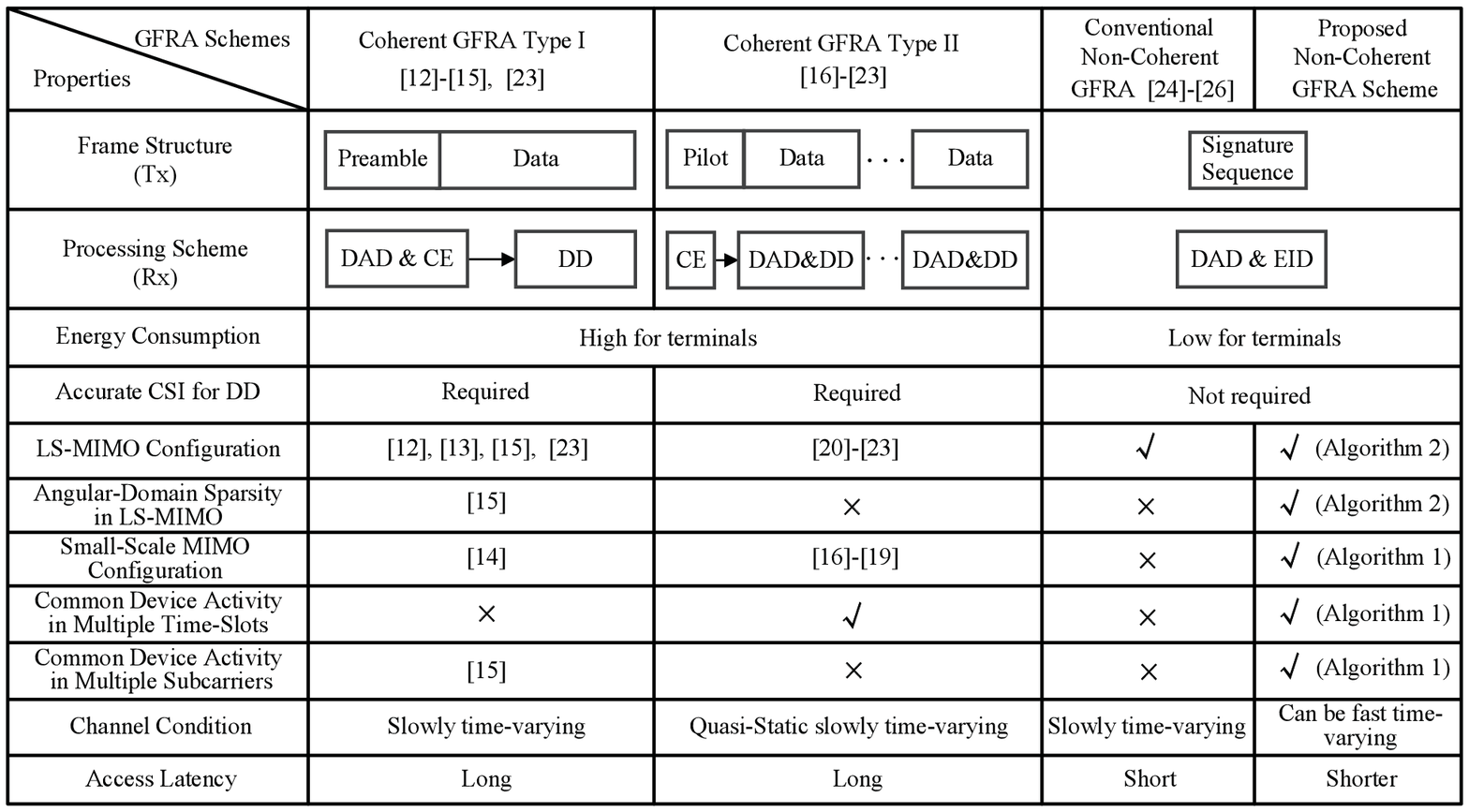}
     \captionsetup{font={footnotesize, color = {black}}, singlelinecheck = off, justification = raggedright,name={Fig.},labelsep=period}
     \caption{Comparison of the coherent GFRA type I, the coherent GFRA type II, the conventional non-coherent GFRA scheme, and the proposed non-coherent GFRA scheme, where their properties are presented and compared in detail.}
     \label{fig:FrameStructure}
     \vspace{-5mm}
\end{figure*}

Recently, several coherent grant-free  massive IoT access schemes have been proposed for cellular IoT systems{\color{black}\footnote{\color{black}Note that GFRA protocols for cellular IoT systems can be employed in UAV-ground data collection scenarios, where a UAV serves as an aerial BS and massive IoT devices are located in the coverage area of the UAV\cite{yuanwei-2}. Compared with cellular systems, the UAV-based aerial BS can support massive IoT access in areas without terrestrial networks\cite{UAV-Zeng-New,Xiao,UAV-6G,yuanwei-2,yuanwei-3}, e.g., rainforest, disaster-affected areas, remote areas, etc. In addition, UAVs can provide flexible and convenient IoT access in areas of interest, e.g., providing real-time urban environment monitoring.}}, where the UL transmission block within a coherence time can be typically divided into two successive stages \cite{yuwei18,shaoxiaodan19,shim19,kemalong20}. In the first stage, the active devices send their preambles to the BS for channel estimation (CE) and device activity detection (DAD). In the second stage, data is transmitted and coherent data detection (DD) is performed at the BS based on the previously estimated channel state information (CSI). We refer to this scheme as {\it coherent GFRA type I}, as shown in Fig. \ref{fig:FrameStructure}. To realize effective massive access, the authors of \cite{yuwei18}  proposed an approximate message passing (AMP)-based joint DAD and CE (JDAD-CE) scheme for large-scale (LS) multiple-input multiple-output (MIMO) BS during the first stage. Their results demonstrated that the DAD error asymptotically approaches zero as the number of BS antennas is sufficiently large.
Furthermore, the authors of \cite{shaoxiaodan19} discussed the achievable rates of both the two-stage UL transmission and the subsequent downlink data transmission phase. In addition, by performing iterative approximations, an efficient low-complexity expectation propagation-based algorithm under the Bayesian framework was proposed for JDAD-CE in \cite{shim19}. To further improve the JDAD-CE performance and to reduce the length of preambles (so as the access latency), the authors of \cite{kemalong20} proposed a generalized multiple measurement vector AMP (GMMV-AMP) algorithm by exploiting the structured sparsity of the UL channel matrix in both the spatial and the angular domains. 

Different from the {\it coherent GFRA type I} scheme in \cite{yuwei18,shaoxiaodan19,shim19,kemalong20}, the authors of \cite{wangbichai16,duyang18,shim18-2,weichao17} assumed that the CSI of all the devices are estimated based on pilots in the first stage. Then, in the subsequent stages within the coherence time, the active devices transmit their data and eventually the receiver detects the active devices as well as their associated data. As indicated in Fig. \ref{fig:FrameStructure}, we refer to this scheme as {\it coherent GFRA type II}. Specifically, by exploiting the block sparsity of the data traffic in consecutive time slots, the authors of \cite{wangbichai16} and \cite{duyang18} proposed a structured iterative support detection algorithm and a threshold-aided block sparsity adaptive subspace pursuit algorithm, respectively, to perform joint DAD and DD (JDAD-DD). Furthermore, by considering the finite alphabet constraint of the transmitted data as prior information, a maximum {\it a posteriori} probability-based greedy algorithm was proposed in \cite{shim18-2} to further improve the performance. In addition, the authors of \cite{weichao17} proposed an AMP-based detector, whereby the finite alphabet constraint of the transmitted symbols was considered and the expectation maximization (EM)-based algorithm was adopted to detect the active devices.

To further improve the spectral and energy efficiency of IoT devices, coherent grant-free massive IoT access relying on index modulation (IM) has been considered in more recent works, e.g., \cite{X.Ma-CL,TWC-Qiao,X.Ma-ACCESS,L.Qiao-TVT}. In fact, IM is a highly spectral- and energy-efficient yet simple digital modulation technique, which utilizes the indices of pre-defined codebooks to convey additional information bits \cite{IM2,IM1}. In the literature, there exist several realizations of IM, e.g., spatial modulation (SM), media modulation, SM based on reconfigurable antennas, and, more recently, reconfigurable intelligent surface (RIS)-based modulation, etc. \cite{SM-Marco,RIS-Marco,RIS-IM,SM-Wen}. Specifically, the {\it coherent GFRA type II} scheme with SM and media modulation are investigated in \cite{X.Ma-CL} and \cite{TWC-Qiao,X.Ma-ACCESS,L.Qiao-TVT}, respectively. The authors of \cite{X.Ma-CL} proposed a two-level sparse structure greedy algorithm to detect the spatially modulated symbols at the receiver with perfectly known CSI. By exploiting the sporadic traffic in IoT devices and the structured sparsity of media modulated symbols, greedy algorithms were proposed in \cite{X.Ma-ACCESS,L.Qiao-TVT}, and an AMP-based algorithm was developed in \cite{TWC-Qiao} for JDAD-DD with perfectly known CSI. Also, with the consideration of the finite alphabet constraint of the transmitted symbols, it was shown that the algorithm proposed in \cite{TWC-Qiao} outperforms those in \cite{L.Qiao-TVT} and \cite{X.Ma-ACCESS}. Furthermore, the authors of \cite{TWC-Qiao} proposed a data-aided CE strategy, which can reduce the channel training overhead in media modulation-based massive access. In addition, the authors of \cite{X.Ma-ACCESS} considered the combination of {\it coherent GFRA type I} and IM, where the active devices using media modulation send preambles before the media modulated data transmission, while a greedy algorithm was proposed to perform JDAD-CE at the BS.

{\color{black}It is important to note that in the {\it coherent GFRA type I} and {\it coherent GFRA type II} schemes, the channels are assumed to be quasi-static for a relatively long time and accurate estimation of the CSI plays a pivotal role for achieving better DAD and DD performance. However, the acquisition of accurate CSI is challenging, especially for practical UAV-ground channels due to the complex propagation conditions \cite{UAV-6G}. In addition, the preambles/pilots allocated for CSI acquisition may not be efficient for application to IoT devices with low energy consumption requirements and small data packets \cite{NC-IM-Pro1,NC-IM-Tcom}.}

{\color{black}To support massive IoT access more efficiently, a single-stage {\it non-coherent GFRA} approach was proposed in \cite{NC-IM-Pro1,NC-IM-Tcom}, namely grant-free non-coherent index modulation (NC-IM), for tackling the issues of low-cost and low-energy consumption for massive IoT access. Specifically, each device is allocated with $I=2^r$ unique non-orthogonal signature sequences, which constitute a pre-defined codebook for IM. Hence, given any active device, $r$-bits can be embedded in the index of the transmitted signature sequence. As such, the DAD and embedded information detection (EID) processes can be jointly performed at the BS by detecting the indices of the transmitted signature sequences, without the need of explicit CSI, as indicated in Fig. \ref{fig:FrameStructure}. The authors of \cite{NC-IM-Tcom} proposed an AMP-based algorithm, which treats the structured sparsity pattern observed by a large-number of receive antennas as a CS multiple measurement vector (MMV) problem. To further improve the performance of the DAD-EID, a section-wise AMP algorithm with a non-separable denoiser was proposed in \cite{NC-IM-Access}, where the structured sparsity of NC-IM and the MMV property in the spatial domain were jointly exploited. Thanks to the simple, fast response, requiring no CSI, and low-energy consumption properties, the grant-free NC-IM scheme is suitable for UAV-based wide-area massive IoT access with low-cost IoT devices. {\color{black}However, without obvious MMV property in the spatial domain, the DAD-EID performance of the algorithms proposed in \cite{NC-IM-Tcom} and \cite{NC-IM-Access} heavily deteriorates in small-scale MIMO systems, which are commonly adopted in small rotary-wing UAVs. Furthermore, due to the energy-constraints of UAVs, the access latency of grant-free NC-IM should be as short as possible in UAV-based massive IoT access. However, to the best of our knowledge, there exist no related discussions on this critical issue to date.} In addition, both \cite{NC-IM-Tcom} and \cite{NC-IM-Access} considered the frequency-flat channel fading, which is impractical for UAV-based wide-area massive IoT access \cite{UAV-6G}.}


\vspace{-3mm}
\subsection{Contributions}
This paper investigates the grant-free NC-IM scheme for the UAV-based massive IoT access in future space-air-ground-sea integrated networks. For each active IoT device, one signature sequence from a predefined codebook is transmitted on one subcarrier over several successive orthogonal frequency division multiplexing (OFDM) symbols. As for small-scale MIMO aerial BSs, we propose a space-time-frequency joint activity and blind information detection (STF-JABID) algorithm to improve the DAD-EID performance. As for LS-MIMO aerial BSs, an angular-domain enhanced JABID (AE-JABID) algorithm is proposed for reducing the access latency. Furthermore, over doubly-selective fading channels (i.e., both frequency- and time-selective fading), where the time-selective fading originates from the high mobility of IoT devices and/or UAV-based aerial BSs, we develop a simple yet effective time-frequency spread transmission (TFST) strategy to achieve better JABID performance and reduced access latency. Finally, simulation results are reported to verify the effectiveness of the proposed algorithms and the developed TFST strategy. Our main contributions are summarized as follows:
\begin{itemize}
\item[$\bullet$] {\bf STF-JABID algorithm tailored for small-scale MIMO aerial BSs:} {\color{black}By exploiting the structured sparsity of the equivalent channel matrix in the spatial domain and the  device activity in the space-time-frequency domain as a prior information, we propose an AMP-based STF-JABID algorithm to improve the DAD-EID performance for small-scale MIMO systems.} Furthermore, by utilizing the EM algorithm and considering the structured sparsity in the spatial domain, the active/inactive status of the devices and the CSI parameters can be adaptively obtained with enhanced accuracy.

\item[$\bullet$] {\bf AE-JABID algorithm developed for LS-MIMO aerial BSs:} In LS-MIMO systems, the channel matrix usually exhibits clustered sparsity in the virtual angular domain \cite{VAD_Channel}. {\color{black}By exploiting the clustered sparsity and the IM structured sparsity of the virtual angular-domain equivalent channel matrix, we propose an AE-JABID algorithm to further reduce the access latency with improved DAD-EID performance.}%

\item[$\bullet$] {\color{black}{\bf Proposed TFST strategy to conquer doubly-selective fading channels:} The proposed TFST strategy spreads each signature sequence in several successive subcarriers and OFDM symbols (i.e., time slots), rather than only in one subcarrier and more time slots, which reduces the access latency. Consequently, the time-selective fading suffered by the whole signature sequence is mitigated, which improves the DAD-EID performance at the BS over doubly-selective fading channels.}
\end{itemize}

\textit {Notation}: Boldface lower and upper-case symbols denote column vectors and matrices, respectively. For a matrix ${\bf A}$, ${\bf A}^T$, ${\bf A}^H$, ${\bf A}^*$, ${\left\| {\bf{A}} \right\|_F}$, $[{\bf{A}}]_{m,n}$ denote the transpose, Hermitian transpose, conjugate, Frobenius norm, the $m$-th row and $n$-th column element of ${\bf{A}}$, respectively. $[{\bf{A}}]_{\Omega,:}$ ($[{\bf{A}}]_{:,\Omega}$) is the sub-matrix containing the rows (columns) of ${\bf{A}}$ indexed in the ordered set $\Omega$. For a vector ${\bf x}$, $\| {\bf x} \|_p$ and $[{\bf x}]_{m}$ denote the ${l_p}$ norm and $m$-th element of ${\bf x}$, respectively. $|\Gamma|_c$ denotes the cardinality of the ordered set $\Gamma$. $\lceil \cdot \rceil$ rounds each element to the nearest integer greater than or equal to that element. The marginal distribution $p\left([{\bf x}]_m\right)$ is denoted as $p\left([{\bf x}]_m\right)={\int}_{\backslash [{\bf x}]_m} p\left({\bf x}\right)$ and $\mathcal{CN}(x; \mu, \nu)$ (or $x\sim\mathcal{CN}(\mu, \nu)$) denotes the complex Gaussian distribution of a random variable $x$ with mean $\mu$ and variance $\nu$. $[K]$ denotes the set $\{1,2,...,K\}$. Finally, for $y\neq 0$ and $x>0$ ($x<0$), ${\rm mod}(x,y)=x-\lfloor x/y \rfloor y$ (${\rm mod}(x,y)=x+\lceil x/y \rceil y$), whereas ${\rm mod}(x,y)=y$ if $x-\lfloor x/y \rfloor y=0$ ($x+\lceil x/y \rceil y=0$).

\vspace{-2mm}
\section{System Model}
In this section, we first present the adopted system model and the associated channel model. Then, we introduce the grant-free NC-IM-based massive IoT access problem. 

\begin{figure*}[t]
\vspace{-10mm}
\centering
\subfigure[]{
    \begin{minipage}[t]{0.49\linewidth}
        \centering
\label{fig:systemModel-SS-MIMO}
        \includegraphics[width = 0.77\columnwidth,keepaspectratio]
        {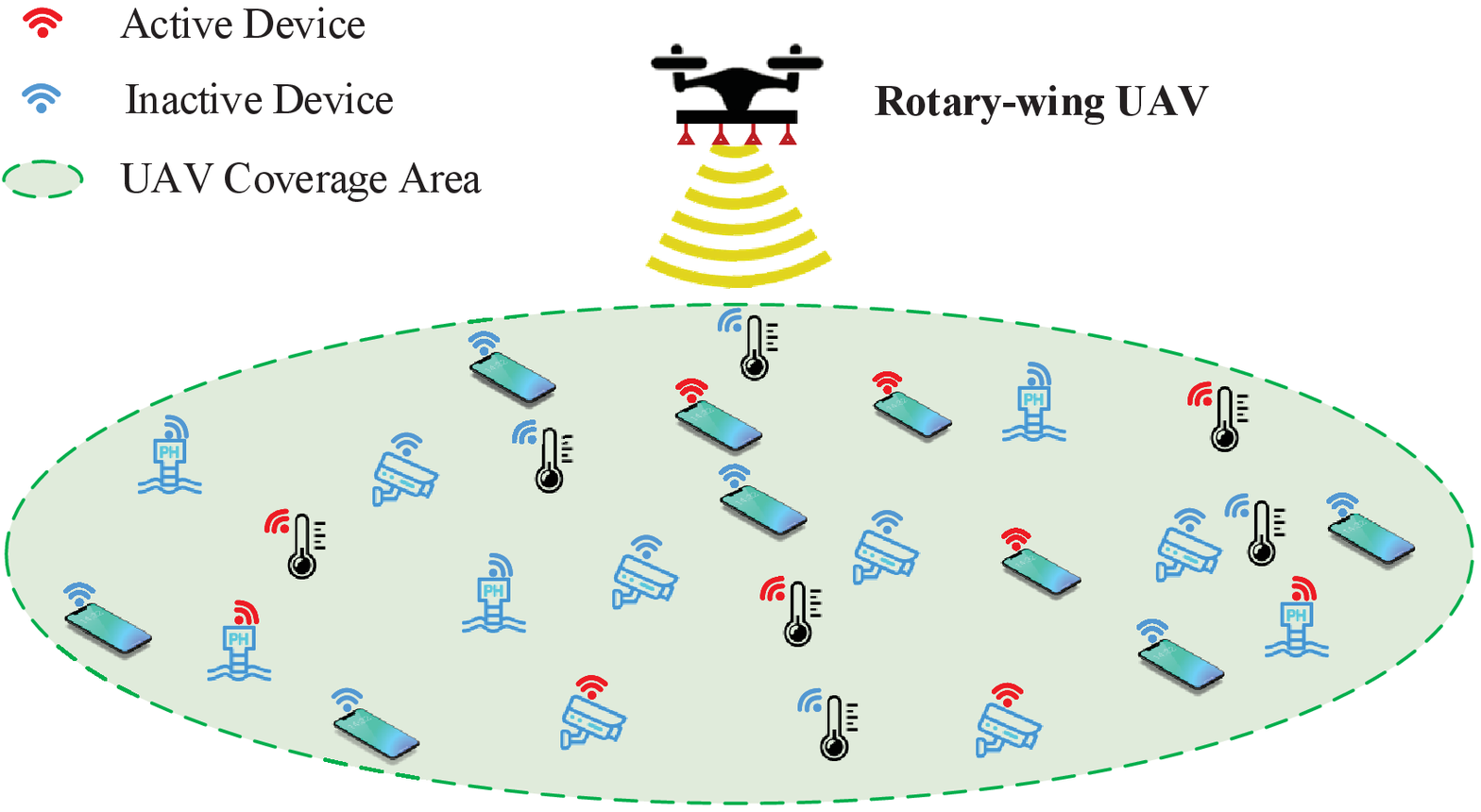}\\
  \end{minipage}}%
\subfigure[]{
    \begin{minipage}[t]{0.49\linewidth}
        \centering
\label{fig:systemModel-LS-MIMO}
        \includegraphics[width = 0.77\columnwidth,keepaspectratio]
        {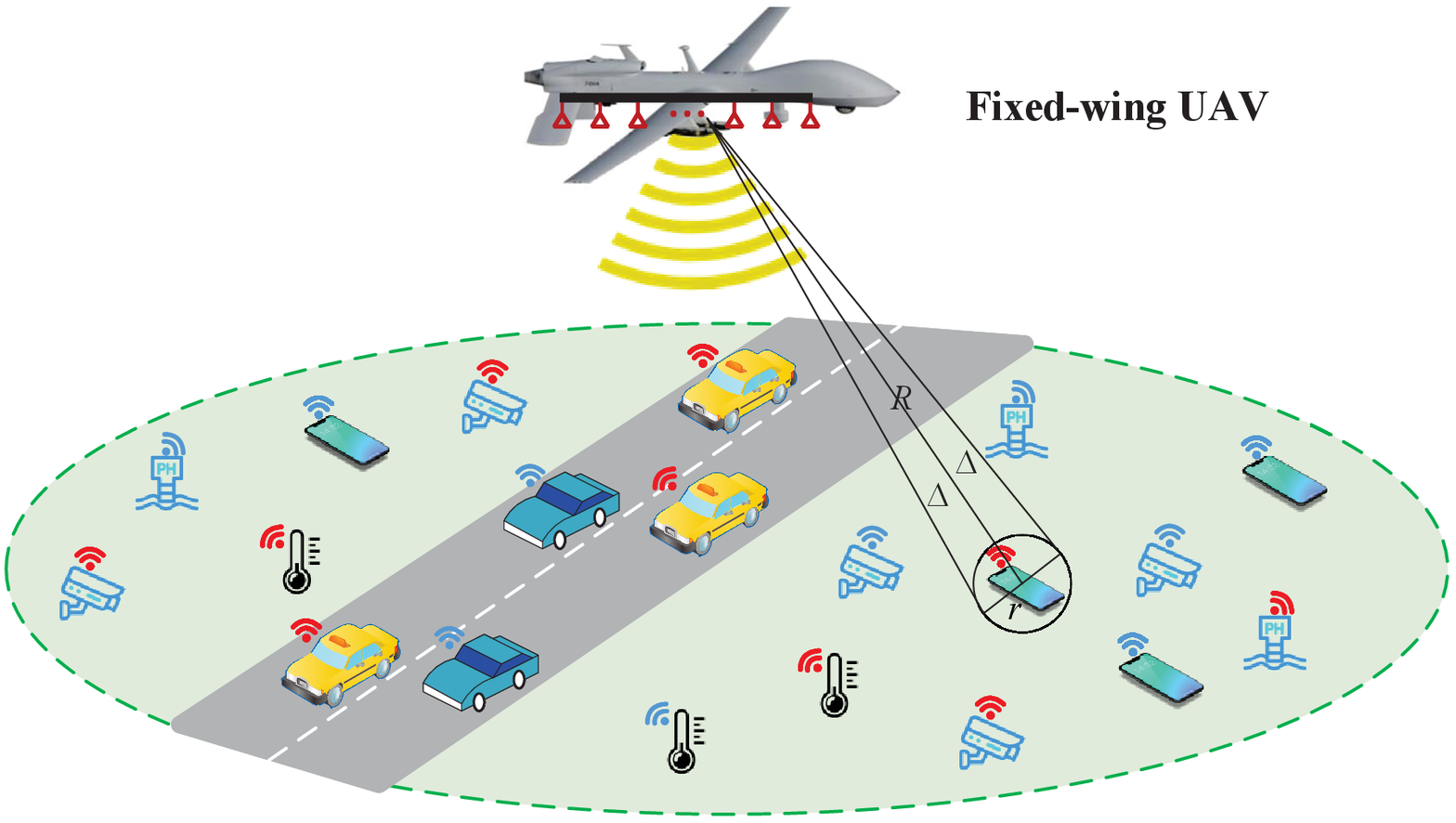}\\
    \end{minipage}%
}%
\centering
\setlength{\abovecaptionskip}{-1mm}
\captionsetup{font={footnotesize}, singlelinecheck = off, justification = raggedright,name={Fig.},labelsep=period}
\caption{System model of grant-free NC-IM-based massive IoT access, where a UAV serves as an aerial BS for data collection: (a) Small-scale MIMO at the aerial BS; (b) LS-MIMO at the aerial BS, where the mobile IoT devices are also considered. }
\label{fig:systemModel}
\vspace{-6mm}
\end{figure*}

\vspace{-3.6mm}
\subsection{Channel Model}
{\color{black}We consider a UAV-based aerial BS, equipped with a uniform linear array (ULA) and $M$ receive antennas\footnote{\color{black}Note that there are two main types of UAVs widely used in practice, i.e., the rotary-wing UAVs and the fixed-wing UAVs\cite{UAV-Zeng-New}. The rotary-wing UAV typically has a relatively low load capacity and limited space, which can restrict the use of massive antenna arrays\cite{UAV-Zeng-New,Xiao}. Hence, we consider small-scale MIMO at the rotary-wing UAV. While, with a greater payload capacity and larger space\cite{UAV-Zeng-New,Xiao}, a fixed-wing UAV can be equipped with LS-MIMO.}, collecting the information from $K$ ground single-antenna IoT devices within its coverage area\footnote{\color{black}The proposed scheme can be applied to efficiently collect the information of marine buoys or other maritime sensors. The channel conditions in maritime communications are different from those in terrestrial communications due to the special environment over the sea surface\cite{Xiao,UAV-Ploss1,WFeng2}. This study is left to a future research work.}, as shown in Fig. \ref{fig:systemModel}.} Due to the large number $K$ IoT devices and their typical small data-packets, we adopt the grant-free NC-IM access scheme for the UL. {\color{black}According to \cite{yuanwei-2}, in complicated practical scenarios, the UL signals can reach the aerial BS via direct line-of-sight (LoS) links or via reflected links created by multiple scatterers, i.e., the non-line-of-sight (NLoS) links. Note that the delay spreads of the NLoS paths can be several times larger than the system sampling period, which results in the frequency-selective channel fading\cite{MIMO-OFDM}.} Hence, we assume that there are $P$ scattering paths and each scattering path corresponds to one angle of arrival (AoA) and one time delay instant. {\color{black}We denote the AoA of the $p$-th path of the $k$-th device by $ \theta_{k,p}$, $\forall p\in[P]$, $\forall k\in[K]$, and the corresponding steering vector (contains the channel correlation property among antennas) at the aerial BS is defined as
\begin{equation}\label{eq:SVec}
\begin{array}{l}
{\bf a}(\theta_{k,p})=\dfrac{1}{\sqrt{M}}\left[{1,e^{j2\pi\frac{d{\rm sin}(\theta_{k,p})}{\lambda}},...,e^{j2\pi (M-1)\frac{d{\rm sin}(\theta_{k,p})}{\lambda}}}\right]^T,
\vspace{-1mm}
\end{array}
\end{equation}
due to the adopted ULA geometry, where $\lambda$ is the carrier wavelength and $d=\lambda/2$ is the antenna spacing at the aerial BS.} 

To combat the frequency-selective channel fading, the OFDM technique with $N$ subcarriers is adopted \cite{MIMO-OFDM}. Hence, based on a geometric channel model\cite{MIMO-OFDM}, the UL channels of the $n$-th subcarrier between the aerial BS and the $k$-th user in the $t$-th time slot, $\forall n\in[N]$, $\forall k, t$, can be denoted by
\begin{equation}\label{eq:CMod}
\begin{array}{l}
{\bf h}_{k,n}^t\!\!=\!\!\sqrt{\dfrac{M}{P}}\!\!\sum\limits_{p=1}^{P}h_{k,p}^t e^{j2\pi \nu_{k,p}^t t }{\bf a}(\theta_{k,p}^t)e^{-j2\pi \tau_{k,p}^t(-\frac{B_s}{2}+\frac{B_s(n-1)}{N})},
\vspace{-1mm}
\end{array}
\end{equation}
where $h_{k,p}^t\sim\mathcal{CN}(0,1)$, $ \nu_{k,p}^t$, $\theta_{k,p}^t$, and $\tau_{k,p}^t$ represent the channel gain, the Doppler shift, the AoA, and the delay for the $p$-th path of the $k$-th device in the $t$-th time slot, respectively, and $B_s$ denotes the double-sided signal bandwidth. The parameters $ \nu_{k,p}^t$, $\theta_{k,p}^t$, $\tau_{k,p}^t$, and $B_s$ will be discussed in detail in the simulation setup of Section V-A.

\subsection{Grant-Free NC-IM-Based Massive IoT Access}
To carry $r$ information bits embedded in each UL transmission, $I=2^r$ unique non-orthogonal signature sequences with length $L$ ($L\ll K$) are allocated for each IoT device. The codebook of the $k$-th device, $\forall k$, is denoted by ${\bf \Phi}_k=[{\bm \phi}_{k,1},...,{\bm \phi}_{k,I}]^T\in\mathbb{C}^{L\times I}$, whose elements are generated by sampling an independent and identically distributed (i.i.d.) symmetric Bernoulli distribution\footnote{It has been proved that signature sequences
based on the Bernoulli distribution provide better performance than those based on the Gaussian distribution. Also, the probability that two devices have identical signature sequences can be neglected in massive IoT access\cite{NC-IM-Tcom}.}. {\color{black}Specifically,  ${\bm \phi}_{k,i}\triangleq [ \phi_{k,i,1},...,\phi_{k,i,L}]^T\in\mathbb{C}^{L\times 1}$, where $\phi_{k,i,l}\in\left\{(\pm1\pm j)/\sqrt{2L}\right\}$ denotes the $l$-th element of ${\bm \phi}_{k,i}$, $\forall k\in[K]$, $\forall i\in[I]$, and $\forall l\in[L]$.}

In the UL, due to the sporadic traffic of IoT devices, only $K_a$ ($K_a\ll K$) devices are simultaneously active. For the active devices, a signature sequence is activated from each of their codebook based on the embedded $r$ information bits to be delivered. Then, the selected sequence is transmitted from each active device on one subcarrier of $L$ successive OFDM symbols. Hence, a total of $rN$ bits can be transmitted via $N$ subcarriers. {\color{black}Note that all the active devices are allocated with the same time-frequency resources.} We refer to the $L$ successive OFDM symbols as a sub-frame. The device activity is invariant within a frame ($J$ sub-frames) and the CSI within each sub-frame is typically assumed to be unchanged in slowly time-varying IoT scenarios\footnote{{\color{black}In this section, we assume that the rotary-wing UAV hovers in the air when it collects data from IoT devices \cite{UAV-Zeng-New}. In addition, the time-selective channel fading resulted by the rapid movement of IoT devices (e.g., automobile sensors) and/or fixed-wing UAV is discussed in Section IV-C. The focus of the paper is the design of the uplink massive IoT access and the related physical-layer signal processing techniques. On the other hand, the design of the trajectory of the UAVs is beyond the scope of this paper.}} \cite{JSAC-Editor}. Hence, the received signal ${\bf Y}_n^j\in\mathbb{C}^{L\times M}$ of the $n$-th subcarrier in the $j$-th sub-frame, $\forall j\in[J]$, can be expressed as
\vspace{-1mm}
\begin{equation}\label{eq:SMod1}
\begin{split}
{\bf Y}_n^j&=\sum\nolimits_{k=1}^{K}a_k{\bf \Phi}_k{\bf e}_{k,n}^j({\bf h}_{k,n}^j)^T+{\bf N}_n^j\\
&=\sum\nolimits_{k=1}^{K}{\bf \Phi}_k{\bf X}_{k,n}^j+{\bf N}_n^j={\bf \Phi}{\bf X}_{n}^j+{\bf N}_n^j,
\vspace{-1mm}
\end{split}
\end{equation}
where the activity indicator $a_k^j\in\{0,1\}$ is equal to one (zero) if the $k$-th device is active (inactive), {\color{black}${\bf{e}}_{k,n}^j\in\{0,1\}^{I\times 1}$, ${\bf h}_{k,n}^j\in\mathbb{C}^{M\times 1}$, and ${\bf X}_{k,n}^j=a_k{\bf e}_{k,n}^j({\bf h}_{k,n}^j)^T\in\mathbb{C}^{I\times M}$ denote the sequence selection vector, the channel vector, and the equivalent channel matrix of the $k$-th device in the $j$-th sub-frame and the $n$-th subcarrier, respectively}, ${\bf N}_n^j$ is the Gaussian noise whose components are i.i.d. complex Gaussian random variables, i.e., ${\cal CN}(0,\sigma_n^2)$, ${\bf \Phi}=[{\bf \Phi}_1,...,{\bf \Phi}_K]\in \mathbb{C}^{L\times KI}$, and ${\bf X}_{n}^j=[({\bf X}_{1,n}^j)^T,...,({\bf X}_{K,n}^j)^T]^T\in \mathbb{C}^{KI\times M}$. {\color{black}Note that only one element of ${\bf e}_{k,n}^j$ is one and the others are equal to zero, i.e., $\| {\bf{e}}_{k,n}^j\|_0=1$ and $\| {\bf{e}}_{k,n}^j\|_2=1$.}

\begin{figure*}[t]
\vspace{-10mm}
\centering
\subfigure[]{
    \begin{minipage}[t]{0.48\linewidth}
        \centering
\label{fig:sparsity}
        \includegraphics[width = 0.7\columnwidth,keepaspectratio]
        {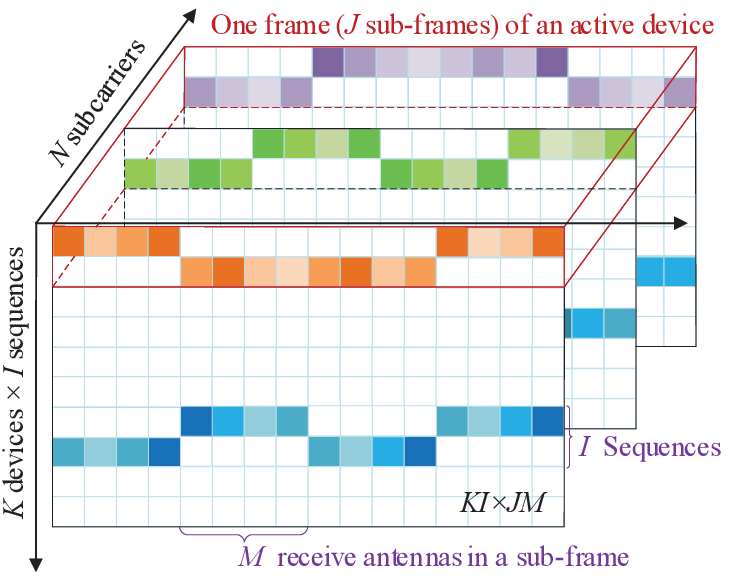}\\
    \end{minipage}%
}%
\subfigure[]{
    \begin{minipage}[t]{0.48\linewidth}
        \centering
\label{fig:FactorGraph}
        \includegraphics[width = 0.7\columnwidth,keepaspectratio]
        {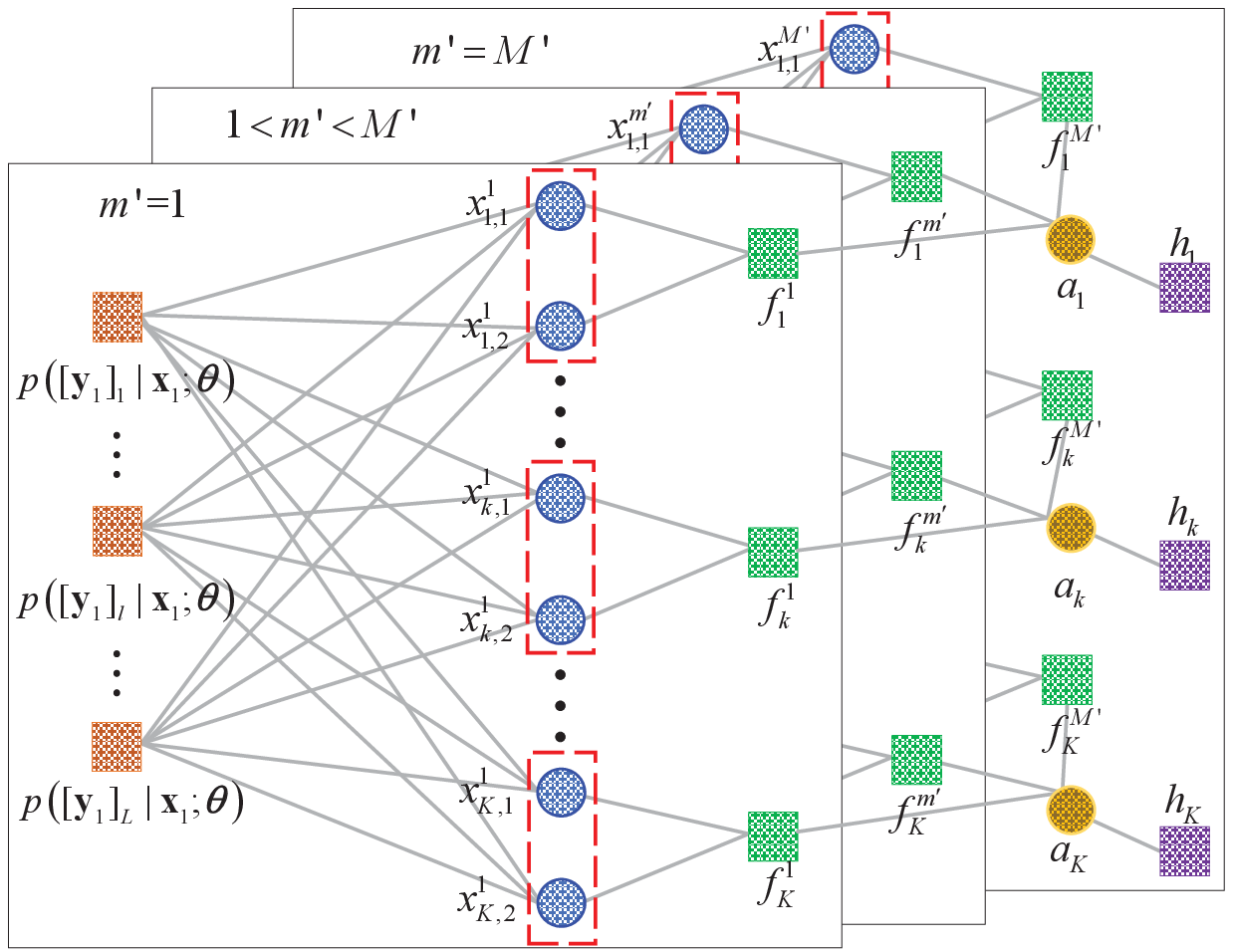}\\
    \end{minipage}%
}%
\centering
\setlength{\abovecaptionskip}{-1mm}
\captionsetup{font={footnotesize}, singlelinecheck = off, justification = raggedright,name={Fig.}, labelsep=period}
\caption{(a) Schematic of the equivalent channel matrix ${\bf{X}}_n$, $1 \le n \le N$, of NC-IM-based massive IoT access over multiple subcarriers within a frame; (b) Factor graph of the joint posterior distribution $p\left({\bf x}_{m'}| {\bf{y}}_{m'},{\bf a};{\bm \theta}\right)$, where $f_k^{m'}=f\left({\bf x}_k^{m'} |a_k;{\bm \theta}\right)$ and $h_k=h\left( a_k;{\bm \theta} \right)$, $\forall k$, $\forall m'$. The circles represent variable nodes and the squares represent factor nodes. }
\label{fig:Lsce1}
\vspace{-6mm}
\end{figure*}

Given any frame (with a total $J$ sub-frames), the received signal of the $n$-th subcarrier ${\bf Y}_n\in\mathbb{C}^{L\times JM}$ can be further expressed in a compact form, i.e.,
\vspace{-1mm}
\begin{equation}\label{eq:SMod2}
\begin{array}{l}
{\bf{Y}}_n={\bf \Phi}{\bf{X}}_n +{\bf{N}}_n,
\vspace{-1mm}
\end{array}
\end{equation}
where ${\bf{Y}}_n=[{\bf Y}_n^1,{\bf Y}_n^2,...,{\bf Y}_n^J]\in\mathbb{C}^{L\times JM}$, ${\bf{X}}_n=[{\bf X}_{n}^1,{\bf X}_{n}^2,...,{\bf X}_{n}^J]\in \mathbb{C}^{KI\times JM}$, and ${\bf{N}}_n=[{\bf N}_n^1,{\bf N}_n^2,...,{\bf N}_n^J]\in \mathbb{C}^{L\times JM}$. Furthermore, the equivalent channel matrix ${\bf{X}}_n$ of a frame, $1 \le n \le N$, is indicated in Fig. \ref{fig:sparsity}. {\color{black}Hence, the target of the massive IoT access, i.e., the DAD-EID problem, focuses on the estimation of the non-zero indices of ${\bf{X}}_n$ based on ${\bf{Y}}_n$, where the estimation of the exact values of ${\bf{X}}_n$ is unnecessary.}

\section{Proposed STF-JABID Algorithm for Aerial BS with Small-Scale MIMO}%
{\color{black}For rotary-wing UAV BSs equipped with small-scale MIMO, the DAD-EID performance of conventional algorithms in \cite{NC-IM-Tcom,NC-IM-Access} is limited by the small number of receive antennas. Hence, how to further improve the DAD-EID performance in small-scale MIMO systems is of great concern.} In this section, we exploit the common device activity in the space-time-frequency domain to further improve the DAD-EID performance. In particular, we first introduce the {\it space-time-frequency structured sparsity} of the equivalent channel matrix to facilitate the signal processing design. Then, a message passing algorithm on a factor graph is derived by leveraging the {\it space-time-frequency structured sparsity}. Furthermore, the unknown parameters are iteratively estimated under the EM framework, and the STF-JABID algorithm is proposed for further improving the DAD-EID performance. 

\subsection{Problem Formulation}
Due to the intermittent traffic of IoT devices, only a fraction of devices are active within a frame, i.e., $K_a\ll K$, which can be denoted as
\begin{equation}\label{eq:SF2}
\begin{array}{l}
\left\vert{\rm supp}\left\{{\bf X}_{n}^j \right\}\right\vert_c=K_a\ll K,~~ \forall n\in[N], ~\forall j\in[J],
\vspace{-1mm}
\end{array}
\end{equation}
where ${\rm supp}\left\{\cdot\right\}$ denotes the support (i.e., the indices of non-zero rows) of its argument.  

Furthermore, owing to the same sparsity of the sequence selection vector observed at multiple receive antennas within a sub-frame, the structured sparsity in the spatial domain can be expressed as
\begin{equation}\label{eq:SF1}
\begin{array}{l}
{\rm supp}\left\{{\bf{e}}_{k,n}^j \right\}={\rm supp}\left\{\left[{\bf X}_{k,n}^j\right]_{:,m} \right\},~~ \forall m\in[M].
\vspace{-1mm}
\end{array}
\end{equation}

Then, due to the common device activity in multiple sub-frames and multiple subcarriers, the structured sparsity in the time-frequency domain can be expressed by
\begin{equation}\label{eq:SF3}
\begin{array}{l}
\left\lceil{\rm supp}\left\{{\bf X}_{1}^1 \right\}/I\right\rceil=\left\lceil{\rm supp}\left\{{\bf X}_{n}^j \right\}/I\right\rceil,~~ \forall n\in[N], ~\forall j\in[J].
\vspace{-1mm}
\end{array}
\end{equation}

Accordingly, we collectively refer to (\ref{eq:SF1}) and (\ref{eq:SF3}) as the {\it space-time-frequency structured sparsity}, which is also illustrated in Fig. \ref{fig:sparsity}. 

To exploit the {\it space-time-frequency structured sparsity} in $\widetilde{N}$ ($1<\widetilde{N}\le N$) contiguous subcarriers and $J$ sub-frames (one frame), we can rewrite (\ref{eq:SMod2}) in a compact matrix form, i.e.,
\begin{equation}\label{eq:Sce1SMod}
\begin{array}{l}
{\bf Y}={\bf \Phi}{\bf{X}} +{\bf{N}},
\vspace{-1mm}
\end{array}
\end{equation}
where ${\bf{Y}}\!=\!\!\left[ {\bf{Y}}_1,\!\!{\bf{Y}}_2,\!...,\!\!{\bf{Y}}_{\widetilde{N}}\right]\!\!\in\!\mathbb{C}^{L\times JM\widetilde{N}}$, ${\bf{X}}\!\!=\!\!\left[ {\bf{X}}_1,\!{\bf{X}}_2,\!...,\!{\bf{X}}_{\widetilde{N}}\right]\!\!\in\!\! \mathbb{C}^{KI\times JM\widetilde{N}}$, and ${\bf{N}}\!\!=\!\!\left[ {\bf{N}}_1,{\bf{N}}_2,...,{\bf{N}}_{\widetilde{N}}\right]\!\!\in\!\! \mathbb{C}^{KI\times JM\widetilde{N}}$. Define $M'=JM\widetilde{N}$ and the $m'$-th column of ${\bf{X}}$, $\forall m'\!\!\in\!\![M']$, is denoted as ${\bf{x}}_{m'}$, where ${\bf{x}}_{m'}=[({\bf{x}}_{1}^{m'})^T,({\bf{x}}_{2}^{m'})^T,...,({\bf{x}}_{K}^{m'})^T]^T\in \mathbb{C}^{KI\times 1}$ and ${\bf{x}}_{k}^{m'}=[x_{k,1}^{m'},...,x_{k,I}^{m'}]^T\in \mathbb{C}^{I\times 1}$, $\forall m', k$. Similarly, we denote the $m'$-th column of ${\bf{Y}}$ and ${\bf{N}}$ as ${\bf{y}}_{m'}$ and ${\bf{n}}_{m'}$, respectively. {\color{black}Note that the sequence selection vectors of different sub-frames (time domain) and different sub-carriers (frequency domain) can be varied according to their embedded information. Hence, employing the {\it space-time-frequency structured sparsity} of ${\bf X}$ in (\ref{eq:Sce1SMod}) is more complicated than utilizing the conventional MMV property, which has been widely used in massive access problems \cite{JSAC-Editor}. }

To solve the DAD-EID problem of (\ref{eq:Sce1SMod}), we aim to minimize the mean square error between ${\bf Y}$ and ${\bf \Phi X}$, which is equivalent to calculating the posterior mean of ${\bf X}$ under the Bayesian framework.
In (\ref{eq:Sce1SMod}), the posterior mean of ${x}_{k,i}^{m'}$, $\forall k\in [K]$, $\forall i\in [I]$, $\forall m'\in[M']$, can be expressed as
\begin{align}\label{eq:Pmean}
\widehat{x}_{k,i}^{m'}= {\int} x_{k,i}^{m'} p\left(x_{k,i}^{m'}| {\bf{y}}_{m'}\right)dx_{k,i}^{m'},
\end{align}
where $p(x_{k,i}^{m'}| {\bf{y}}_{m'})$ is the marginal distribution of $p({\bf x}_{m'}| {\bf{y}}_{m'})$ and it can be written as
\begin{equation}\label{eq:Marginal}
\begin{array}{l}
p\left(x_{k,i}^{m'}| {\bf{y}}_{m'}\right)= \mathlarger{\int}_{\backslash x_{k,i}^{m'}} p\left({\bf x}_{m'}| {\bf{y}}_{m'}\right).
\vspace{-1mm}
\end{array}
\end{equation}

Based on the Bayes’s theorem \cite{Kay}, the posterior distribution $p({\bf x}_{m'}| {\bf{y}}_{m'})$ can be expressed as
\begin{equation}
\begin{split}\label{eq:Baysian}
&p\left({\bf x}_{m'}| {\bf{y}}_{m'},{\bf a};{\bm \theta}\right)\!\!=\!\!\frac{1}{{p\left({\bf{y}}_{m'}\right)}}\!{p\left({\bf{y}}_{m'}|{\bf x}_{m'};{\bm \theta}\right)\!\!f\left({\bf x}_{m'} |{\bf a};{\bm \theta}\right)\!\!h({\bf a};{\bm \theta})}\\
&=\dfrac{1}{p\left({\bf{y}}_{m'}\right)}\prod\limits_{l=1}^L p\left([{\bf{y}}_{m'}]_l|{\bf x}_{m'};{\bm \theta}\right)\prod\limits_{k=1}^K f\left({\bf x}_k^{m'} |a_k;{\bm \theta}\right)h(a_k;{\bm \theta}),
\vspace{-1mm}
\end{split}
\end{equation}
where the likelihood function is denoted as
\begin{equation}\label{eq:likelihood}
\begin{array}{l}
p\left([{\bf{y}}_{m'}]_l|{\bf x}_{m'};{\bm \theta}\right)\!\!=\!\!\dfrac{1}{\pi \sigma_n^2}{\rm exp}\left( \!\!\!-\!\dfrac{1}{\sigma_n^2}\left| [{\bf{y}}_{m'}]_l\!\!-\!\!\!\!\sum\limits_{k=1}^K\!\!\left[ {\bf \Phi}_k {\bf x}_k^{m'}\right]_l \right|^2\!\right).
\vspace{-1mm}
\end{array}
\end{equation}

By leveraging the {\it space-time-frequency structured sparsity}, the prior distribution of ${\bf x}_k^{m'}$, $\forall k, m'$, can be expressed as
\begin{align}\label{eq:prior}
f\left({\bf x}_k^{m'} |a_k;{\bm \theta}\right)&=(1-a_k)\sum\nolimits_{i=1}^{I}\delta\left( x_{k,i}^{m'}\right)\nonumber \\
&+a_k\sum\nolimits_{i=1}^{I}{\cal CN}\left(x_{k,i}^{m'};\mu_0,\tau_0 \right)\prod\nolimits_{g\neq i} \delta\left( x_{k,g}^{m'}\right),
\vspace{-1mm}
\end{align}
where the distribution of the activity indicator is denoted as
\begin{equation}\label{eq:ACTprior}
\begin{array}{l}
h\left( a_k;{\bm \theta} \right)=(1-\lambda_k)\delta\left( a_k \right)+\lambda_k\delta\left( a_k-1 \right),
\vspace{-1mm}
\end{array}
\end{equation}
the parameter set is denoted as ${\bm \theta}=\{\mu_0,\!\tau_0, \!\sigma_n^2,\! \lambda_1,\!...,\!\lambda_K \}$ , ${\bf a}\!=\!\![a_1,a_2,...,a_K]^T$, and $\delta(\cdot)$ is the Dirac delta function. The joint posterior distribution $p\left({\bf X}| {\bf{Y}}\right)=\prod\nolimits_{m'=1}^{M'}p\left({\bf x}_{m'}| {\bf{y}}_{m'},{\bf a};{\bm \theta}\right)$ is presented on the factor graph as illustrated in Fig. \ref{fig:FactorGraph}.

\subsection{Message Passing on the Factor Graph}
{\color{black}Calculating the marginal distribution $p\left(x_{k,i}^{m'}| {\bf{y}}_{m'}\right)$ in (\ref{eq:Marginal}) from $p\left({\bf x}_{m'}| {\bf{y}}_{m'}\right)$ is extremely challenging in massive IoT access, due to the very large number of IoT devices $K$. To address this issue, the AMP algorithm can be adopted to obtain the approximate marginal distributions with relatively low complexity \cite{AMP,yuwei18,kemalong20}. Furthermore, to exploit the {\it space-time-frequency structured sparsity}, we perform message passing on the rest of the factor graph, where the prior distributions (\ref{eq:prior}) and (\ref{eq:ACTprior}) are considered.}


According to the AMP algorithm, we can approximately decouple (\ref{eq:Sce1SMod}) into $KIM'$ ($M'=JM\widetilde{N}$) scalar problems as
\vspace{-2mm}
\begin{equation}\label{eq:Decoupling}
\begin{array}{l}
{\bf Y}={\bf \Phi}{\bf{X}} +{\bf{N}}\; \to \;r_{k,i}^{m'} = x_{k,i}^{m'} + {\widetilde n_{k,i}^{m'}},
\vspace{-2mm}
\end{array}
\end{equation}
where $k\in[K]$, $i\in [I]$, $m'\in [M']$, $r_{k,i}^{m'}$ is the mean of $x_{k,i}^{m'}$ that is estimated by the AMP algorithm, and ${\widetilde n_{k,i}^{m'}}\sim {\cal CN}({\widetilde n_{k,i}^{m'}};0,\varphi_{k,i}^{m'})$ is the associated noise with zero mean and variance $\varphi _{k,i}^{m'}$ \cite{AMPmeng}. 

Specifically, in the $t$-th AMP iteration, $r_{k,i}^{m'}$ and $\varphi_{k,i}^{m'}$ are updated on the variable nodes $\{ x_{k,i}^{m'}\}$,  $\forall k,i,m'$, which can be denoted as
\begin{align}
\label{eq:UpdateSigma} \varphi_{k,i}^{m'}&=\left({\sum\nolimits_{l=1}^{L}\dfrac{\left|[{\bf \Phi}_k]_{l,i}\right|^2}{\sigma_n^2+V_{l}^{m'}}}\right)^{-1},
\end{align}
\vspace{-3mm}
\begin{align}
\label{eq:UpdateR} r_{k,i}^{m'}&={\widehat{{x}}_{k,i}^{m'}}+\varphi_{k,i}^{m'}\sum\nolimits_{l=1}^{L}\dfrac{[{\bf \Phi}^*_k]_{l,i}\left({\left[{\bf y}_{m'}\right]_l-Z_{l}^{m'}}\right)}{\sigma_n^2+V_{l}^{m'}},
\end{align}
where $V_{l}^{m'}$ and $Z_{l}^{m'}$, $\forall l,m'$, are updated at the factor nodes $\{ p\left( [{\bf y}_{m'}]_l|{\bf x}_{m'};{\bm \theta}\right)\}$ of the factor graph as
\begin{align}
\label{eq:UpdateV} V_{l}^{m'}&=\sum\nolimits_{k=1}^K\left|[{\bf \Phi}_k]_{l,:}\right|^2 \widehat{{\bf v}}_{k}^{m'},\\
\label{eq:UpdateZ} Z_{l}^{m'}&=\sum\nolimits_{k=1}^K [{\bf \Phi}_k]_{l,:}\widehat{{\bf x}}_{k}^{m'}-V_{l}^{m'}\dfrac{\left[{\bf y}_{m'}\right]_l-\left(Z_{l}^{m'}\right)^{t-1}}{\sigma_n^2+\left(V_{l}^{m'}\right)^{t-1}},
\end{align}
$\widehat{\bf{x}}_{k}^{m'}=[{\widehat x}_{k,1}^{m'},...,{\widehat x}_{k,I}^{m'}]^T$ and $\widehat{\bf{v}}_{k}^{m'}=[{\widehat v}_{k,1}^{m'},...,{\widehat v}_{k,I}^{m'}]^T$ are the posterior mean and posterior variance,
 respectively, $(Z_{l}^{m'})^{t-1}$ and $(V_{l}^{m'})^{t-1}$ correspond to the $(t-1)$-th iteration, and the index $t$ is omitted for simplicity. Detailed derivations of the AMP update rules in (\ref{eq:UpdateSigma})$-$(\ref{eq:UpdateZ}), i.e., the decoupling step of AMP, we invite the interested readers to consult \cite{AMPmeng}.

Then, we perform the sum-product algorithm on the rest of the factor graph to acquire the posterior distribution of $x_{k,i}^{m'}$ \cite{FactorGraph}. Specifically, we start with the message through the path $\{x_{k,i}^{m'}\} \to \{f_k^{m'}\} \to \{a_k\}$. Furthermore, we calculate the backward message through the path $\{h_k\}\to \{a_k\} \to \{f_k^{m'}\} \to \{x_{k,i}^{m'}\}$. {\color{black}By using the sum-product rule, the prior information, i.e., the {\it space-time-frequency structured sparsity} can be fully exploited. }

{\color{black}Furthermore, by considering the messages on the whole factor graph, we can get the approximate posterior distribution of $x_{k,i}^{m'}$, $\forall k, i, m'$, as}
\begin{equation}\label{eq:post}
\begin{split}
q(x_{k,i}^{m'}| {\bf{y}}_{m'})=(1-\pi_{k,i}^{m'})\delta(x_{k,i}^{m'})+\pi_{k,i}^{m'}{\cal CN}\left(x_{k,i}^{m'};\overline{\mu}_{k,i}^{m'},\overline{\tau}_{k,i}^{m'}\right),
\end{split}
\end{equation}
where we have
\begin{align}
\label{eq:post1}\overline{\mu}_{k,i}^{m'}&=\left(\mu_0 \varphi_{k,i}^{m'}+\tau_0 r_{k,i}^{m'}\right) \big / \left(\varphi_{k,i}^{m'}+r_{k,i}^{m'}\right),\\
\label{eq:post2}\overline{\tau}_{k,i}^{m'}&=\left(\tau_0\varphi_{k,i}^{m'}\right) \big/ \left(\varphi_{k,i}^{m'}+r_{k,i}^{m'}\right),\\
\label{eq:post3}{\cal L}_{k,i}^{m'}&={\rm ln}\dfrac{\varphi_{k,i}^{m'}}{\tau_0+\varphi_{k,i}^{m'}}-\dfrac{\left( r_{k,i}^{m'}-\mu_0\right)^2}{\tau_0+\varphi_{k,i}^{m'}}+\dfrac{\left| {r_{k,i}^{m'}} \right|^2}{\varphi_{k,i}^{m'}},\\
\label{eq:post4}\pi_{k,i}^{m'}&=\dfrac{\lambda_k{\rm exp}({\cal L}_{k,i}^{m'})}{\sum\limits_{i=1}^{I}{\rm exp}({\cal L}_{k,i}^{m'})\left[ \lambda_k+I^{M'}(1-\lambda_k)\prod\limits_{m'=1}^{M'}{}\dfrac{1}{\sum\limits_{i=1}^{I}{\rm exp}({\cal L}_{k,i}^{m'})}\right]}.
\end{align}
%
The details of the derivations of (\ref{eq:post})-(\ref{eq:post4}) can be found in Appendix A.

Finally, the posterior mean and variance of $x_{k,i}^{m'}$, $\forall k,i,m'$, are, respectively, denoted as
\begin{align}
\label{eq:postmean}{\widehat x}_{k,i}^{m'}&={\int}x_{k,i}^{m'}q(x_{k,i}^{m'}| {\bf{y}}_{m'})d x_{k,i}^{m'}=\pi_{k,i}^{m'}\overline{\mu}_{k,i}^{m'},
\end{align}
\vspace{-5mm}
\begin{align}
\label{eq:postvar}{\widehat v}_{k,i}^{m'}&={\int}\left(x_{k,i}^{m'}-{\widehat x}_{k,i}^{m'}\right)^2q(x_{k,i}^{m'}| {\bf{y}}_{m'})d x_{k,i}^{m'}\nonumber\\
&=\pi_{k,i}^{m'}\left( \left| \overline{\mu}_{k,i}^{m'}\right|^2+\overline{\tau}_{k,i}^{m'}\right)-\left( {\widehat x}_{k,i}^{m'}\right)^2.
\end{align}
{\color{black}By using (\ref{eq:postmean}) and (\ref{eq:postvar}) as the denoising step of the AMP scheme, near minimum mean square error (MMSE) reconstruction performance can be achieved in massive IoT access\cite{Meng-JSAC}.}

\subsection{Parameters Estimation}

It is noteworthy that the unknown parameters, denoted as ${\bm \theta}=\{\mu_0,\tau_0, \sigma_n^2, \lambda_1,...,\lambda_K \}$, can be obtained by using the EM algorithm. The EM algorithm updates the parameter set ${\bm \theta}$ as follows
\begin{align}
\label{eq:EM1} Q\left({{\bm \theta},{\bm \theta}^t}\right)&=\mathbb{E}\left\{{{\rm ln}~p\left({{\bf X, Y;{\bm \theta}}}\right)|{\bf Y};{\bm \theta}^t}\right\},\\
\label{eq:EM2} {\bm \theta}^{t+1}&={\rm arg}\max\limits_{{\bm \theta}}Q\left({{\bm \theta},{\bm \theta}^t}\right),
\end{align}
where ${\bm \theta}^t$ denotes the estimated parameters in the $t$-th iteration, $\mathbb{E}\{ \cdot|{\bf Y};{\bm \theta}^t\}$ represents the expectation conditioned on the received signal ${\bf Y}$ under ${\bm \theta}^t$ \cite{EM}. 
{\color{black}Similar to \cite{kemalong20}, we can obtain the EM update rules of the prior variance $\tau_0=\frac{\sum\nolimits_{m'=1}^{M'}\sum\nolimits_{k=1}^{K}\sum\nolimits_{i=1}^{I}\pi_{k,i}^{m'}\left[\left(\mu_0-\overline{\mu}_{k,i}^{m'} \right)^2+\overline{\tau}_{k,i}^{m'} \right]}{\sum\nolimits_{m'=1}^{M'}\sum\nolimits_{k=1}^{K}\sum\nolimits_{i=1}^{I}\pi_{k,i}^{m'}}$ and the prior mean $\mu_0=\frac{\sum\nolimits_{m'=1}^{M'}\sum\nolimits_{k=1}^{K}\sum\nolimits_{i=1}^{I}\pi_{k,i}^{m'}\overline{\mu}_{k,i}^{m'}}{\sum\nolimits_{m'=1}^{M'}\sum\nolimits_{k=1}^{K}\sum\nolimits_{i=1}^{I}\pi_{k,i}^{m'}}$ for the prior distribution in (\ref{eq:prior}). Furthermore, the parameters indicating the devices' activity $\lambda_k$, $\forall k$, can be denoted as
\begin{align}
\label{eq:EMACT}  \lambda_k&=\dfrac{1}{M'}\sum\nolimits_{m'=1}^{M'}{1}\big/\left({1+{1}\big/{\sum\nolimits_{i=1}^{I} \frac{\pi_{k,i}^{m'}}{1-\pi_{k,i}^{m'}}   }}\right),
\end{align}
where the  device activity in the space-time-frequency domain is exploited for improving DAD performance. The derivations of the EM update rules can be found in Appendix B.}

\begin{algorithm}[t]
\algsetup{linenosize=\small} \small
\color{black}
\caption{Proposed STF-JABID Algorithm}\label{Algorithm:1}
\begin{algorithmic}[1]
\raggedright
\REQUIRE The received signals ${\bf{Y}}\!\!=\!\!\left[ {\bf{y}}_1,{\bf{y}}_2,...,{\bf{y}}_{M'}\right]\!\!\in\!\! \mathbb{C}^{L\times M'}$, the pre-allocated sequences ${\bf \Phi}\!\!=\!\![{\bf \Phi}_1,...,{\bf \Phi}_K]\!\!\in\!\! \mathbb{C}^{L\times KI}$, the noise variance ${\sigma_n^2}$, the maximum iteration number $T_0$, the damping parameter $\kappa$, the DAD threshold $T_{h1}$, and the termination threshold $\epsilon$.
\ENSURE The estimated equivalent channel matrix $\widehat{\bf{X}}=[ \widehat{\bf{x}}_1,\widehat{\bf{x}}_2,...,$\\
$\widehat{\bf{x}}_{M'}]\in \mathbb{C}^{KI\times M'}$, the set of active devices $\Omega$, and the support of ${\bf e}_{k,\widetilde{n}}^j$, $\widetilde{n}\!\in\![\widetilde{N}]$, where ${\bf e}_{k,\widetilde{n}}^j$ is defined in (\ref{eq:SMod1}).
\STATE ${\forall k,i,l,m'}$: We initialize the iterative index $t$~$=$~1, the activity indicator $\lambda^1_k=\lambda_0=\frac{L}{KI}\left\{\mathop{\rm max}\limits_{c>0} \frac{1-2KI\left[ \left(1+c^2 \right)\Psi(-c)-c\psi(c)\right]/L}{1+c^2-2\left[\left(1+c^2 \right)\Psi(-c)-c\psi(c) \right]}\right\}$, the prior mean $\mu_0^1=0$, the prior variance $\tau_0^1=\frac{\left\| {\bf{Y}} \right\|_F-L\sigma_n^2}{\left\| {\bf{\Phi}} \right\|_F\lambda_0}$, $(Z^{m'}_{l})^1=\left[{\bf y}_{m'}\right]_l$, $(V^{m'}_{l})^1=1$, $({\widehat x}_{k,i}^{m'})^1=0$, and $({\widehat v}_{k,i}^{m'})^1=1$;
\label{A1:initial}
\FOR {$t=2$ to $ T_0$}
\label{A1:T0}
\STATE \textbf{\textbf{\%}AMP operation:}
\label{A1:AMP-S}
\STATE ${\forall k,i,l,m'}$: Compute $(V^{m'}_{l})^t$, $(Z^{m'}_{l})^t$, $(\varphi_{k,i}^{m'})^t$, and $(r_{k,i}^{m'})^t$ by using (\ref{eq:UpdateV}), (\ref{eq:UpdateZ}), (\ref{eq:UpdateSigma}), and (\ref{eq:UpdateR}), respectively;~~\{Decoupling step\}
\label{A1:decoupling}
\STATE ${\forall l,m'}$: $(V^{m'}_{l})^t=\kappa(V^{m'}_{l})^{t-1}+(1-\kappa)(V^{m'}_{l})^t$, $(Z^{m'}_{l})^t=\kappa(Z^{m'}_{l})^{t-1}+(1-\kappa)(Z^{m'}_{l})^t$;
\label{A1:damping}
\STATE ${\forall k,i,m'}$: Compute $({\widehat x}_{k,i}^{m'})^t$ and $({\widehat v}_{k,i}^{m'})^t$ by using (\ref{eq:postmean}) and (\ref{eq:postvar}), respectively;~~\{Denoising step\}
\label{A1:denoising}
\STATE \textbf{\textbf{\%}EM operation:}
\label{A1:EM-S}
\STATE ${\forall k}$: $(\mu_0)^t=\frac{\sum\nolimits_{k=1}^{K}\sum\nolimits_{i=1}^{I}\sum\nolimits_{m'=1}^{M'}(\pi_{k,i}^{m'})^{t-1}(\overline{\mu}_{k,i}^{m'})^{t-1}}{\sum\nolimits_{k=1}^{K}\sum\limits_{i=1}^{I}\sum\nolimits_{m'=1}^{M'}(\pi_{k,i}^{m'})^{t-1}}$,\\
~$(\tau_0)^t=\frac{\sum\limits_{k=1}^{K}\sum\limits_{i=1}^{I}\sum\nolimits_{m'=1}^{M'}(\pi_{k,i}^{m'})^{t-1}\left[\left((\mu_0)^t-(\overline{\mu}_{k,i}^{m'})^{t-1} \right)^2+(\overline{\tau}_{k,i}^{m'})^{t-1} \right]}{\sum\limits_{k=1}^{K}\sum\limits_{i=1}^{I}\sum\nolimits_{m'=1}^{M'}(\pi_{k,i}^{m'})^{t-1}}$;
\STATE ${\forall k}$: Update devices' activity parameter $\lambda^t_k$ by using (\ref{eq:EMACT});
\label{A1:EM}
\IF{$\left\| \widehat{\bf{X}}^t-\widehat{\bf{X}}^{t-1} \right\|_F/\left\| \widehat{\bf{X}}^{t-1} \right\|_F<\epsilon$}
\label{A1:if}
\STATE ${\bf break}$;~~\{End the iteration\}
\label{A1:break}
\ENDIF
\label{A1:endif}
\ENDFOR
\label{A1:endfor}
\STATE The estimated equivalent channel matrix $\widehat{\bf{X}}=\widehat{\bf{X}}^t$;
\label{A1:Xest}
\STATE \textbf{\textbf{\%}Extract the active devices:}
\STATE ${\forall k\in [K]}$: The set of active devices $\Omega=\{k|\lambda_k^t>T_{h1}\}$;
\label{A1:AUD}
\STATE \textbf{\textbf{\%}Extract the embedded information of active devices:}
\STATE ${\forall k\in\Omega, \widetilde{n}\!\in\![\widetilde{N}], j}$: ${\rm supp}\{ {\bf e}_{k,\widetilde{n}}^j\}\!\!=\!\mathop{\!\!\rm argmax}\limits_{i\in [I]}\left\{\!\! \sum\limits_{m=1}^{M}\left| \left[ \!\widehat{\bf X}_{k,\widetilde{n}}^j\!\right]_{i,m} \right|^2\!\right\}$, where ${\bf X}_{k,\widetilde{n}}^j\!=\!\!a_k{\bf e}_{k,\widetilde{n}}^j({\bf h}_{k,\widetilde{n}})^T$ is denoted in (\ref{eq:SMod1}).
\label{A1:BID}
\end{algorithmic}
\end{algorithm}

\subsection{Proposed STF-JABID Algorithm}

Based on (\ref{eq:UpdateSigma})-(\ref{eq:UpdateZ}), (\ref{eq:post1})-(\ref{eq:postvar}), and (\ref{eq:EMACT}), we summarize the proposed STF-JABID algorithm in {\bf Algorithm} 1. The details are explained as follows.

In line \ref{A1:initial}, to avoid stucking at a local extremum of the EM algorithm, the activity indicator $\lambda_k^1$, $\forall k$, is initialized to $\frac{L}{KI}\left\{\mathop{\rm max}\limits_{c>0} \frac{1-2KI\left[ \left(1+c^2 \right)\Psi(-c)-c\psi(c)\right]/L}{1+c^2-2\left[\left(1+c^2 \right)\Psi(-c)-c\psi(c) \right]}\right\}$ according to \cite{AMPwusheng},  where $\Psi(-c)$ and $\psi(c)$ denote the cumulative distribution and the probability density function of the standard normal distribution. In addition, the prior variance $\tau_0$ is initialized to $\frac{\left\| {\bf{Y}} \right\|_F-L\sigma_n^2}{\left\| {\bf{\Phi}} \right\|_F\lambda_k}$. The iteration starts in line \ref{A1:T0}. In particular, lines \ref{A1:AMP-S}-\ref{A1:denoising} denote the AMP operation, where $(V^{m'}_{l})^t$, $(Z^{m'}_{l})^t$, $(\varphi_{k,i}^{m'})^t$, and $(r_{k,i}^{m'})^t$, $\forall k,i,m',l$, are calculated according to (\ref{eq:UpdateV}), (\ref{eq:UpdateZ}), (\ref{eq:UpdateSigma}), and (\ref{eq:UpdateR}), respectively, in the $t$-th iteration of the AMP decoupling step (line \ref{A1:decoupling}). In addition, a damping parameter $\kappa$ is
used in line \ref{A1:damping} to prevent the algorithm from diverging\cite{kemalong20}. Furthermore, in the AMP denoising step (line \ref{A1:denoising}), we calculate the posterior mean $({\widehat x}_{k,i}^{m'})^t$ and the associated posterior variance $({\widehat v}_{k,i}^{m'})^t$ of the $t$-th iteration by using (\ref{eq:postmean}) and (\ref{eq:postvar}), respectively. Accordingly, the EM operation updates the hyper-parameters  $\mu_0^t$, $\tau_0^t$, and $\lambda^t_k$ in lines \ref{A1:EM-S}-\ref{A1:EM}. Then, we have $t=t+1$ and the iteration restarts in line \ref{A1:AMP-S} until the maximum iteration number $T_0$ is reached. If line \ref{A1:if} is triggered, i.e., the normalized mean square error (NMSE) between $\widehat{\bf{X}}^t$ and $\widehat{\bf{X}}^{t-1}$ is smaller than the predefined $\epsilon$, we also end the iteration. After the iteration, we acquire the estimated equivalent channel matrix $\widehat{\bf{X}}=\widehat{\bf{X}}^t$ in line \ref{A1:Xest} and the set of active devices $\Omega=\{k|\lambda_k^t>T_{h1}, \forall k\}$ in line \ref{A1:AUD}. Based on the estimated $\widehat{\bf{X}}$, in line \ref{A1:BID}, we can acquire the index of the selected signature sequence, i.e., ${\rm supp}\{ {\bf e}_{k,\widetilde{n}}^j\}$, of the $k$-th device in the $\widetilde{n}$-th subcarrier for the $j$-th time slot, ${\forall k\in\Omega, \widetilde{n}\in[\widetilde{N}], \forall j}$, by selecting the row of $\widehat{\bf X}_{k,\widetilde{n}}^j$ that has the maximum power.  Hence, we acquire the embedded information by exploiting the  device activity observed at multiple receive antennas\footnote{\color{black}For single antenna-based aerial BSs, the EID performance of the proposed STF-JABID algorithm may degrade. The same is true for conventional algorithms in \cite{NC-IM-Tcom, NC-IM-Access}. However, the DAD performance of the proposed STF-JABID algorithm can be improved even in single antenna systems, due to the exploitation of the time-frequency structured sparsity (\ref{eq:SF3}).}. By performing line \ref{A1:AUD} and line \ref{A1:BID}, the DAD-EID process is achieved in a non-coherent approach without the need of explicit CSI. Algorithm complexity will be discussed in the simulation section.

\section{Proposed AE-JABID Algorithm for Aerial BS with LS-MIMO}%
{\color{black}Different from the rotary-wing UAV BSs with small-scale MIMO, the DAD-EID performance of the fixed-wing UAV BSs with LS-MIMO can be significantly improved by using the MMV property observed at a large number of receive antennas \cite{NC-IM-Tcom,NC-IM-Access}. However, how to further improve the DAD-EID performance with reduced access latency under LS-MIMO BSs has never been discussed in the literature.} In this section, we first formulate the grant-free NC-IM-based massive IoT access problem in the virtual angular domain and illustrate the {\it angular-domain structured sparsity} of the equivalent massive access channel matrix. Furthermore, by exploiting the {\it angular-domain structured sparsity}, an AMP-based AE-JABID algorithm is proposed for improving DAD-EID performance and reducing access latency. In addition, a TFST strategy is proposed for overcoming doubly-selective fading channels, also with reduced access latency.

\vspace{-2.6mm}
\subsection{Problem Formulation}
The aerial BS is usually deployed at a high altitude location with few scatterers around, whereas the IoT devices are typically located at low elevation in a rich scattering environment far from the aerial BS. Assuming that the device is located at a distance $R$ from the aerial BS and is surrounded by rich scatterers within a radius of $r$, where $r\ll R$. Hence, the angular spread is within $\vartriangle\approx{\rm arctan}(r/R)$, as shown in Fig. \ref{fig:systemModel-LS-MIMO}. This scenario can be modeled by using the classical one-ring channel model, which leads to the clustered sparsity of LS-MIMO channels in the virtual angular domain \cite{OneRing}. Particularly, the virtual angular-domain LS-MIMO channel associated with the $k$-th device and the $n$-th subcarrier, $\forall k,n$, can be represented as
\begin{equation}\label{eq:AD1}
\begin{array}{l}
\widetilde{\bf h}_{k,n}={\bf A}_R^H{\bf h}_{k,n},
\vspace{-1mm}
\end{array}
\end{equation}
where the transformation matrix ${\bf A}_R\in \mathbb{C}^{M\times M}$ at the aerial BS is a unitary matrix. Here, ${\bf A}_R$ depends on the geometry of the adopted array, and becomes the discrete Fourier transform (DFT) matrix for a ULA with $d=\lambda/2$ \cite{VAD_Channel}. In the LS-MIMO scenario, due to the small angular spread $\vartriangle$, the virtual angular-domain channel vector $\widetilde{\bf h}_{k,n}^j$ has clustered sparsity, $\forall k, n$, i.e.,
\begin{equation}\label{eq:AD2}
\begin{array}{l}
\left\vert{\rm supp}\left\{\widetilde{\bf h}_{k,n} \right\}\right\vert_c \ll M.
\vspace{-1mm}
\end{array}
\end{equation}

Furthermore, the equivalent virtual angular-domain channel matrix of the $k$-th device in the $j$-th sub-frame and $n$-th subcarrier can be expressed as ${\bf W}_{k,n}^j=a_k{\bf e}_{k,n}^j\left(\widetilde{\bf h}_{k,n}\right)^T\in \mathbb{C}^{I\times M}$, and we have ${\bf W}_{n}^j=\left[({\bf W}_{1,n}^j)^T,...,({\bf W}_{K,n}^j)^T\right]^T\in \mathbb{C}^{KI\times M}$. 

Due to the structured sparsity of the sequence selection vector, i.e., $\| {\bf{e}}_{k,n}^j\|_0=1$, $\forall k, n, j$, we have the structured sparsity of the virtual angular-domain channel ${\bf W}_{k,n}^j$ as shown in Fig. \ref{fig:AngleChannel}, denoted as
\begin{equation}\label{eq:AD3}
\begin{array}{l}
{\rm supp}\left\{{\bf{e}}_{k,n}^j \right\}={\rm supp}\left\{{\bf W}_{k,n}^j \right\}.
\vspace{-1mm}
\end{array}
\end{equation}

We collectively refer to the clustered sparsity in (\ref{eq:AD2}) and the structured sparsity in (\ref{eq:AD3}) as the {\it angular-domain structured sparsity}, which is a prominent feature in aerial LS-MIMO BS based massive IoT access. Hence, we focus on the DAD-EID problem in any given sub-frame and subcarrier, and transform the system model (\ref{eq:SMod1}) into the virtual angular domain as
\begin{equation}\label{eq:ADSMod1}
\begin{array}{l}
{\bf R}_n^j={\bf Y}_n^j{\bf A}_R^*={\bf \Phi}{\bf X}_{n}^j{\bf A}_R^*+{\bf N}_n^j{\bf A}_R^*={\bf \Phi}{\bf W}_{n}^j+\overline{\bf N}_n^j,
\end{array}
\end{equation}
where ${\bf A}_R\in \mathbb{C}^{M\times M}$ is the DFT matrix, ${\bf R}_n^j\in \mathbb{C}^{L\times M}$, ${\bf W}_{n}^j\in \mathbb{C}^{KI\times M}$, and $\overline{\bf N}_n^j\in \mathbb{C}^{L\times M}$ are the equivalent received signal, channel matrix, and noise of the $n$-th subcarrier and the $j$-th sub-frame, $\forall n, j$, in the virtual angular domain, respectively. In addition, elements in $\overline{\bf N}_n^j$ are i.i.d. complex Gaussian random variables with zero mean and variance $\sigma_{\overline n}^2$.

For the sake of simplicity, we omit the index of the subcarrier $n$ and the index of the sub-frame $j$, then we have
\begin{equation}\label{eq:ADSMod2}
\begin{array}{l}
{\bf R}={\bf \Phi}{\bf W}+\overline{\bf N},
\end{array}
\end{equation}
with ${\bf W}=\left[({\bf W}_{1})^T,...,({\bf W}_{K})^T\right]^T\in \mathbb{C}^{KI\times M}$ and ${\bf W}_{k}=a_k{\bf e}_{k}\left(\widetilde{\bf h}_{k}\right)^T\in \mathbb{C}^{I\times M}$. We denote the $m$-th column of ${\bf W}_{k}$ as ${\bf{w}}_{k}^{m}=[w_{k,1}^{m},...,w_{k,I}^{m}]^T\in \mathbb{C}^{I\times 1}$, $\forall m\in[M]$ and $\forall k\in[K]$. In addition, we denote the $m$-th column of  ${\bf{R}}$ and ${\bf W}$ as ${\bf r}_m$ and ${\bf w}_m$, respectively.

\vspace{-3.5mm}
\subsection{Proposed AE-JABID Algorithm}
Firstly, we assume that the elements of ${\bf W}$ follow the Bernoulli Gaussian distribution, i.e., 
\begin{equation}\label{eq:ADprior}
\begin{array}{l}
g\left(w_{k,i}^m; \rho_{k,i}^m\right)=(1-\rho_{k,i}^m)\delta\left( w_{k,i}^m\right)+\rho_{k,i}^m{\cal CN}\left( w_{k,i}^m;\widetilde{\mu}_0, \widetilde{\tau}_0 \right),
\end{array}
\end{equation}
where $\rho_{k,i}^m\in \{0,1\}$ is the activity indicator of the element $w_{k,i}^m$, $\forall k, i, m$. 

\begin{figure*}[t]
\vspace{-5mm}
\centering
\subfigure[]{
    \begin{minipage}[t]{0.49\linewidth}
        \centering
\label{fig:AngleChannel}
        \includegraphics[width = 0.73\columnwidth,keepaspectratio]
        {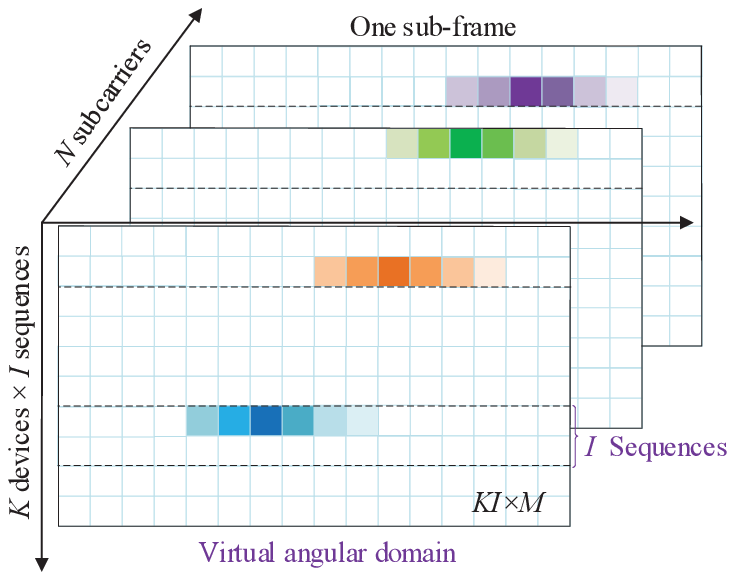}\\
    \end{minipage}%
}%
\subfigure[]{
    \begin{minipage}[t]{0.48\linewidth}
        \centering
\label{fig:FactorGraph2}
        \includegraphics[width = 0.7\columnwidth,keepaspectratio]
        {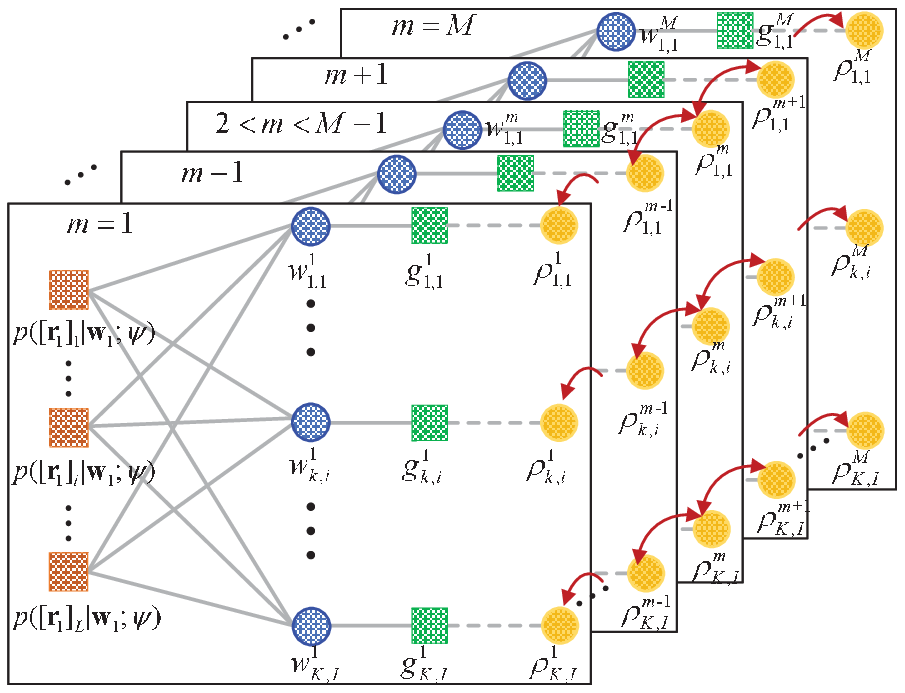}\\
    \end{minipage}%
}%
\centering
\setlength{\abovecaptionskip}{-1mm}
\captionsetup{font={footnotesize}, singlelinecheck = off, justification = raggedright,name={Fig.},labelsep=period}
\caption{(a) Schematic of the equivalent channel matrix for multiple subcarriers within a sub-frame in the virtual angular domain; (b) Factor graph of the joint posterior distribution $p({\bf W}| {\bf{R}})$, where $g_{k,i}^{m}=g\left(w_{k,i}^m; \rho_{k,i}^m\right)$, $\forall k, i, m$. The circles represent variable nodes and the squares represent factor nodes. }
\label{fig:Lsce1}
\vspace{-5mm}
\end{figure*}

Then, based on Bayes's theorem \cite{Kay}, the posterior distribution $p({\bf W}| {\bf{R}})$ is expressed as
\begin{equation}\label{eq:Baysian2}
\begin{split}
p\left({\bf W}| {\bf{R}};{\bm \psi}\right)&=\prod\nolimits_{m=1}^{M}p\left({\bf w}_{m}| {\bf{r}}_{m};{\bm \psi}\right)\\
&=\prod\nolimits_{m=1}^{M}\dfrac{1}{p\left({\bf{r}}_{m}\right)}p\left({\bf{r}}_{m}|{\bf w}_{m};{\bm \psi}\right)p\left({\bf w}_{m};{\bm \psi}\right)\\
&=\prod\limits_{m=1}^{M}\dfrac{1}{p\left({\bf{r}}_{m}\right)}\prod\limits_{l=1}^L p\left([{\bf{r}}_{m}]_l|{\bf w}_{m};{\bm \psi}\right)\\
&~~~~~~~~\times\prod\limits_{k=1}^K\prod\limits_{i=1}^I g\left({w}_{k,i}^{m};\rho_{k,i}^m\right),
\vspace{-1mm}
\end{split}
\end{equation}
where the likelihood function is denoted as
\begin{equation}\label{eq:likelihood2}
\begin{array}{l}
p\left([{\bf{r}}_{m}]_l|{\bf w}_{m};{\bm \psi}\right)\!=\!\dfrac{1}{\pi \sigma_{\overline n}^2}{\rm exp}\left(\!\!-\dfrac{1}{\sigma_{\overline n}^2} \left| [{\bf{r}}_{m}]_l\!\!-\!\!\!\sum\limits_{k=1}^K\!\!\left[ {\bf \Phi}_k {\bf w}_k^{m}\right]_l \right|^2\right),
\vspace{-1mm}
\end{array}
\end{equation}
and the parameter set is ${\bm \psi}=\{\widetilde{\mu}_0, \widetilde{\tau}_0,\sigma_{\overline n}^2,\rho_k, k\in[K] \}$. As shown in Fig. \ref{fig:FactorGraph2}, the joint posterior distribution $p({\bf W}| {\bf{R}})$ is illustrated on the factor graph.

Similar to the AMP decoupling step in (\ref{eq:Decoupling}), by using AMP algorithm, we can approximately decouple (\ref{eq:ADSMod2}) into $KIM$ scalar problems, i.e., 
\begin{equation}\label{eq:Decoupling2}
\begin{array}{l}
{\bf R}={\bf \Phi}{\bf W}+\overline{\bf N}\; \to \;\widetilde{r}_{k,i}^{m} = w_{k,i}^{m} + { \overline{n}_{k,i}^{m}},
\end{array}
\end{equation}
where $k\in[K]$, $i\in [I]$, $m\in [M]$, $\widetilde{r}_{k,i}^{m}$ is the mean of $w_{k,i}^{m}$, and ${\overline n_{k,i}^{m}}\sim {\cal CN}({\overline n_{k,i}^{m}};0,\xi_{k,i}^{m})$ is the associated noise with zero mean and variance $\xi_{k,i}^{m}$. In the $t$-th AMP iteration, $\widetilde{r}_{k,i}^{m}$ and $\xi_{k,i}^{m}$, $\forall k,i,m$, are updated as
\begin{align}
\label{eq:UpdateSigma2} \xi_{k,i}^{m}&=\left({\sum\limits_{l=1}^{L}\dfrac{\left|[{\bf \Phi}_k]_{l,i}\right|^2}{\sigma_{\overline n}^2+{\widetilde V}_{l}^{m}}}\right)^{-1},\\
\label{eq:UpdateR2} {\widetilde r}_{k,i}^{m}&={\widehat{{ w}}_{k,i}^{m}}+\xi_{k,i}^{m}\sum\limits_{l=1}^{L}\dfrac{[{\bf \Phi}^*_k]_{l,i}\left({\left[{\bf r}_{m}\right]_l-{\widetilde Z}_{l}^{m}}\right)}{\sigma_{\overline n}^2+{\widetilde V}_{l}^{m}},
\end{align}
where ${\widetilde V}_{l}^{m}$ and ${\widetilde Z}_{l}^{m}$, $\forall l,m$, are updated as
\begin{align}
\label{eq:UpdateV2} {\widetilde V}_{l}^{m}&=\sum\limits_{k=1}^K\left|[{\bf \Phi}_k]_{l,:}\right|^2 \widehat{{\bf u}}_{k}^{m},\\
\label{eq:UpdateZ2} {\widetilde Z}_{l}^{m}&=\sum\limits_{k=1}^K [{\bf \Phi}_k]_{l,:}\widehat{{\bf w}}_{k}^{m}-{\widetilde V}_{l}^{m}\dfrac{\left[{\bf r}_{m}\right]_l-\left({\widetilde Z}_{l}^{m}\right)^{t-1}}{\sigma_{\overline n}^2+\left({\widetilde V}_{l}^{m}\right)^{t-1}},
\end{align}
$\widehat{\bf{w}}_{k}^{m}=[{\widehat w}_{k,1}^{m},...,{\widehat w}_{k,I}^{m}]^T$ and $\widehat{\bf{u}}_{k}^{m}=[{\widehat u}_{k,1}^{m},...,{\widehat u}_{k,I}^{m}]^T$ are the posterior mean and posterior variance, respectively, $({\widetilde Z}_{l}^{m})^{t-1}$ and $({\widetilde V}_{l}^{m})^{t-1}$ correspond to the $(t-1)$-th iteration,  and the iteration index $t$ is omitted for simplicity. 

Then, according to Bayes's theorem and the AMP algorithm, the posterior distribution can be approximated as
\begin{equation}\label{eq:post2}
\begin{split}
p(w_{k,i}^{m}| {\bf{r}}_{m};{\bm \psi})&\approx q(w_{k,i}^{m}| {\widetilde r}_{k,i}^m;{\bm \psi})\\
&=\frac{1}{ p\left({\widetilde r}_{k,i}^{m} \right)}p\left({\widetilde r}_{k,i}^m| w_{k,i}^{m} ;{\bm \psi}\right) g\left(w_{k,i}^{m}; \rho_{k,i}^m \right)\\
&=(1-{\widetilde \pi}_{k,i}^{m})\delta(w_{k,i}^{m})\!+\!{\widetilde \pi}_{k,i}^{m}{\cal CN}\left(w_{k,i}^{m};\widetilde{\mu}_{k,i}^{m},\widetilde{\tau}_{k,i}^{m}\right),
\end{split}
\end{equation}
where we have
\begin{align}
\label{eq:post1-2}\widetilde{\mu}_{k,i}^{m}&=\left({\widetilde \mu}_0 \xi_{k,i}^{m}+{\widetilde \tau}_0 {\widetilde r}_{k,i}^{m}\right) \big / \left(\xi_{k,i}^{m}+{\widetilde r}_{k,i}^{m}\right),\\
\label{eq:post2-2}\widetilde{\tau}_{k,i}^{m}&=\left({\widetilde \tau}_0\xi_{k,i}^{m}\right) \big /\left(\xi_{k,i}^{m}+{\widetilde r}_{k,i}^{m}\right),\\
\label{eq:post3-2}{\widetilde{\cal L}}_{k,i}^{m}&={\rm ln}\dfrac{\xi_{k,i}^{m}}{{\widetilde \tau}_0+\xi_{k,i}^{m}}-\dfrac{\left( {\widetilde r}_{k,i}^{m}-{\widetilde \mu}_0\right)^2}{{\widetilde \tau}_0+\xi_{k,i}^{m}}+\dfrac{\left| {{\widetilde r}_{k,i}^{m}} \right|^2}{\xi_{k,i}^{m}},\\
\label{eq:post4-2}{\widetilde \pi}_{k,i}^{m}&=\rho_{k,i}^m \big /\left[{\rho_{k,i}^m+(1-\rho_{k,i}^m){\rm exp}\left({{\widetilde{\cal L}}_{k,i}^{m}} \right)}\right].
\end{align}

Furthermore, we can obtain the AMP denoising step, i.e., the posterior mean and variance of $w_{k,i}^{m}$, $\forall k,i,m$, respectively, denoted as
\begin{align}
\label{eq:postmean2}{\widehat w}_{k,i}^{m}&={\int}w_{k,i}^{m}p(w_{k,i}^{m}| {\bf{r}}_{m};{\bm \psi})d w_{k,i}^{m}={\widetilde \pi}_{k,i}^{m}\widetilde{\mu}_{k,i}^{m},\\
\label{eq:postvar2}{\widehat u}_{k,i}^{m}&={\int}\left(w_{k,i}^{m}-{\widehat w}_{k,i}^{m}\right)^2p(w_{k,i}^{m}| {\bf{r}}_{m};{\bm \psi})d w_{k,i}^{m}\nonumber\\
&={\widetilde  \pi}_{k,i}^{m}\left( \left| \widetilde {\mu}_{k,i}^{m}\right|^2+\widetilde{\tau}_{k,i}^{m}\right)-\left( {\widehat w}_{k,i}^{m}\right)^2.
\end{align}


Similarly, by using the EM algorithm, we can obtain the update rules $\widetilde\mu_0=\frac{\sum\nolimits_{m=1}^{M}\sum\nolimits_{k=1}^{K}\sum\nolimits_{i=1}^{I}\widetilde\pi_{k,i}^{m}\overline{\widetilde\mu}_{k,i}^{m}}{\sum\nolimits_{m=1}^{M}\sum\nolimits_{k=1}^{K}\sum\nolimits_{i=1}^{I}\widetilde\pi_{k,i}^{m}}$ and $\widetilde\tau_0=\frac{\sum\nolimits_{m=1}^{M}\sum\nolimits_{k=1}^{K}\sum\nolimits_{i=1}^{I}\widetilde\pi_{k,i}^{m}\left[\left(\widetilde\mu_0-\widetilde{\mu}_{k,i}^{m} \right)^2+\widetilde{\tau}_{k,i}^{m} \right]}{\sum\nolimits_{m=1}^{M}\sum\nolimits_{k=1}^{K}\sum\nolimits_{i=1}^{I}\widetilde\pi_{k,i}^{m}}$ for the prior distribution (\ref{eq:ADprior}). Furthermore, the EM update rule for the activity indicator $\rho_{k,i}^m$, $\forall k, i, m$, is denoted as
\begin{align}
\label{eq:EMACT2} \rho_{k,i}^{m}=\widetilde\pi_{k,i}^{m},
\end{align}
which reveals that $\widetilde\pi_{k,i}^{m}$ in (\ref{eq:post2}) indicates the activity of the scalar variable $w_{k,i}^{m}$.

To exploit the clustered sparsity (\ref{eq:AD2}) in the virtual angular domain, we update $\rho_{k,i}^{m}$ by using its neighboring $2Q$ $(Q>1)$ activity indicators, i.e.,
\begin{align}
\label{eq:ActUpdate}\left( \rho_{k,i}^{m}\right)^{t+1}=\dfrac{1}{\sum\nolimits_{(\widetilde{m},\zeta)\in {\cal Q}} \zeta}\sum\nolimits_{(\widetilde{m},\zeta)\in {\cal Q}}\zeta\left( \rho_{k,i}^{\widetilde{m}}\right)^{t},
\end{align}
where the set ${\cal Q}$ contains the indices of neighboring indicators and their weight coefficients, i.e.,
\begin{equation}
\begin{split}
\label{eq:Naighbors}{\cal Q}&=\{ \left({\rm mod}(m\pm q,M),\zeta_{m\pm q}\right), 1\leq q \leq Q\},
\end{split}
\end{equation}
$\zeta_{m\pm q}\in (0,1]$. If $q_1<q_2$, we set $\zeta_{m\pm q_1}>\zeta_{m\pm q_2}$, i.e., we let the weight coefficient of the closer neighbors larger, to depict the correlation of elements in the virtual angular domain. The aforementioned update rules (\ref{eq:ActUpdate}) and (\ref{eq:Naighbors}) are illustrated in Fig. \ref{fig:FactorGraph2}.

Given any active device, we observe that the maximum channel gain in the virtual angular domain is much larger than that in the spatial domain, which is briefly discussed in {\em Remark 1} and motivates the DAD-EID design in the virtual angular domain. Furthermore, we notice that the estimated activity indicator tends to converge to $1$, if its associated element takes sufficiently large value in the angular domain. Hence, for DAD, we utilize the estimated activity indicators to acquire the set of active devices, i.e.,
\vspace{1mm}
\begin{align}
\label{eq:ACTset} \Omega=\{ k | \mathop{\rm max}\limits_{m\in [M]}\rho_{k,i}^m>T_h, k\in[K], i\in[I]\},
\end{align}
\vspace{1mm}
where $T_h\in(0.5,1)$ is a predefined threshold. In addition, for EID, we obtain the support of the sequence selection vector ${\bf e}_k$ as
\begin{align}
\label{eq:EID}{\rm supp}\{ {\bf e}_{k}\}=\mathop{\rm argmax}\limits_{i\in [I]}\left\{  \mathop{\rm max}\limits_{m\in [M]}(\rho_{k,i}^m)\right\}, ~~\forall k\in\Omega.
\end{align}
\vspace{1mm}

{\em Remark 1:} Given any path, whose AoA is assumed to be on the grid. Then, the maximum channel gain in the angular domain is $M$ times larger than that in the spatial domain, where $M$ is the number of receive antennas. The detailed derivation is given in Appendix C.

%
\begin{algorithm}[t]
\algsetup{linenosize=\small} \small
\color{black}
\caption{Proposed AE-JABID Algorithm}\label{Algorithm:1}
\begin{algorithmic}[1]
\REQUIRE The equivalent received signal ${\bf{R}}\!\!\in\!\! \mathbb{C}^{L\times M}$, the pre-allocated sequences ${\bf \Phi}=[{\bf \Phi}_1,...,{\bf \Phi}_K]\in \mathbb{C}^{L\times KI}$, the noise variance ${\sigma_{\overline n}^2}$, the maximum iteration number $T_0$, the DAD threshold $T_{h2}$, the damping parameter $\kappa$, and the termination threshold $\epsilon$.
\ENSURE The estimated equivalent channel matrix in the virtual angular domain $\widehat{\bf{W}}\!\!=\!\!\left[ \widehat{\bf{w}}_1,\widehat{\bf{w}}_2,...,\widehat{\bf{w}}_{M}\right]\!\!\in\!\! \mathbb{C}^{KI\times M}$, the set of active devices $\Omega$, and the support of ${\bf e}_{k}$, where ${\bf e}_{k}$ is defined in (\ref{eq:ADSMod2}).
\STATE ${\forall k,i,l,m}$: We initialize the iterative index $t$=1, the activity indicator $(\rho^m_{k,i})^1=\rho_0=\frac{L}{KI}\left\{\mathop{\rm max}\limits_{c>0} \frac{1-2KI\left[ \left(1+c^2 \right)\Psi(-c)-c\psi(c)\right]/L}{1+c^2-2\left[\left(1+c^2 \right)\Psi(-c)-c\psi(c) \right]}\right\}$, the prior mean $\widetilde\mu_0^1=0$, the prior variance $\widetilde\tau_0^1=\frac{\left\| {\bf{R}} \right\|_F-L\sigma_{\overline n}^2}{\left\| {\bf{\Phi}} \right\|_F\rho_0}$, $(\widetilde Z^{m}_{l})^1=\left[{\bf r}_{m}\right]_l$, $(\widetilde V^{m}_{l})^1=1$, $({\widehat w}_{k,i}^{m})^1=0$, and $({\widehat u}_{k,i}^{m})^1=1$;
\label{A2:initial}
\FOR {$t=2$ to $ T_0$}
\label{A2:T0}
\STATE \textbf{\textbf{\%}AMP operation:}
\label{A2:AMP-S}
\STATE ${\forall k,i,l,m}$: Compute $(\widetilde V^{m}_{l})^t$, $(\widetilde Z^{m}_{l})^t$, $(\xi_{k,i}^{m})^t$, and $(\widetilde r_{k,i}^{m})^t$ by using (\ref{eq:UpdateV2}), (\ref{eq:UpdateZ2}), (\ref{eq:UpdateSigma2}), and (\ref{eq:UpdateR2}), respectively;~~\{Decoupling step\}
\label{A2:decoupling}
\STATE ${\forall l,m'}$: $(\widetilde V^{m}_{l})^t=\kappa(\widetilde V^{m}_{l})^{t-1}+(1-\kappa)(\widetilde V^{m}_{l})^t$,\\
 $(\widetilde Z^{m}_{l})^t=\kappa(\widetilde Z^{m}_{l})^{t-1}+(1-\kappa)(\widetilde Z^{m}_{l})^t$;
\label{A2:damping}
\STATE ${\forall k,i,m}$: Compute $({\widehat w}_{k,i}^{m})^t$ and $({\widehat u}_{k,i}^{m})^t$ by using (\ref{eq:postmean2}) and (\ref{eq:postvar2}), respectively;~~\{Denoising step\}
\label{A2:denoising}
\STATE \textbf{\textbf{\%}EM operation:}
\label{A2:EM-S}
\STATE ${\forall k}$: $(\widetilde\mu_0)^t=\frac{\sum\nolimits_{k=1}^{K}\sum\nolimits_{i=1}^{I}\sum\nolimits_{m=1}^{M}(\widetilde\pi_{k,i}^{m})^{t-1}(\overline{\widetilde\mu}_{k,i}^{m})^{t-1}}{\sum\limits_{k=1}^{K}\sum\limits_{i=1}^{I}\sum\limits_{m=1}^{M}(\widetilde\pi_{k,i}^{m})^{t-1}}$,\\
~~~$(\widetilde\tau_0)^t=\frac{\sum\limits_{k=1}^{K}\sum\limits_{i=1}^{I}\sum\limits_{m=1}^{M}(\widetilde\pi_{k,i}^{m})^{t-1}\left[\left((\widetilde\mu_0)^t-(\widetilde{\mu}_{k,i}^{m})^{t-1} \right)^2+(\widetilde{\tau}_{k,i}^{m})^{t-1} \right]}{\sum\limits_{k=1}^{K}\sum\limits_{i=1}^{I}\sum\limits_{m=1}^{M}(\widetilde\pi_{k,i}^{m})^{t-1}}$;
\label{A2:EMmeanvar}
\STATE ${\forall k, i, m}$: Update the activity indicator $(\rho^m_{k,i})^t$ by using (\ref{eq:EMACT2}) and (\ref{eq:ActUpdate});
\label{A2:EMactivity}
\IF{$\left\| \widehat{\bf{W}}^t-\widehat{\bf{W}}^{t-1} \right\|_F/\left\| \widehat{\bf{W}}^{t-1} \right\|_F<\epsilon$}
\label{A2:IF}
\STATE ${\bf break}$;~~\{End the iteration\}
\label{A2:Break}
\ENDIF
\label{A2:ENDIF}
\ENDFOR
\label{A2:endfor}
\STATE The estimated equivalent channel matrix $\widehat{\bf{W}}=\widehat{\bf{W}}^t$;
\label{A2:Xest}
\STATE \textbf{\textbf{\%}Extract the active devices:}
\STATE ${\forall k,i,m}$: Acquire the set of active devices $\Omega$ by using (\ref{eq:ACTset}) based on $(\rho_{k,i}^m)^t$;
\label{A2:AUD}
\STATE \textbf{\textbf{\%}Extract the embedded information of active devices:}
\STATE ${\forall k\in\Omega}$: Acquire the support of the sequence selection vector by using (\ref{eq:EID}) based on $(\rho_{k,i}^m)^t$.
\label{A2:BID}
\end{algorithmic}
\end{algorithm}

Finally, based on (\ref{eq:UpdateSigma2})-(\ref{eq:UpdateZ2}), (\ref{eq:post1-2})-(\ref{eq:postvar2}), and (\ref{eq:EMACT2})-(\ref{eq:ACTset}), we summarize the proposed AE-JABID algorithm in {\bf Algorithm} 2. The details are explained as follows. In line \ref{A1:initial}, the hyper-parameters are initialized in the same way as for Algorithm 1. The iteration containing the AMP operation and the EM operation starts in line \ref{A2:T0}. Specifically, lines \ref{A2:AMP-S}-\ref{A2:denoising} denote the AMP operation, where $(\widetilde V^{m}_{l})^t$, $(\widetilde Z^{m}_{l})^t$, $(\xi_{k,i}^{m})^t$, and $(\widetilde r_{k,i}^{m})^t$, $\forall k,i,m,l$, are calculated according to (\ref{eq:UpdateV2}), (\ref{eq:UpdateZ2}), (\ref{eq:UpdateSigma2}), and (\ref{eq:UpdateR2}), respectively, in the $t$-th iteration of the AMP decoupling step (line \ref{A2:decoupling}). In line \ref{A2:damping}, the damping technique is also used to prevent the algorithm from diverging. Then, in the AMP denoising step of line \ref{A2:denoising}, we calculate the posterior mean $({\widehat w}_{k,i}^{m})^t$ and the associated posterior variance $({\widehat u}_{k,i}^{m})^t$ of the $t$-th iteration by using (\ref{eq:postmean2}) and (\ref{eq:postvar2}), respectively. Furthermore, $\widetilde\mu_0^t$ and $\widetilde\tau_0^t$ are updated in line \ref{A2:EMmeanvar}. The activity indicator is updated by using (\ref{eq:EMACT2}) and (\ref{eq:ActUpdate}) in line \ref{A2:EMactivity}.  
Then, we have $t=t+1$ and the iteration restarts in line \ref{A2:AMP-S} until the maximum iteration number $T_0$ is achieved. If line \ref{A2:IF} is satisfied, i.e., the NMSE between $\widehat{\bf{W}}^t$ and $\widehat{\bf{W}}^{t-1}$ is smaller than the predefined $\epsilon$, the iteration stops. After that, we can acquire the estimated equivalent channel matrix in the virtual angular domain as $\widehat{\bf{W}}=\widehat{\bf{W}}^t$ in line \ref{A2:Xest} and the set of active devices $\Omega$ in line \ref{A1:AUD} according to (\ref{eq:ACTset}). In line \ref{A2:BID}, we can acquire the embedded information of active devices by using (\ref{eq:EID}).


\subsection{Proposed TFST Strategy for NC-IM to Overcome Doubly-Selective Fading Channels}
\begin{figure*}[t]
     \centering
     \includegraphics[width = 1.55 \columnwidth,keepaspectratio]
     {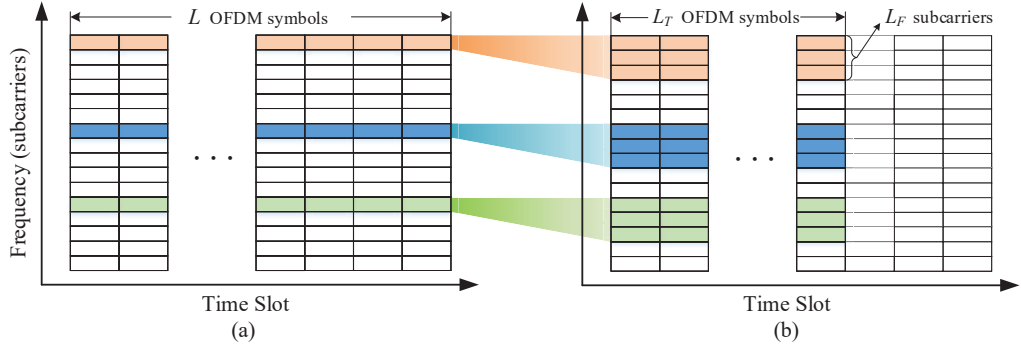}
     \captionsetup{font={footnotesize}, singlelinecheck = off, justification = raggedright,name={Fig.},labelsep=period}
     \caption{The time-frequency resources allocated for signature sequences in grant-free NC-IM-based massive IoT access: (a) Each sequence is transmitted in a single subcarrier within $L$ OFDM symbols; (b) Proposed TFST strategy, where each sequence is transmitted in $L_F$ successive subcarriers within $L_T$ OFDM symbols and $L=L_FL_T$.}
     \label{fig:spreading}
    \vspace{-3mm}
\end{figure*}

In the aforementioned signal models, i.e., (\ref{eq:Sce1SMod}) and (\ref{eq:ADSMod2}), an $L$-length signature sequence of NC-IM is transmitted using one single subcarrier in $L$ continuous OFDM symbols (a sub-frame), as shown in Fig. \ref{fig:spreading}(a). In this subsection, the high correlation among successive OFDM subcarriers within the coherence bandwidth motivates us to spread the signature sequence in both time and frequency for reduced access latency, which is particularly suitable for the doubly-selective fading channels with limited coherence time. 

Specifically, as shown in Fig. \ref{fig:spreading}(b), in the proposed TFST strategy, one $L$-length signature sequence spreads in $L_F$ subcarriers and $L_T$ OFDM symbols\footnote{Without loss of generality, we assume that $L=L_TL_F$, where $L$, $L_T$, and $L_F$ are integers.} ($L_T$ time slots). Note that $L_F$ can be predefined according to the coherence bandwidth of the channel. By using such a strategy, the transmission time overhead can be reduced from $L$ to $L_T$ time slots, i.e., it is reduced by a factor of $L_F$. Hence, the severe time-selective fading suffered by the whole signature sequence can be mitigated, which improves the DAD-EID performance in doubly-selective fading channels. The signal model in doubly-selective fading channels is illustrated as follows.

In doubly-selective fading channels, without loss of generality, the CSI of the $l_t$-th OFDM symbol and the $l_f$-th subcarrier between the $k$-th device and the aerial BS, $l_t\in[L_T], l_f\in [L_F]$, can be rewritten as
\vspace{1.8mm}
\begin{align}
\label{eq:CSI_DS} {\bf h}_{k,l_f}^{l_t}\!\!&=\!\!\sqrt{\dfrac{M}{P}}\sum\nolimits_{p=1}^{P}h_{k,p} e^{j2\pi \nu_{k,p}\Delta{T}}{\bf a}(\theta_{k,p})\nonumber\\
&~~~~~~~~~~\times e^{-j2\pi \tau_{k,p}(-\frac{B_s}{2}+\frac{B_s(l_f-1)}{N})}\in \mathbb{C}^{M\times 1},
\end{align}
\vspace{2mm}
where $\Delta{T}=((l_t-1)N+l_tN_{\rm CP}+l_f)T_s$, $N_{\rm CP}$ is the length of the cyclic prefix, $T_s=\frac{1}{B_s}$ is the sampling period, and other parameters are the same as that in (\ref{eq:CMod}). Given the $k$-th device, $\forall k\in[K]$, the transmitted signature sequence ${\bf s}_k\in \mathbb{C}^{L\times 1}$ is denoted as
\begin{align}
\label{eq:sequence} {\bf s}_k=\left[ {\bf s}_{k,1}^T,...,{\bf s}_{k,l_f}^T,...,{\bf s}_{k,L_F}^T \right]^T={\bf \Phi}_k{\bf e}_k\in \mathbb{C}^{L \times 1},
\end{align}
where ${\bf e}_k\in \mathbb{C}^{I \times 1}$ is the sequence selection vector and ${\bf s}_{k,l_f}\in \mathbb{C}^{L_T \times 1}$ is part of the sequence that transmitted by the $l_f$-th subcarrier.

Hence, by considering the inter-sub-carrier interference (ICI) \cite{ICI-OFDM}, the received signal ${\bf Y}_{l_f}\in \mathbb{C}^{L_T\times M}$ of the $l_f$-th subcarrier, $\forall l_f\in [L_f]$, can be written as
\vspace{-2mm}
\begin{align}
\label{eq:receive} {\bf Y}_{l_f}=\sum\nolimits_{k=1}^{K}a_k\left(\beta {\bf s}_{k,l_f}{\bf 1}_M^T \odot {\bf H}_{k,l_f} + {\bf I}_{k,l_f}\right) + {\bf N}_{l_f},
\end{align}
where $a_k\in\{0,1\}$ is the activity indicator, $\beta\in(0,1)$ is the attenuation factor caused by ICI, ${\bf 1}_M\in \mathbb{C}^{M\times 1}$ is a vector of all ones, $\odot$ is the Hadamard product, ${\bf H}_{k,l_f}\!\!=\!\!\left[ {\bf h}_{k,l_f}^{1}\!,...,\!{\bf h}_{k,l_f}^{l_t}\!,...,\!{\bf h}_{k,l_f}^{L_T}\right]^T\!\!\in\!\! \mathbb{C}^{L_T\times M}$, ${\bf I}_{k,l_f}\!=\!\Big[ {\bf i}_{k,l_f}^{1},...,{\bf i}_{k,l_f}^{l_t},...,$\\
${\bf i}_{k,l_f}^{L_T}\Big]^T\!\!\!\!\in\! \mathbb{C}^{L_T\times M}$\!, and ${\bf N}_{l_f}\!\!\in\!\mathbb{C}^{L_T\times M}$ denote the channel matrix, the ICI term, and the noise of the $k$-th device in the $l_f$-th subcarrier. Note that the ICI term can be characterized by the Doppler frequency and the time duration \cite{ICI-OFDM}. Furthermore, the received signals in $L_F$ successive subcarriers can be expressed as
\begin{align}
\label{eq:receiveAgg} \widetilde{\bf Y}=\sum\nolimits_{k=1}^{K}a_k \left(\beta{\bf s}_{k}{\bf 1}_M^T \odot \widetilde{\bf H}_k + \widetilde{\bf I}_k\right) + \widetilde{\bf N},
\end{align}
where $\widetilde{\bf Y}\!\!=\!\!\left[ ({\bf Y}_{1})^T,...,({\bf Y}_{l_f})^T,...,({\bf Y}_{L_F})^T\right]^T\!\!\in\! \mathbb{C}^{L\times M}$, $\widetilde{\bf H}_k\!=\!\!\left[ ({\bf H}_{k,1})^T,...,({\bf H}_{k,l_f})^T,...,({\bf H}_{k,L_F})^T\right]^T\!\!\in\!\! \mathbb{C}^{L\times M}$, $\widetilde{\bf I}_k\!=\!\left[ ({\bf I}_{k,1})^T,...,({\bf I}_{k,l_f})^T,...,({\bf I}_{k,L_F})^T\right]^T\!\!\in\!\! \mathbb{C}^{L\times M}$, and $\widetilde{\bf N}\!\!=\!\!\left[ ({\bf N}_{1})^T,...,({\bf N}_{l_f})^T,...,({\bf N}_{L_F})^T\right]^T\!\!\in \!\!\mathbb{C}^{L\times M}$. Note that if the CSI variation can be neglected within the selected OFDM time-frequency resource region, i.e., $L_T$ time slots and $L_F$ subcarriers, (\ref{eq:receiveAgg}) becomes (\ref{eq:SMod1}).

To illustrate the effectiveness of the proposed TFST strategy in doubly-selective fading channels, we approximately transform (\ref{eq:receiveAgg}) into (\ref{eq:ADSMod2}) and apply the proposed AE-JABID algorithm for DAD-EID. Simulation results show that the approximation from (\ref{eq:receiveAgg}) to (\ref{eq:ADSMod2}) only slightly degrades the DAD-EID performance with properly chosen $L_F$. Also, simulation results verify the superiority of the TFST strategy. 


\section{Simulation Results}

\subsection{Simulation Setups}
In this section, an extensive simulation investigation is carried out to evaluate the NMSE, the activity detection error rate (ADER), and the total bit error rate (BER$_{\rm total}$) of grant-free NC-IM-based massive IoT access. The NMSE for the proposed {\bf Algorithm} 1 and {\bf Algorithm} 2 are defined as ${\rm NMSE}_{\bf X}=\frac{\left\| \widehat{\bf{X}}-{\bf{X}}\right\|_F}{\left\| {\bf{X}} \right\|_F}$ and ${\rm NMSE}_{\bf W}=\frac{\left\| \widehat{\bf{W}}-{\bf{W}}\right\|_F}{\left\| {\bf{W}} \right\|_F}$, respectively, where ${\bf{X}}$ is the actual equivalent channel matrix, ${\bf{W}}$ is the actual equivalent channel matrix in the virtual angular domain, $\widehat{\bf{X}}$ and $\widehat{\bf{W}}$ are the outputs of the proposed {\bf Algorithm} 1 and {\bf Algorithm} 2, respectively.
The ADER is defined as ${\rm ADER}=\frac{E_m+E_f}{K}$,
%
where $E_m$ is the number of active devices missed to be detected, $E_f$ is the number of inactive devices falsely detected to be active.
In addition, the BER$_{\rm total}$ is defined as ${\rm BER_{total}}=\frac{(E_m+E_f)r+B_d}{(K_a+E_f)r }$,
where $B_d$ denotes the number of error bits for detected active devices, and $r$ is the number of bits embedded in each signature sequence selection.

In the simulations, the number of devices is $K=100$ with $K_a=10$ active devices. {\color{black} The UAV height is $h_u=100$ m. The carrier frequency is $f_c=1$ GHz, the bandwidth is $B_s=10$ MHz, the system employs OFDM with $N=512$ subcarriers and a cyclic prefix of length $N_{\rm CP}=32$ with the subcarrier interval $\Delta f=15$ kHz\cite{kemalong20}. Furthermore, we consider that the number of paths $P$ varies from 8 to 14 and the delay of the $k$-th device for the $p$-th path $\tau_{k,p}$ is randomly and uniformly selected from $\left[0,N_{\rm CP}/B_s\right]$. In addition, the AoAs of the $k$-th device $\theta_{k,p}$, $\forall k\in[K]$, $p\in[P]$, are generated within an angular spread $10^ \circ$, where the center of the AoAs is uniformly distributed within $\left[ -\frac{\pi}{2},\frac{\pi}{2}\right]$. For doubly-selective fading channel model, we denote the maximum radial velocity between IoT devices and the UAV as $v_{\rm max}$, hence the maximum Doppler shift is $\nu_{\rm max}=\frac{v_{\rm max}f_c}{3\times 10^8}$. Then, the Doppler shift $\nu_{k,p}$ of the $k$-th device and the $p$-th path, $\forall k\in[K], p\in[P]$, is uniformly distributed within $\left[ -\nu_{\rm max},\nu_{\rm max}\right]$.} As for the TFST strategy, to consider the ICI, we generate the frequency-domain channel matrix of $N$ subcarriers based on (1)-(3) of \cite{ICI-OFDM}. As for the proposed algorithms, the maximum iteration number is set to $T_0=200$, the damping parameter $\kappa=0.3$, and the termination threshold is set to $\epsilon=10^{-6}$. {\color{black}For Algorithm 1, the DAD threshold $T_{h1}$ is set to 0.7, which can achieve good ADER performance with smaller sequence length $L$.} For Algorithm 2, the DAD threshold $T_{h2}$ is set to 0.9, $Q$ in (\ref{eq:Naighbors}) is set to 4, $\zeta_{m\pm 1}=1$, $\zeta_{m\pm 2}=0.8$, $\zeta_{m\pm 3}=0.6$, and $\zeta_{m\pm 4}=0.4$ for (\ref{eq:Naighbors}) to exploit the clustered sparsity (\ref{eq:AD2}), $\forall m\in [M]$. {\color{black}According to \cite{UAV-Ploss1,UAV-Ploss2}, the signal-to-noise ratio (SNR) in dB is denoted as

\begin{figure*}[t]
\vspace{-9mm}
\centering
\subfigure[]{
    \begin{minipage}[t]{0.33\linewidth}
        \centering
\label{fig:Lsce1NMSE}
        \includegraphics[width = 0.9\columnwidth,keepaspectratio]
        {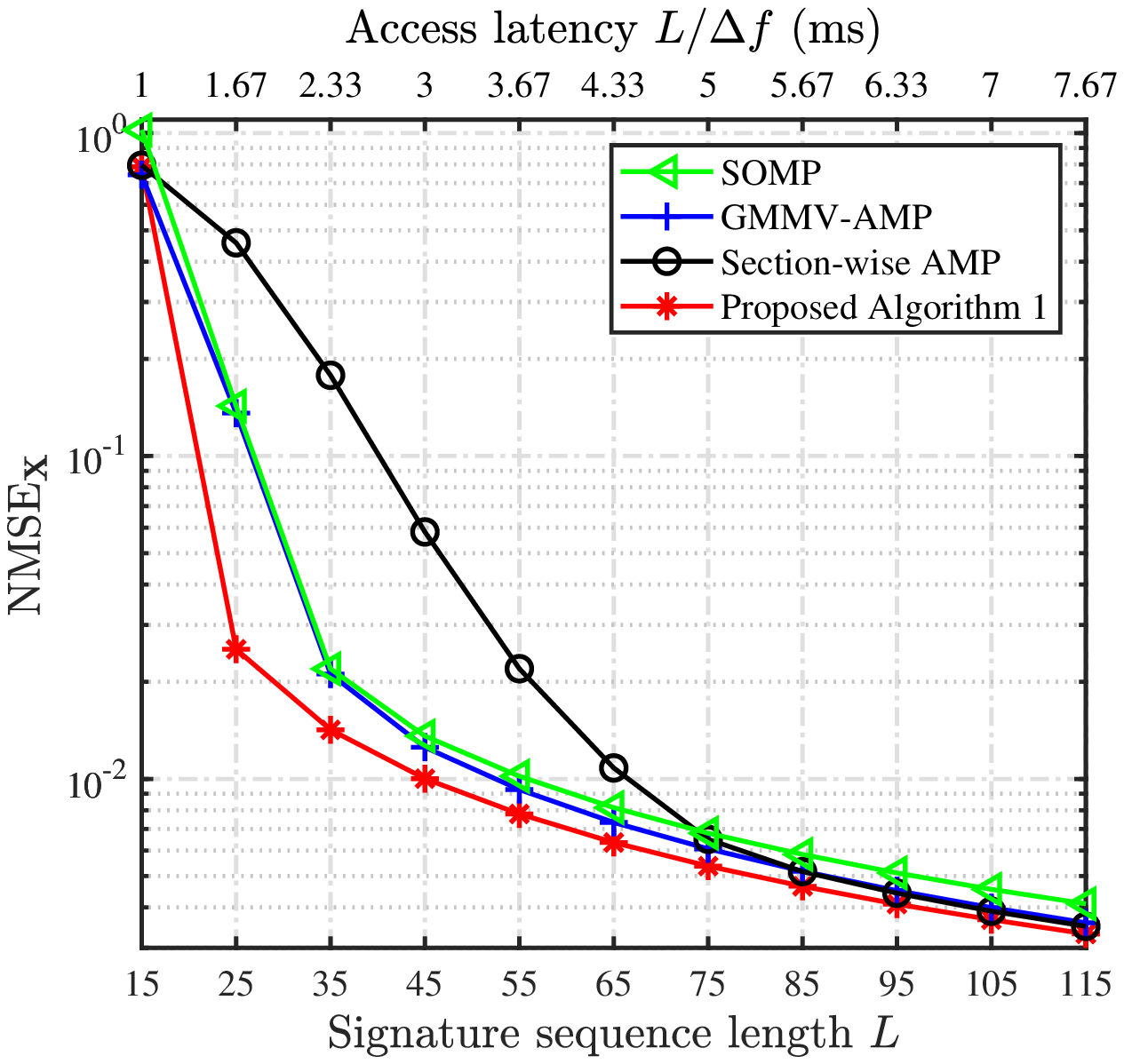}\\
    \end{minipage}%
}%
\subfigure[]{
    \begin{minipage}[t]{0.33\linewidth}
        \centering
\label{fig:Lsce1Pe}
        \includegraphics[width = 0.9\columnwidth,keepaspectratio]
        {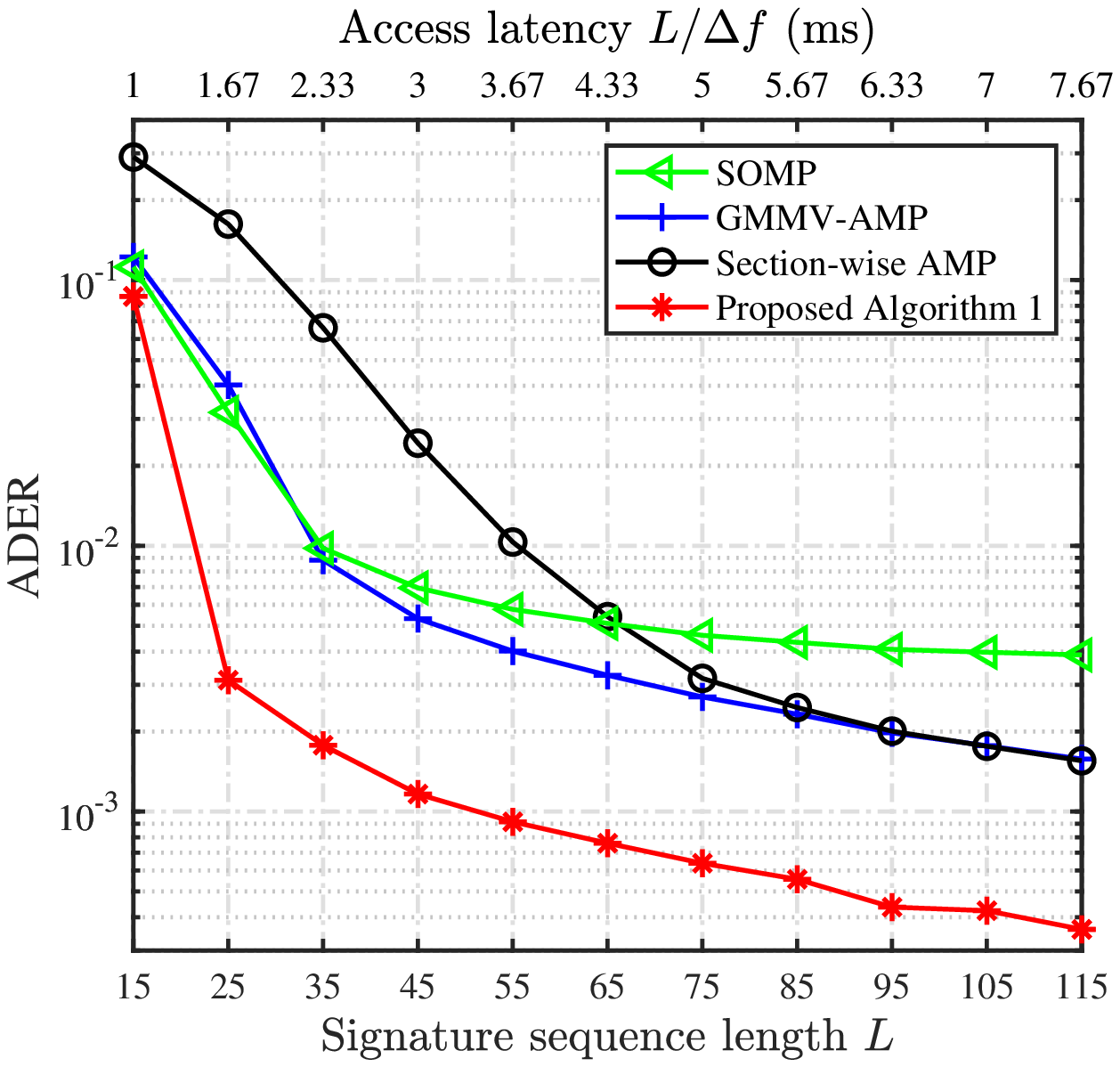}\\
    \end{minipage}%
}%
\subfigure[]{
    \begin{minipage}[t]{0.33\linewidth}
        \centering
\label{fig:Lsce1BER}
        \includegraphics[width = 0.9\columnwidth,keepaspectratio]
        {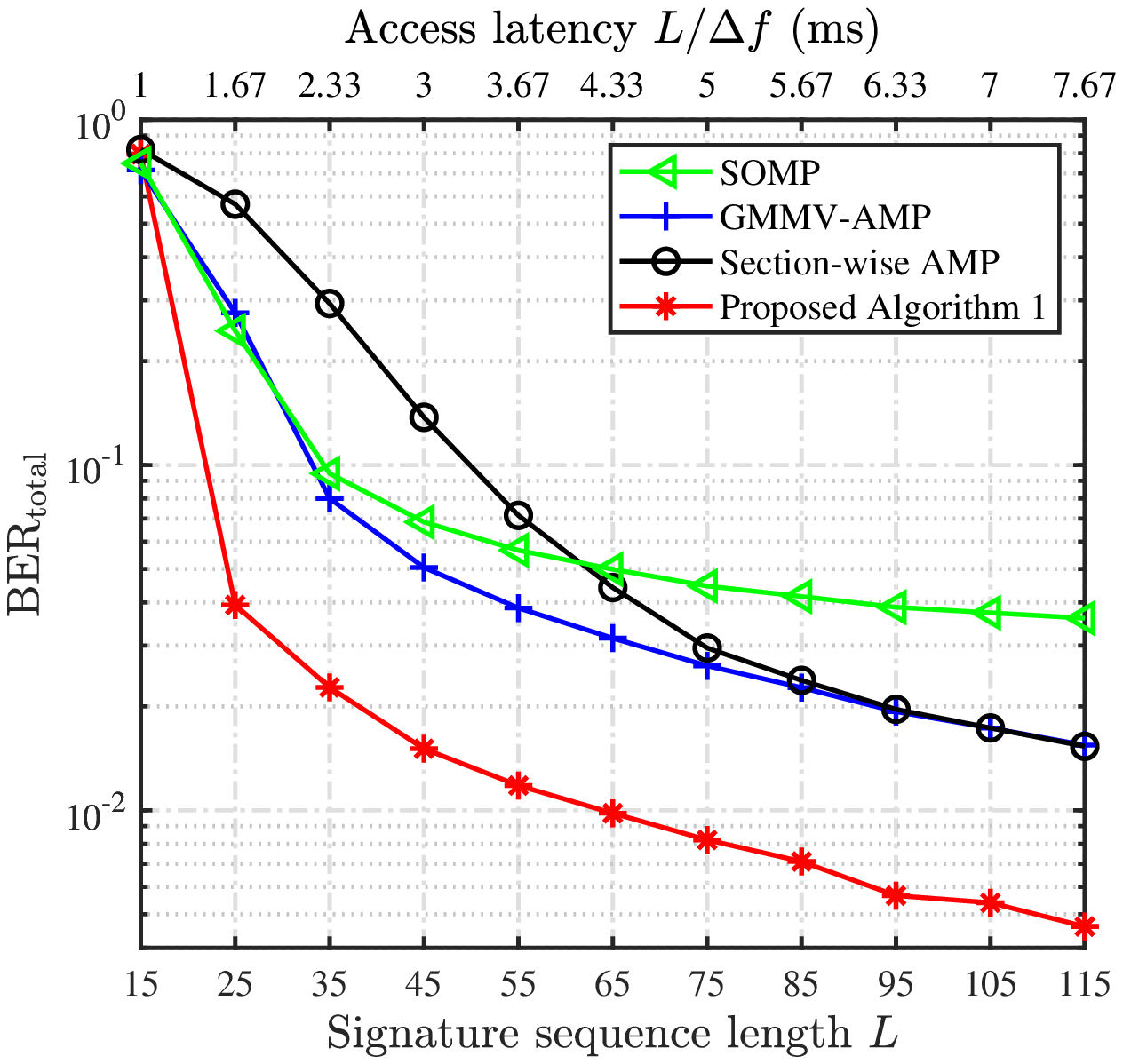}\\
    \end{minipage}%
}%
\centering
\setlength{\abovecaptionskip}{-1mm}
\captionsetup{font={footnotesize, color = {black}}, singlelinecheck = off, justification = raggedright,name={Fig.},labelsep=period}
\caption{Performance comparison of different algorithms versus the signature sequence length $L$ in small-scale MIMO systems: (a) NMSE$_{\bf X}$ performance comparison; (b) ADER performance comparison; (c) BER$_{\rm total}$ performance comparison. }
\label{fig:Lsce1}
\end{figure*}

\begin{figure*}[t]
\vspace{-4mm}
\centering
\subfigure[]{
    \begin{minipage}[t]{0.33\linewidth}
        \centering
\label{fig:SNRsce1NMSE}
        \includegraphics[width = 0.9\columnwidth,keepaspectratio]
        {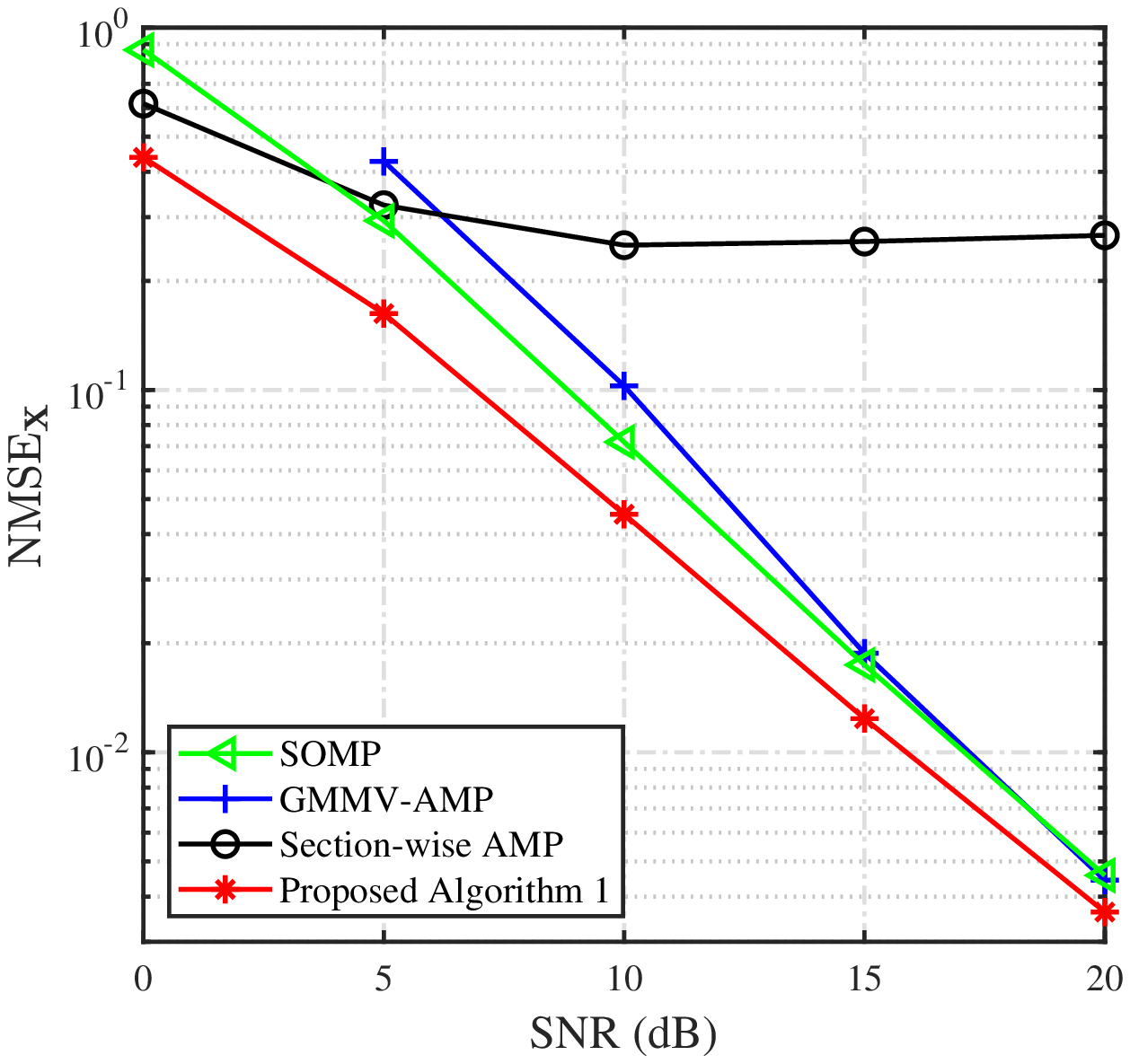}\\
    \end{minipage}%
}%
\subfigure[]{
    \begin{minipage}[t]{0.33\linewidth}
        \centering
\label{fig:SNRsce1ADER}
       \includegraphics[width = 0.9\columnwidth,keepaspectratio]
       {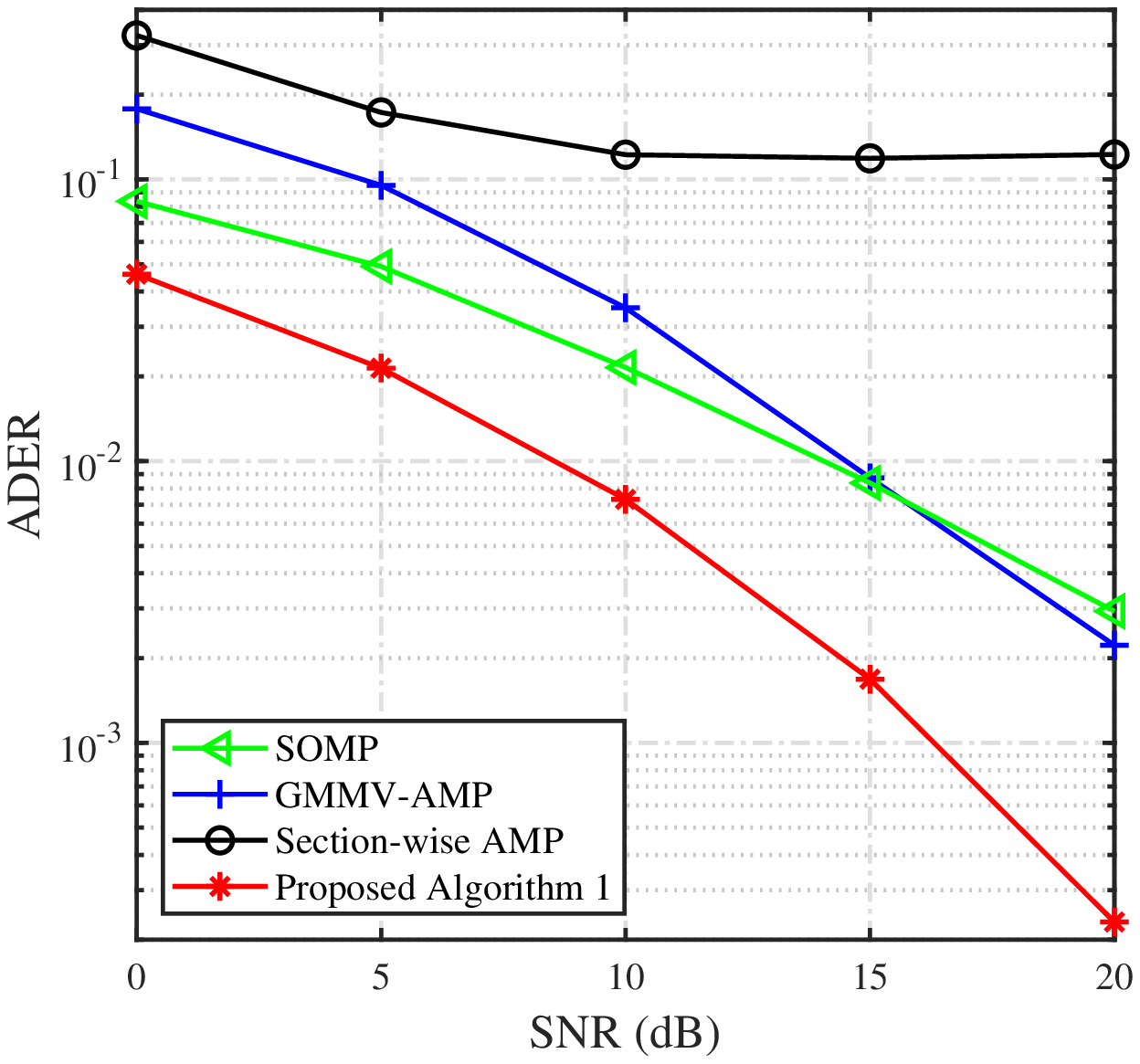}\\
    \end{minipage}%
}%
\subfigure[]{
    \begin{minipage}[t]{0.33\linewidth}
        \centering
\label{fig:SNRsce1BER}
        \includegraphics[width = 0.9\columnwidth,keepaspectratio]
        {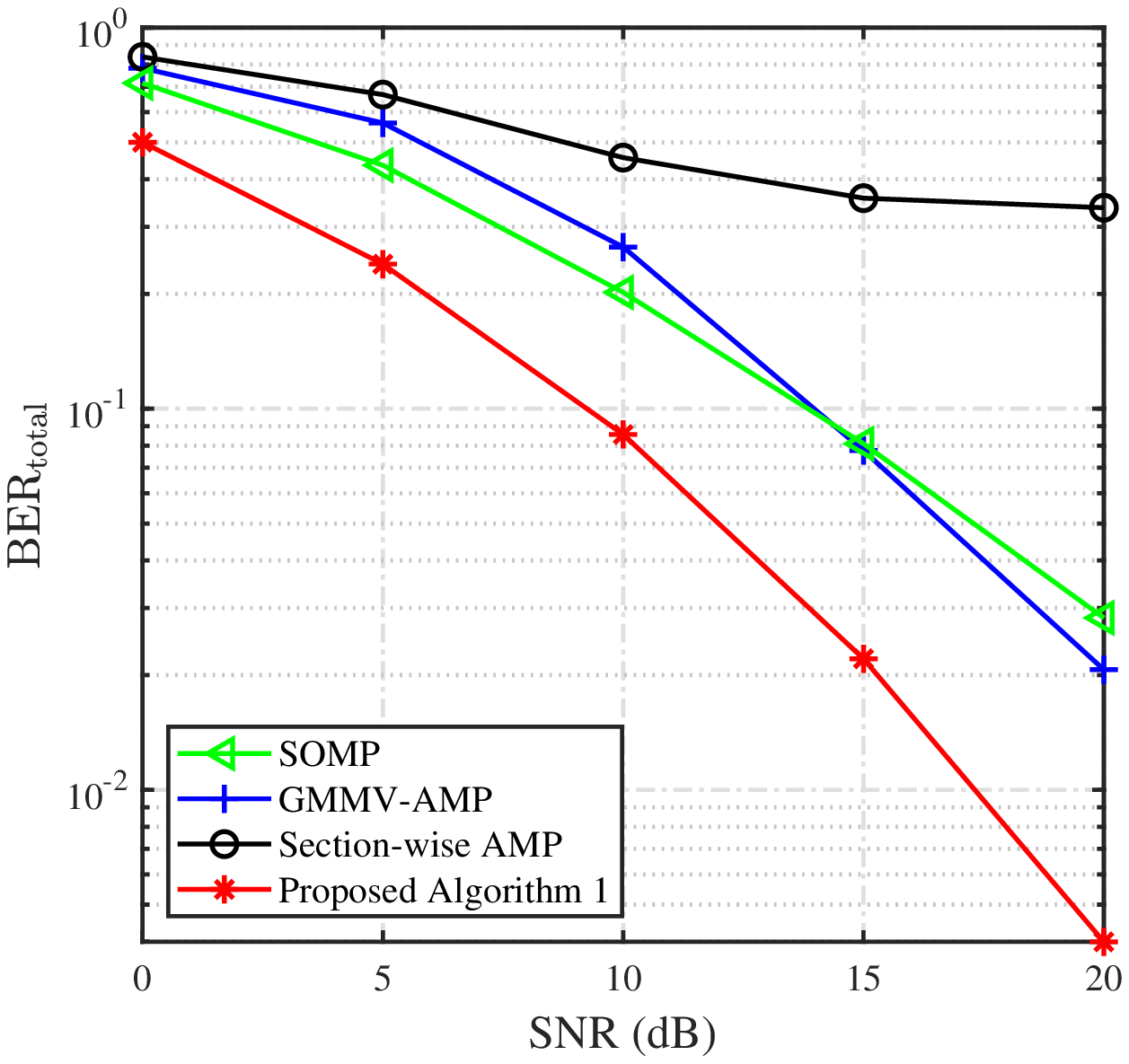}\\
    \end{minipage}%
}%
\centering
\setlength{\abovecaptionskip}{-1mm}
\captionsetup{font={footnotesize, color = {black}}, singlelinecheck = off, justification = raggedright,name={Fig.},labelsep=period}
\caption{Performance comparison of different algorithms versus the SNR in small-scale MIMO systems: (a) NMSE$_{\bf X}$ performance comparison; (b) ADER performance comparison; (c) BER$_{\rm total}$ performance comparison. }
\label{fig:SNRsce1}
\vspace{-4mm}
\end{figure*}

\vspace{-4mm}
\begin{align}
{\rm SNR}~[{\rm dB}] &= P_t - P_L -P_n,\nonumber\\
P_L ~[{\rm dB}] &= A/(1+ae^{-b(c-a)}) + B,\nonumber\\
A&=\eta_{\mathrm{LoS}}-\eta_{\mathrm{NLoS}},\nonumber\\
B&=20{\rm log}_{10}(d)+20{\rm log}_{10}(4 \pi f_c/300)+\eta_{\mathrm{NLoS}},\nonumber\\
c&=180{\rm arcsin}(h_u/d)/\pi,
\end{align}
where $P_t$ in dBm is the transmit power of IoT devices, $P_L$ in dB is the path loss, $P_n=(-174 + 10{\rm log}_{10}B_s)$ dBm is the noise power with $B_s$ in Hz, $f_c$ is the carrier frequency in MHz, $d$ is the distance between the device and the UAV, $\eta_{\mathrm{LoS}}$ ($\eta_{\mathrm{NLoS}}$) is the LoS (NLoS) mean excessive
path loss values, and $a$, $b$ are environment-related constant parameters. For example, according to \cite{UAV-Ploss1,UAV-Ploss2}, assuming that $P_t=14$ dBm\footnote{\color{black}According to \cite{NB-IOT}, the 3GPP Release 14 had added a new low power class at 14 dBm for IoT devices, which aims to achieve smaller size, lower power consumption, and lower cost IoT devices compared to the 20 dBm devices in NB-IoT Release 13.}, the horizontal distance between the UAV and the IoT device at the edge of the coverage aera is $r_u=500$ m, $d=\sqrt{h_u^2+r_u^2}=509.9$ m, $\eta_{\mathrm{LoS}}=2.3$, $\eta_{\mathrm{NLoS}}=34$, $a=5.0188$, and $b=0.3511$, we obtain SNR = 17.84 dB. These calculations help us to set the SNR in the simulations. For simplicity, power control is considered in the system and the same SNR for each IoT device can be achieved at the aerial BS\cite{MIMO-OFDM}.
}

For comparison, we consider the following benchmark schemes. {\bf SOMP}: The simultaneous orthogonal matching pursuit algorithm proposed in \cite{SOMP}, where the iteration stops if the average power of the residual is smaller than the noise power. Note that the noise power is assumed to be known in advance; {\bf GMMV-AMP}: Algorithm 1 proposed in \cite{kemalong20}, where the noise variance is known in advance; {\bf Section-wise AMP}: The state-of-the-art AMP-based algorithm with inseparable denoiser proposed in \cite{NC-IM-Access} for NC-IM over frequency-flat fading channels, where the noise variance and the channel power are known in advance; {\bf Benchmark 1}: A modified AE-JABID algorithm, where the update rule of the activity indicator in line \ref{A2:EMactivity} of {\bf Algorithm} 2 is replaced by $(\rho_{k,i}^m)^{t+1}=\frac{1}{M}\sum\nolimits_{m=1}^{M}(\rho_{k,i}^m)^{t}$, $\forall m\in [M]$. Note that the SOMP algorithm, the GMMV-AMP algorithm, and the section-wise AMP algorithm are suitable for solving (\ref{eq:SMod1}), where the common device activity in multiple receive antennas is exploited. While, benchmark 1 is designed for solving (\ref{eq:ADSMod2}), where the channel sparsity in the virtual angular domain is exploited.

\vspace{-3mm}
\subsection{Simulation Results for Small-Scale MIMO Systems}
Fig. \ref{fig:Lsce1NMSE}, Fig. \ref{fig:Lsce1Pe}, and Fig. \ref{fig:Lsce1BER} compare the NMSE$_{\bf X}$, ADER, and BER$_{\rm total}$ performance of different solutions versus the length $L$ of the signature sequence {\color{black}(also versus the access latency $L/\Delta f$, where $1/\Delta f$ denotes the time duration of one OFDM symbol)}, respectively. Note that the number of signature sequences allocated to each device is $I=2$, the number of receive antennas at the aerial BS is $M=2$, the number of sub-frames within a frame is $J=2$, the SNR is set to 15 dB, and $\widetilde{N}=8$ subcarriers are utilized for joint signal processing in the proposed STF-JABID algorithm (i.e., {\bf Algorithm} 1). It can be seen that the proposed STF-JABID algorithm outperforms the SOMP algorithm, the GMMV-AMP algorithm, and the section-wise AMP algorithm in terms of NMSE$_{\bf X}$, ADER, and BER$_{\rm total}$ performance. Thanks to the joint signal processing of multiple receive antennas, subcarriers and sub-frames, i.e., the exploitation of the space-time-frequency structured sparsity, the proposed STF-JABID algorithm achieves better DAD-EID performance in small-scale MIMO-based UAV systems. In fact, the considered benchmark algorithms only enable to employ the common device activity in multiple antennas, leading to performance degradation.

The superiority of the proposed STF-JABID algorithm can also be verified in Fig. \ref{fig:SNRsce1}, where we compare the NMSE$_{\bf X}$, ADER, and BER$_{\rm total}$ performance versus the SNR. From Fig. \ref{fig:SNRsce1}, we observe that the proposed STF-JABID algorithm consistently outperforms the SOMP algorithm, the GMMV-AMP algorithm, and the section-wise AMP algorithm in all SNRs, in terms of NMSE$_{\bf X}$, ADER, and BER$_{\rm total}$ performance. Here the number of signature sequences allocated to each device is $I=4$, the number of receive antennas is $M=2$, the length of signature sequence is $L=40$, and we integrate $J=2$ sub-frames and $\widetilde{N}=8$ subcarriers for joint signal processing in the proposed STF-JABID algorithm.

\vspace{-2mm}
\subsection{Simulation Results for LS-MIMO Systems}
\begin{figure*}[t]
\vspace{-6mm}
\centering
\subfigure[]{
    \begin{minipage}[t]{0.33\linewidth}
        \centering
\label{fig:Lsce2NMSE}
        \includegraphics[width = 0.95\columnwidth,keepaspectratio]
        {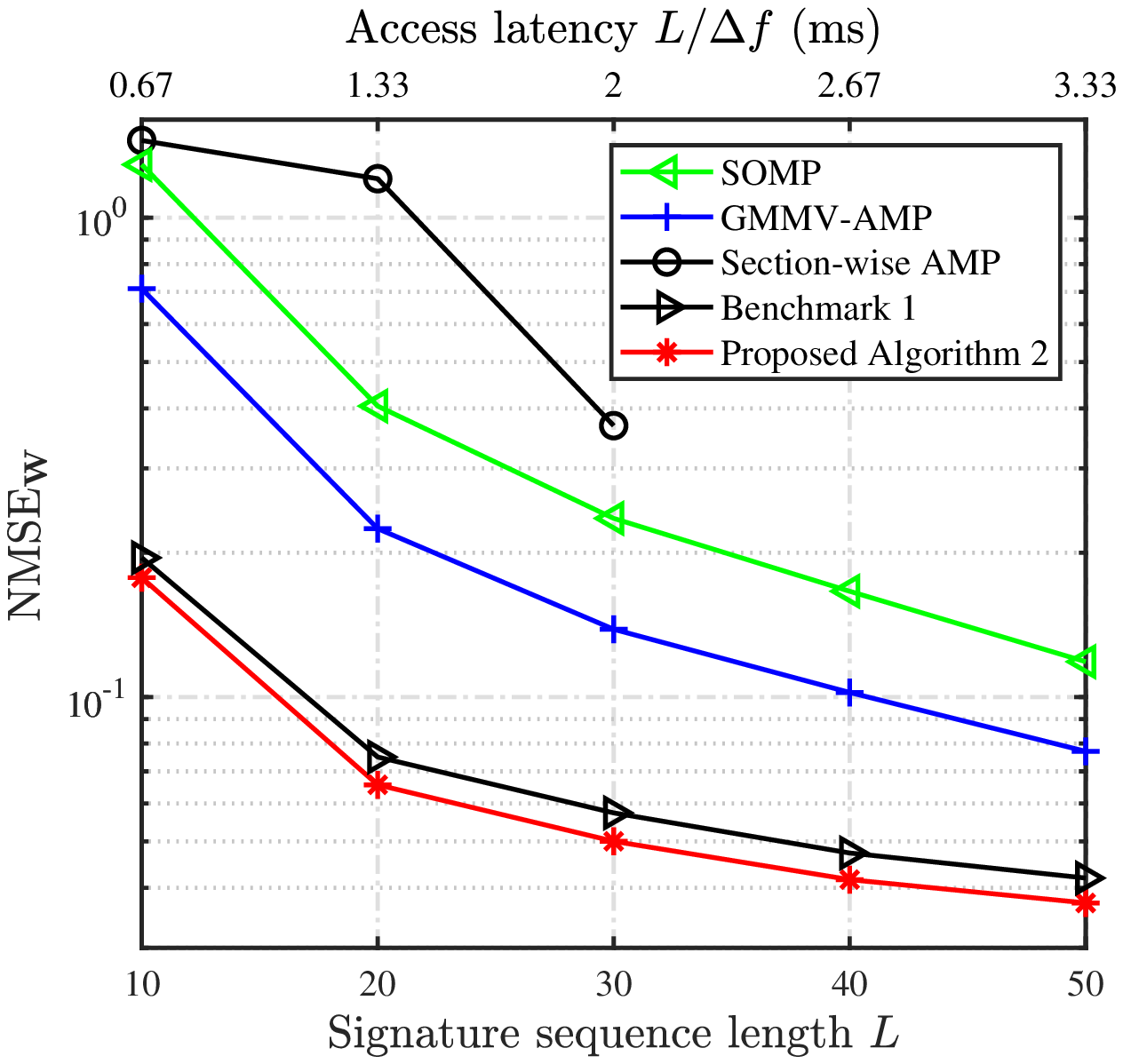}\\
    \end{minipage}%
}%
\subfigure[]{
    \begin{minipage}[t]{0.33\linewidth}
        \centering
\label{fig:Lsce2ADER}
        \includegraphics[width = 0.95\columnwidth,keepaspectratio]
        {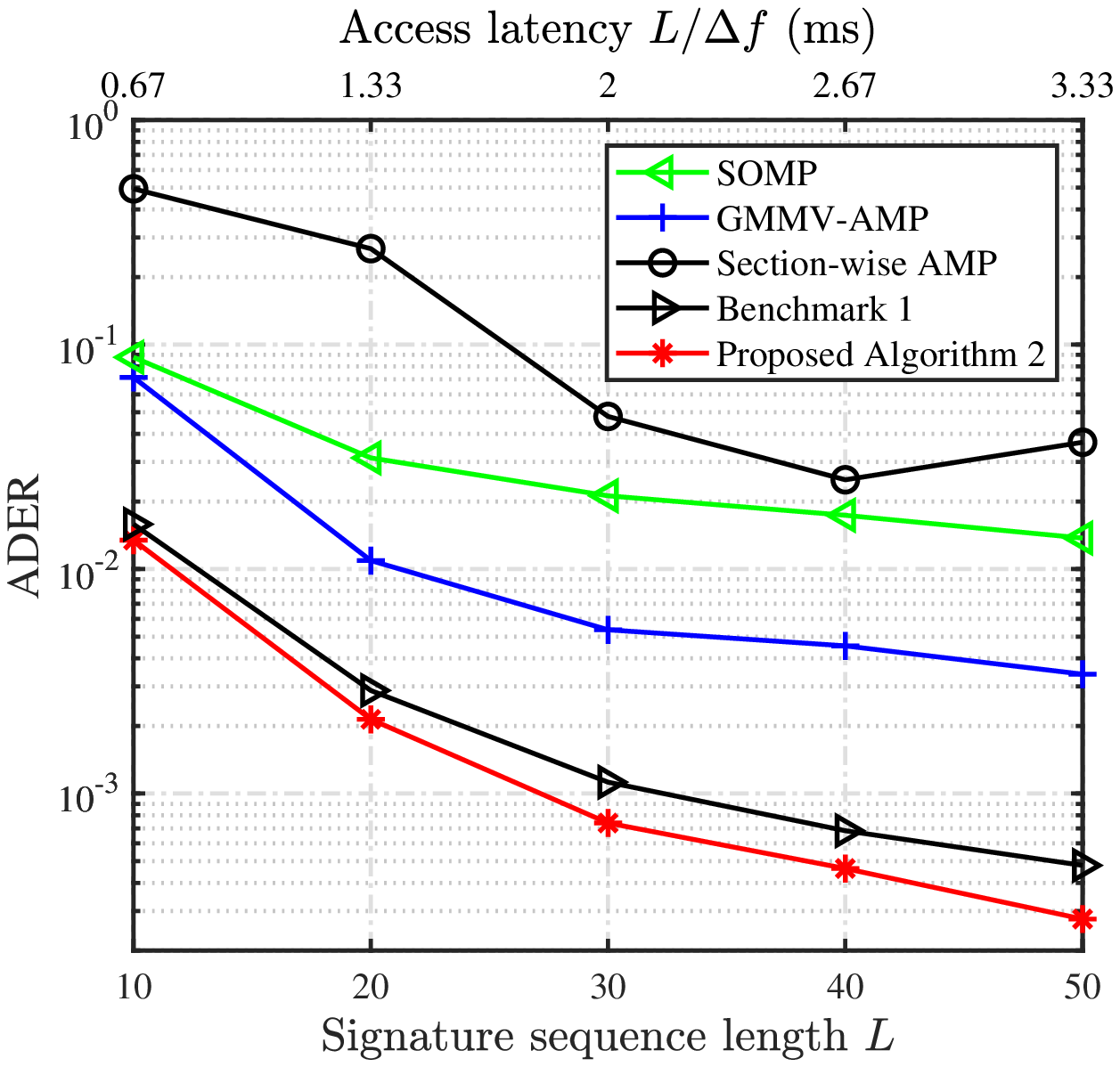}\\
    \end{minipage}%
}%
\subfigure[]{
    \begin{minipage}[t]{0.33\linewidth}
        \centering
\label{fig:Lsce2BER}
        \includegraphics[width = 0.95\columnwidth,keepaspectratio]
        {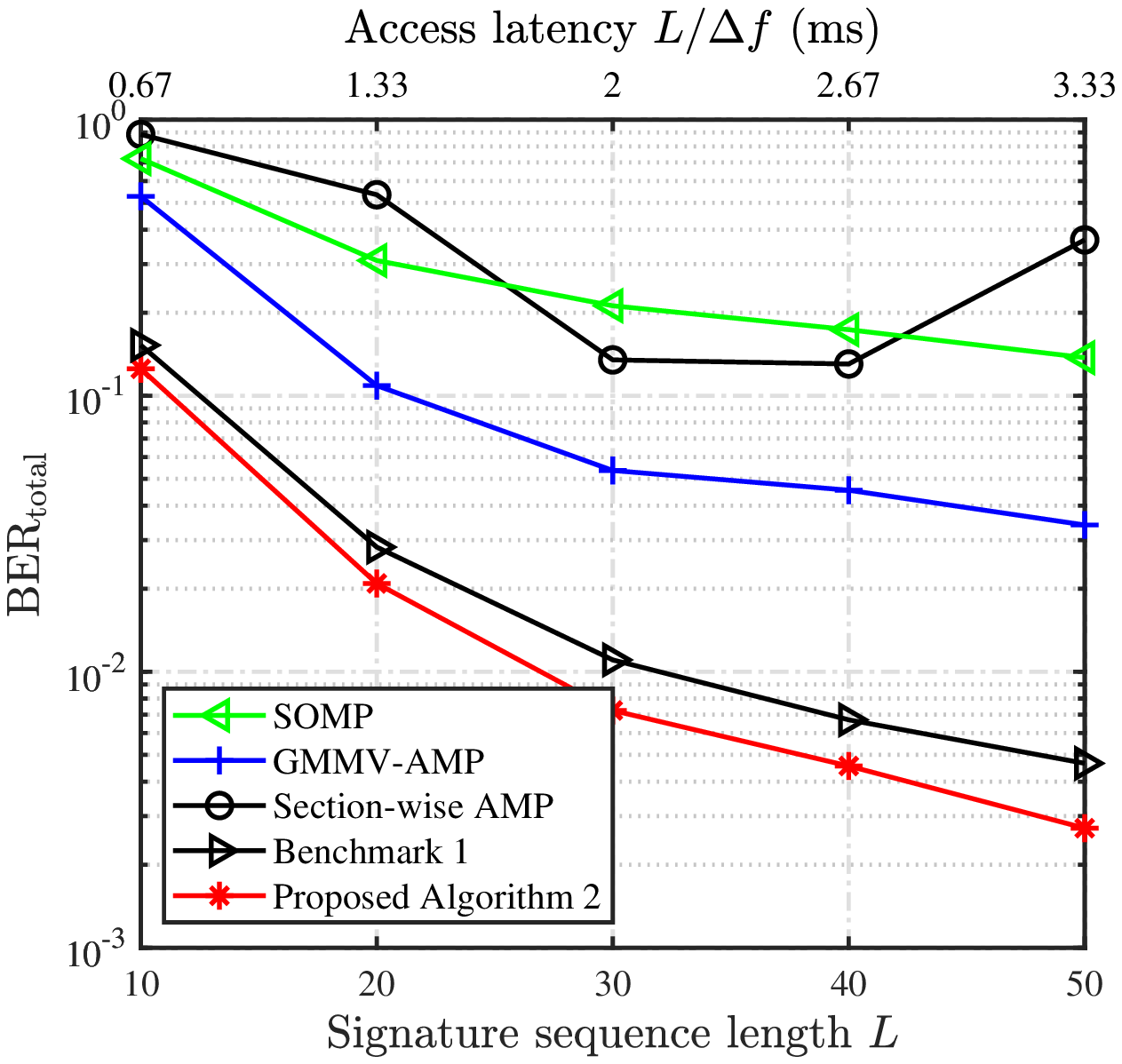}\\
    \end{minipage}%
}%
\centering
\setlength{\abovecaptionskip}{-1mm}
\captionsetup{font={footnotesize, color = {black}}, singlelinecheck = off, justification = raggedright,name={Fig.},labelsep=period}
\caption{Performance comparison of different algorithms versus the signature sequence length $L$ in LS-MIMO systems: (a) NMSE$_{\bf W}$ performance comparison; (b) ADER performance comparison; (c) BER$_{\rm total}$ performance comparison. }
\label{fig:Lsce2}
\end{figure*}

\begin{figure*}[t]
\vspace{-3mm}
\centering
\subfigure[]{
    \begin{minipage}[t]{0.33\linewidth}
        \centering
\label{fig:AntennaSce2NMSE}
        \includegraphics[width = 0.94\columnwidth,keepaspectratio]
        {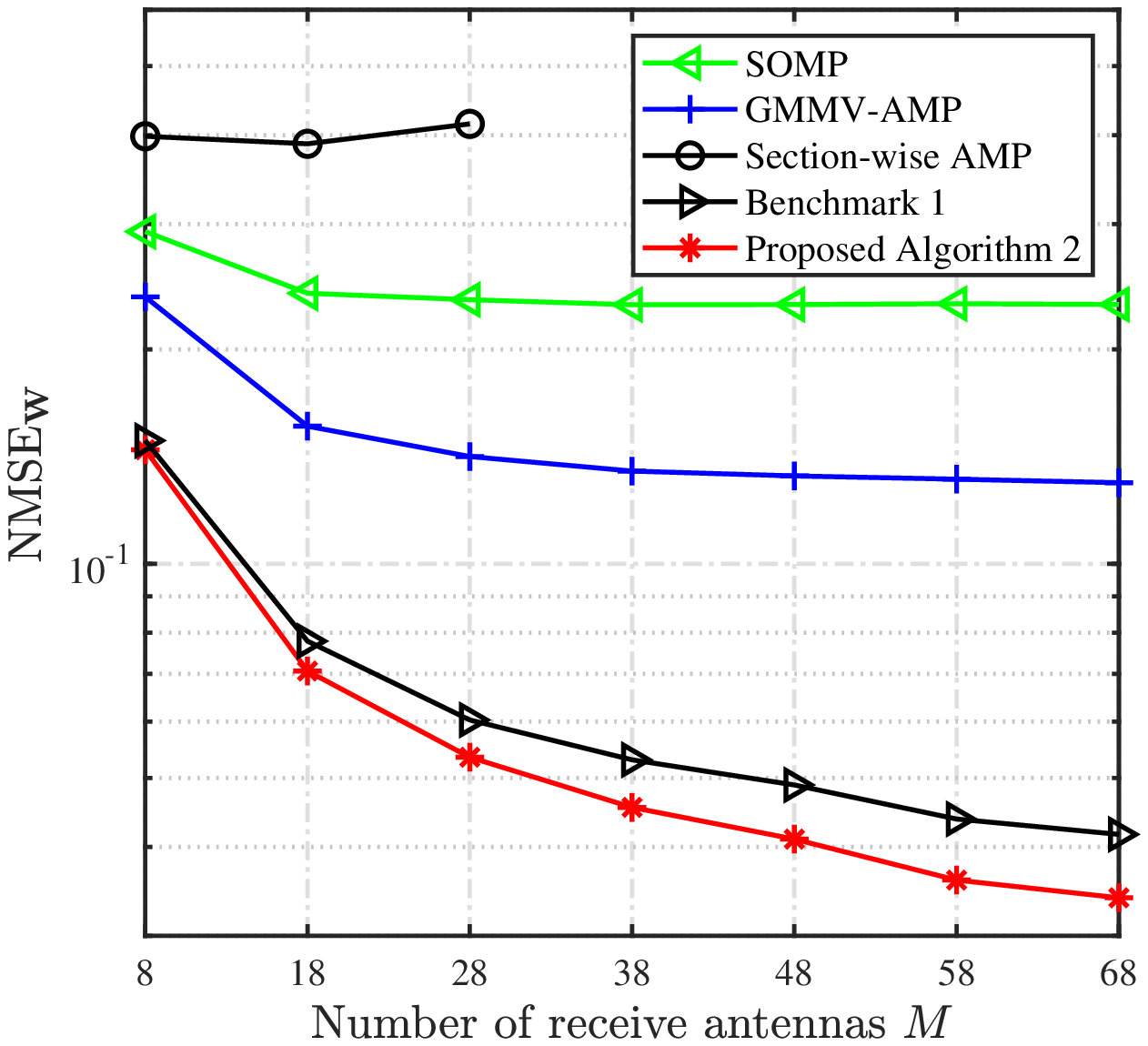}\\
    \end{minipage}%
}%
\subfigure[]{
    \begin{minipage}[t]{0.33\linewidth}
        \centering
\label{fig:AntennaSce2ADER}
       \includegraphics[width = 0.94\columnwidth,keepaspectratio]
       {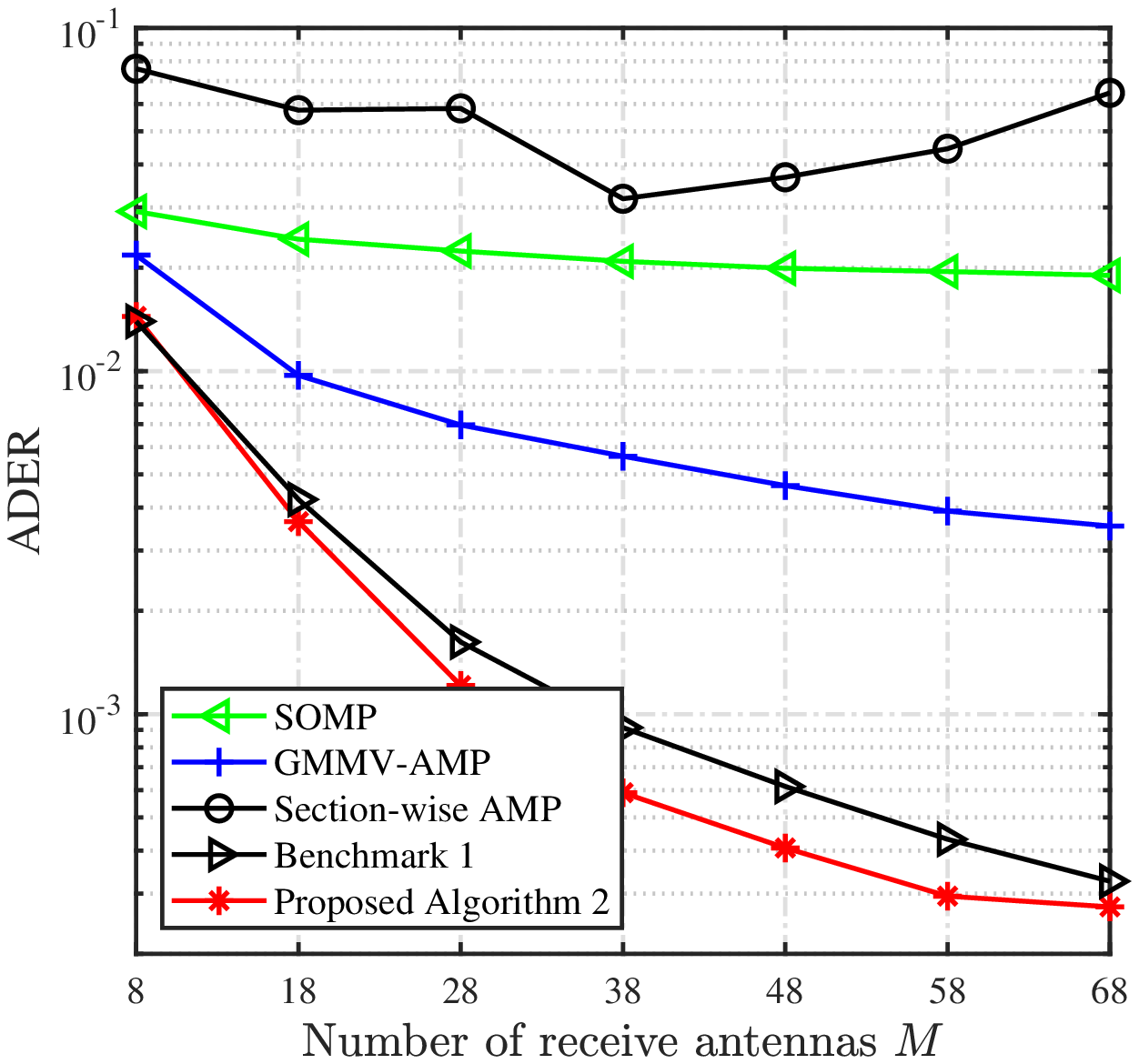}\\
    \end{minipage}%
}%
\subfigure[]{
    \begin{minipage}[t]{0.33\linewidth}
        \centering
\label{fig:AntennaSce2BER}
        \includegraphics[width = 0.94\columnwidth,keepaspectratio]
        {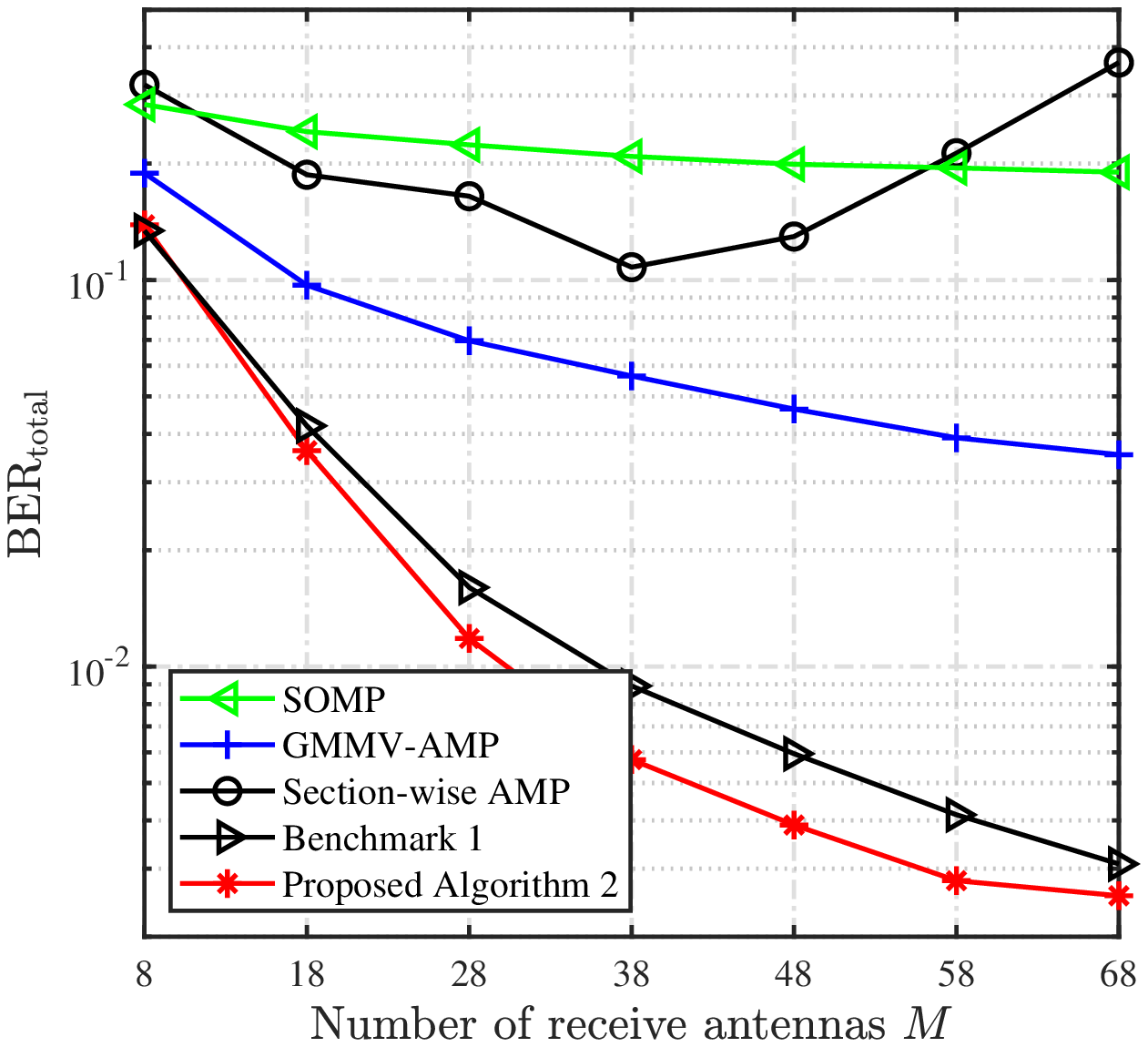}\\
    \end{minipage}%
}%
\centering
\setlength{\abovecaptionskip}{-1mm}
\captionsetup{font={footnotesize}, singlelinecheck = off, justification = raggedright,name={Fig.},labelsep=period}
\caption{Performance comparison of different algorithms versus the number of receive antennas $M$ in LS-MIMO systems: (a) NMSE$_{\bf W}$ performance comparison; (b) ADER performance comparison; (c) BER$_{\rm total}$ performance comparison. }
\label{fig:AntennaSce2}
\vspace{-4mm}
\end{figure*}

Fig. \ref{fig:Lsce2} illustrates the superiority of the proposed AE-JABID algorithm (i.e., {\bf Algorithm 2}) in terms of NMSE$_{\bf W}$, ADER, and BER$_{\rm total}$ performance versus the length of signature sequence $L$ {\color{black}(also versus the access latency $L/\Delta f$)}, given $M=32$, $I=2$, and SNR $= 5$ dB. Thanks to the exploitation of the virtual angular-domain structured sparsity, it can be seen that the proposed AE-JABID algorithm outperforms the SOMP algorithm, the GMMV-AMP algorithm, and the section-wise AMP algorithm in NMSE$_{\bf W}$, ADER, and BER$_{\rm total}$ performance with reduced length of sequence $L$ (also the access latency). The insight is that the virtual angular-domain channel might be more sparse than the spatial-domain channel in LS-MIMO systems, which can be properly exploited by the proposed AE-JABID algorithm for reducing the access latency. Furthermore, owing to the exploitation of the clustered sparsity of angular-domain channel in (\ref{eq:ActUpdate}), the proposed AE-JABID algorithm outperforms benchmark 1.

Fig. \ref{fig:AntennaSce2} provides the NMSE$_{\bf W}$, ADER, and BER$_{\rm total}$ performance of different algorithms versus the number of receive antennas $M$, given $L=30$, $I=2$, SNR $=5$ dB. It can be observed that the performance of all the algorithms improves as the number of receive antennas increases, due to the exploitation of the common device activity among multiple receive antennas. More importantly, thanks to the proposed DAD procedure (\ref{eq:ACTset}), the AE-JABID algorithm can efficiently exploit the common device activity of multiple receive antennas in the virtual angular-domain. Hence, it can be seen that the advantages of the proposed AE-JABID algorithm over other benchmark algorithms enlarge as the number of receive antennas increases.

\begin{figure*}[t]
\vspace{-9mm}
\centering
\subfigure[]{
    \begin{minipage}[t]{0.48\linewidth}
        \centering
\label{fig:DoubleSeleADER}
        \includegraphics[width = 0.7\columnwidth,keepaspectratio]
        {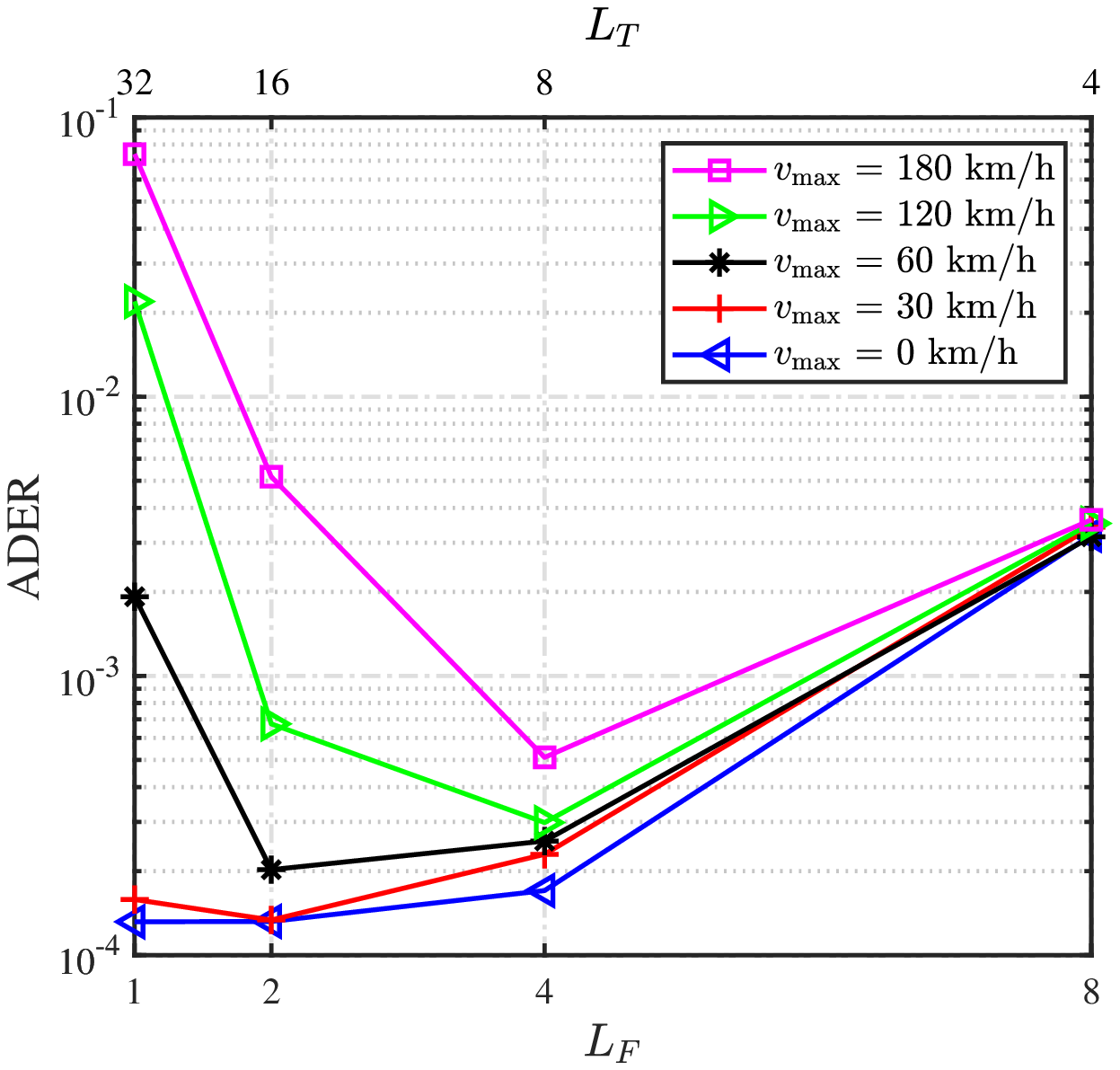}\\
    \end{minipage}%
}%
\subfigure[]{
    \begin{minipage}[t]{0.48\linewidth}
        \centering
\label{fig:DoubleSeleBER}
        \includegraphics[width = 0.7\columnwidth,keepaspectratio]
        {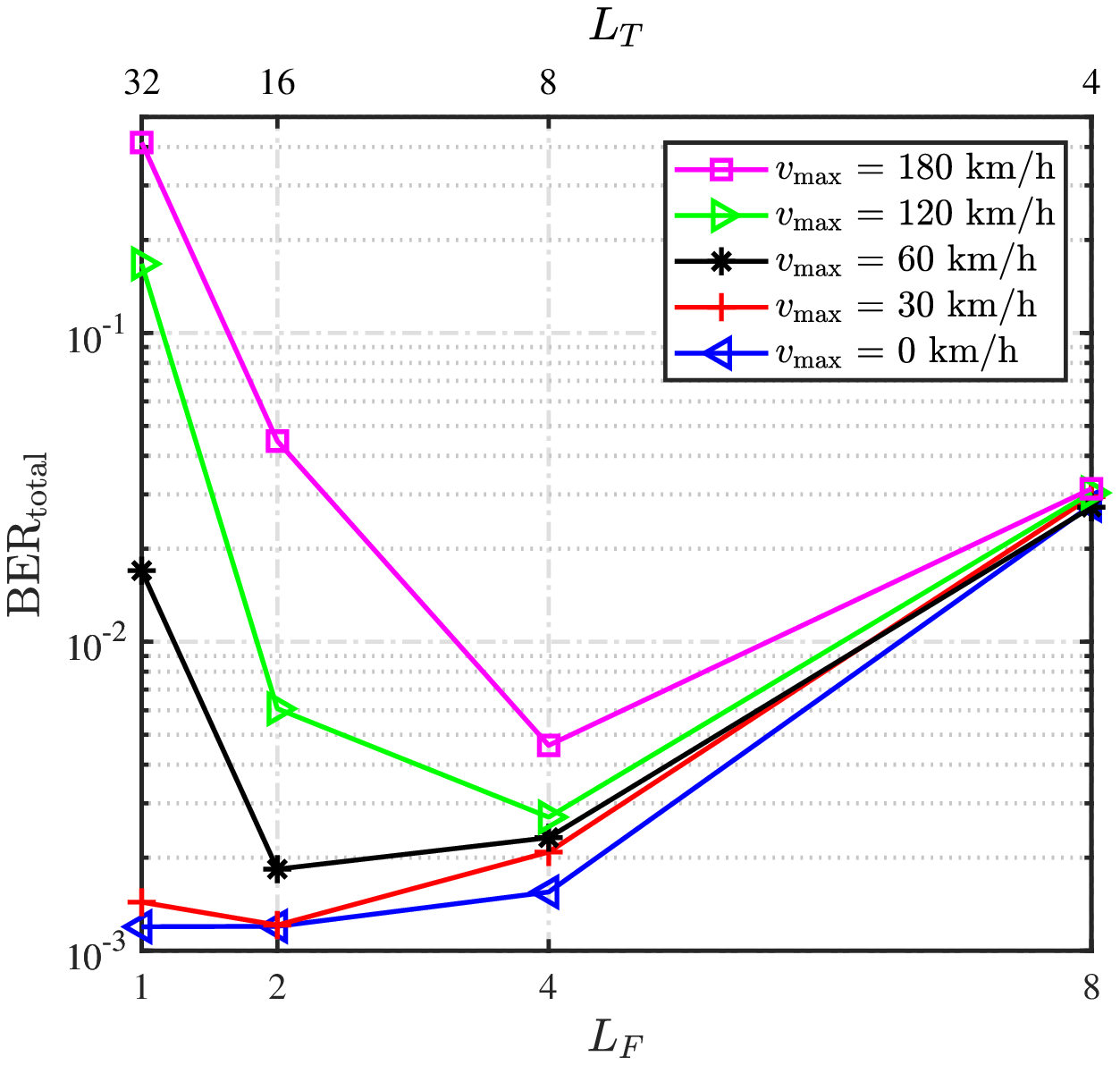}\\
    \end{minipage}%
}%
\centering
\setlength{\abovecaptionskip}{-1mm}
\captionsetup{font={footnotesize, color = {black}}, singlelinecheck = off, justification = raggedright,name={Fig.},labelsep=period}
\caption{Performance comparison of the proposed AE-JABID algorithm versus $L_F$ under different maximum radial velocity between IoT devices and the UAV $v_{\rm max}$: (a) ADER performance comparison; (b) BER$_{\rm total}$ performance comparison. }
\label{fig:DoubleSele}
\end{figure*}

To depict the superiority of the proposed TFST strategy in Section IV-C, Fig. \ref{fig:DoubleSele} shows the performance of the proposed AE-JABID algorithm versus $L_F$ {\color{black}(also versus $L_T=L/L_F$)}, i.e., $L_F=1, 2, 4, 8$ subcarriers, under different $v_{\rm max}$, the length of signature sequence is fixed to $L=32$, the number of signature sequences allocated for each device is $I=2$, the number of receive antennas is $M=32$, and SNR $=10$ dB.
It is observed that the performance of the blue curve ($v_{\rm max}=0$ km/h) is the best among all the curves and degrades with the increase of $L_F$ due to frequency-selective fading. While, the ADER and BER$_{\rm total}$ performance under $v_{\rm max}=0$ km/h is almost unchanged if $L_F$ is small (i.e., $L_F\le 4$), due to the CSI correlations among successive subcarriers within the coherence bandwidth. {\color{black}As a comparison, non-TFST strategy is used if $L_F=1$ (i.e., $L_T=32$). It can be seen that, under the non-TFST strategy, the performance of AE-JABID algorithm deteriorates dramatically as $v_{\rm max}$ increases from 0 km/h to 180 km/h, due to the faster variation in time-selective fading. In contrast, by using the proposed TFST strategy, i.e., by properly enlarging $L_F$ for signature sequence transmission, the ADER and BER$_{\rm total}$ performance can be improved significantly even with high $v_{\rm max}$. For example, in Fig. \ref{fig:DoubleSele}, thanks to the exploitation of the CSI correlation among continuous subcarriers, the ADER and BER$_{\rm total}$ performance of AE-JABID algorithm at $v_{\rm max}=180$ km/h improved by nearly 20 dB if using $L_F=4$ rather than $L_F=1$.}

\begin{table*}[!t]
\vspace{-2mm}
\scriptsize
\centering
\captionsetup{font = {normalsize, color = {black}}, labelsep = period} 
\caption*{Table I: Computational complexity comparison of different algorithms for DAD-EID of NC-IM-based massive IoT access using $J$ sub-frames and $\widetilde{N}$ subcarriers for transmission}
\color{black}
\begin{threeparttable}
\begin{tabular}{|p{4cm}|p{10cm}|}
\Xhline{1.2pt}
\makecell[c]{\bf Algorithms} & \makecell[c]{\bf Computational complexity\tnote{1}}\\
\Xhline{1.2pt}
\makecell[c]{Section-wise AMP} & \makecell[c]{$\mathcal{O}[J\widetilde{N}T(2KILM)]$} \\
\hline
\makecell[c]{AE-JABID} &\makecell[c]{${\cal O}[J\widetilde{N}T(4KILM+7\frac{3}{4}KIM+\frac{1}{2}QKIM)]$} \\
\hline
\makecell[c]{Benchmark 1}&\makecell[c]{${\cal O}[J\widetilde{N}T(4KILM+8KIM)]$}\\
\hline
\makecell[c]{GMMV-AMP} &\makecell[c]{${\cal O}[J\widetilde{N}T(4KILM+8KIM)]$ } \\
\hline
\makecell[c]{STF-JABID} &\makecell[c]{${\cal O}[T(4KILM'+7\frac{1}{4}KIM'+1\frac{1}{4}KI^2M'+\frac{3}{4}KI^2(M')^2)]$ }\\
\hline
\makecell[c]{SOMP} & \makecell[c]{$\mathcal{O}\{J\widetilde{N}[K_aKILM+\sum\nolimits_{t=1}^{K_a}(t^3+2Lt^2+2LMt)]\}$}\\
\Xhline{1.2pt}
\end{tabular}
\begin{tablenotes}
\scriptsize\item[1]For the sake of simplicity, a real-valued multiplication is assumed to require a complexity that is equal to a quarter of a complex-valued multiplication \cite{kemalong20}. In addition, $M'=JM\widetilde{N}$, and $T$ is the number of iterations.
\end{tablenotes}
\end{threeparttable}
\vspace{-6mm}
\end{table*}

\begin{figure*}[t]
\vspace{-7mm}
\centering
\subfigure[]{
    \begin{minipage}[t]{0.33\linewidth}
        \centering
\label{fig:compEffi1}
        \includegraphics[width = 0.95\columnwidth,keepaspectratio]
        {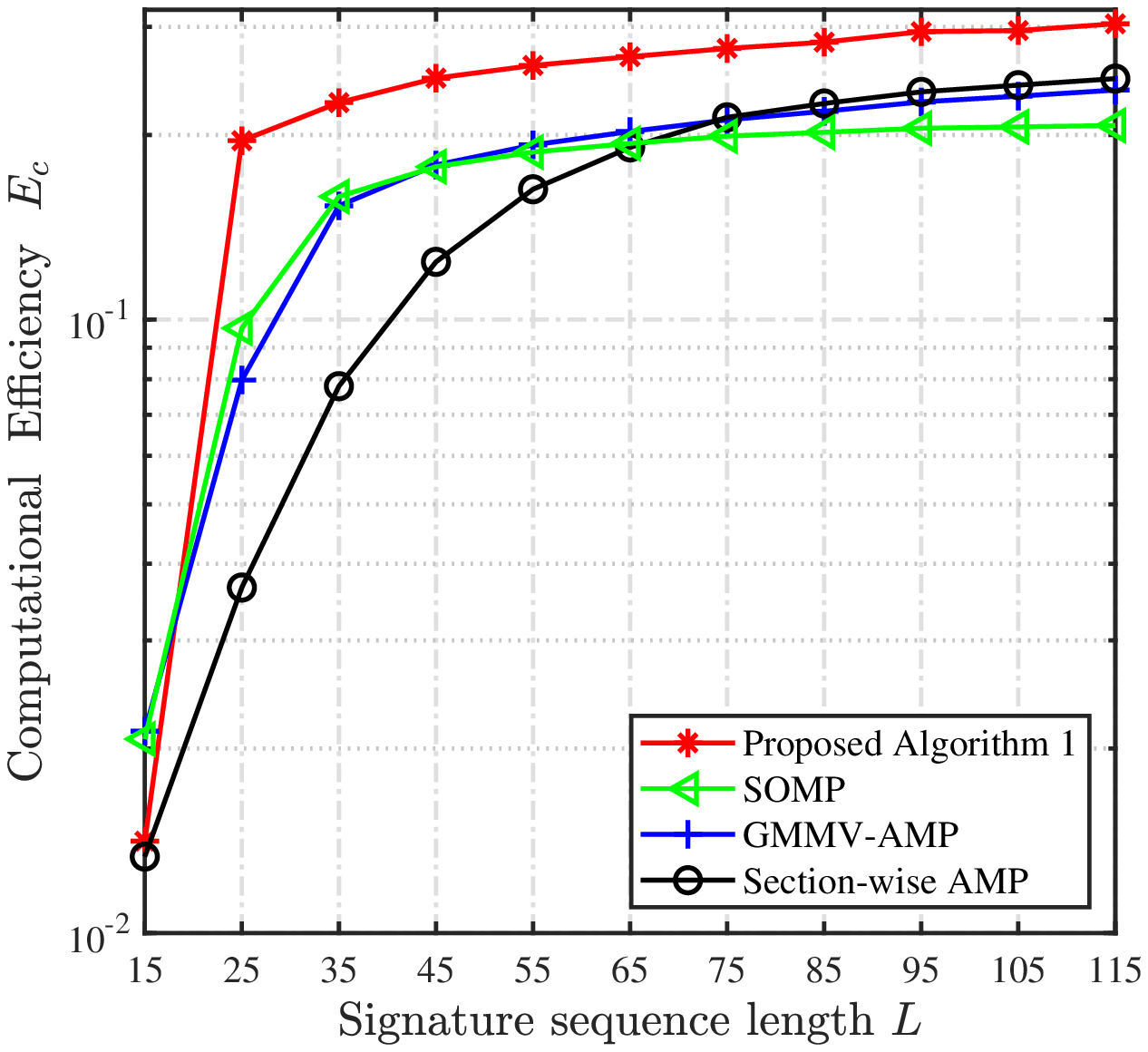}\\
    \end{minipage}%
}%
\subfigure[]{
    \begin{minipage}[t]{0.33\linewidth}
        \centering
\label{fig:compEffi2}
       \includegraphics[width = 0.95\columnwidth,keepaspectratio]
       {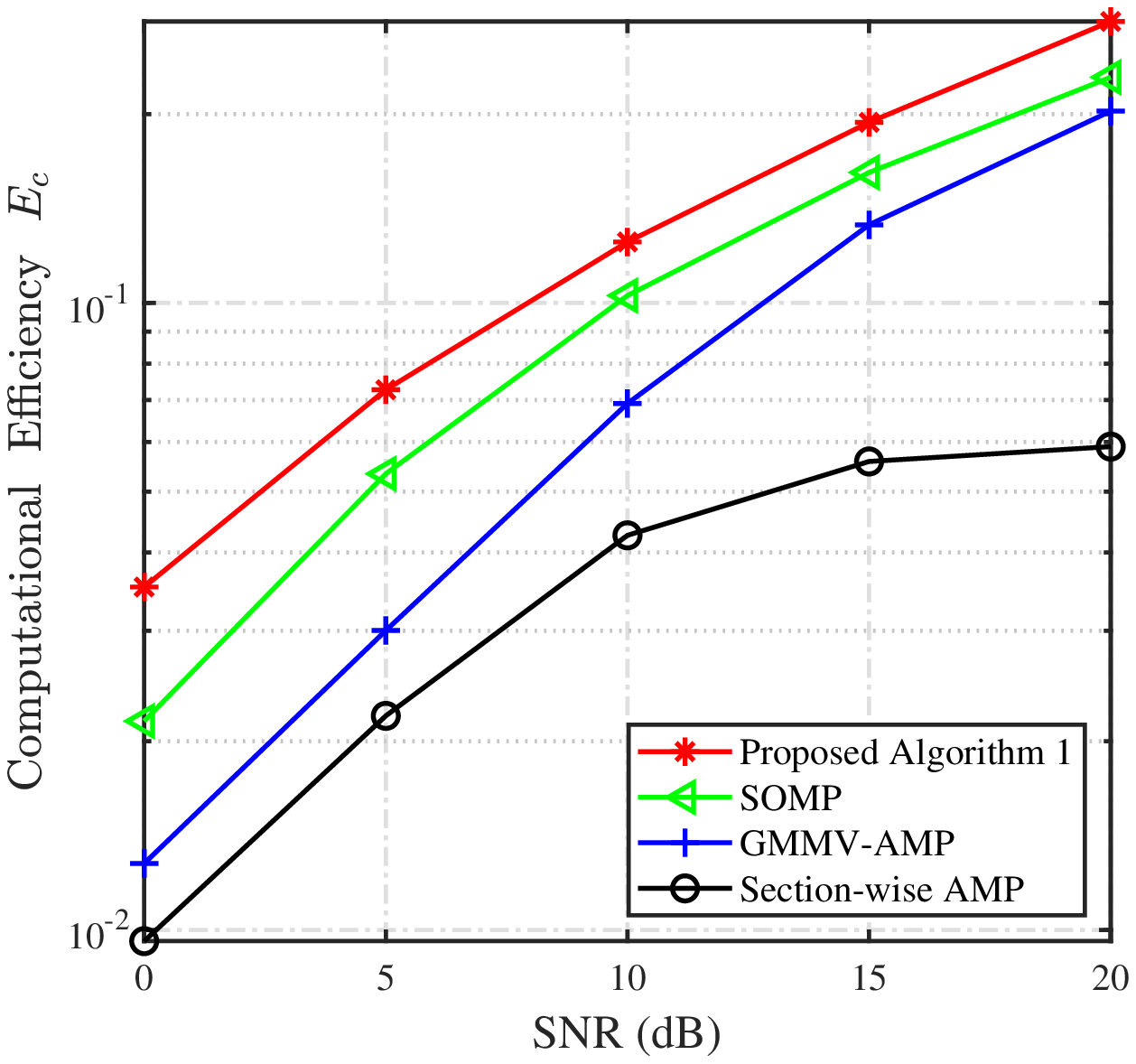}\\
    \end{minipage}%
}%
\subfigure[]{
    \begin{minipage}[t]{0.33\linewidth}
        \centering
\label{fig:compEffi3}
        \includegraphics[width = 0.95\columnwidth,keepaspectratio]
        {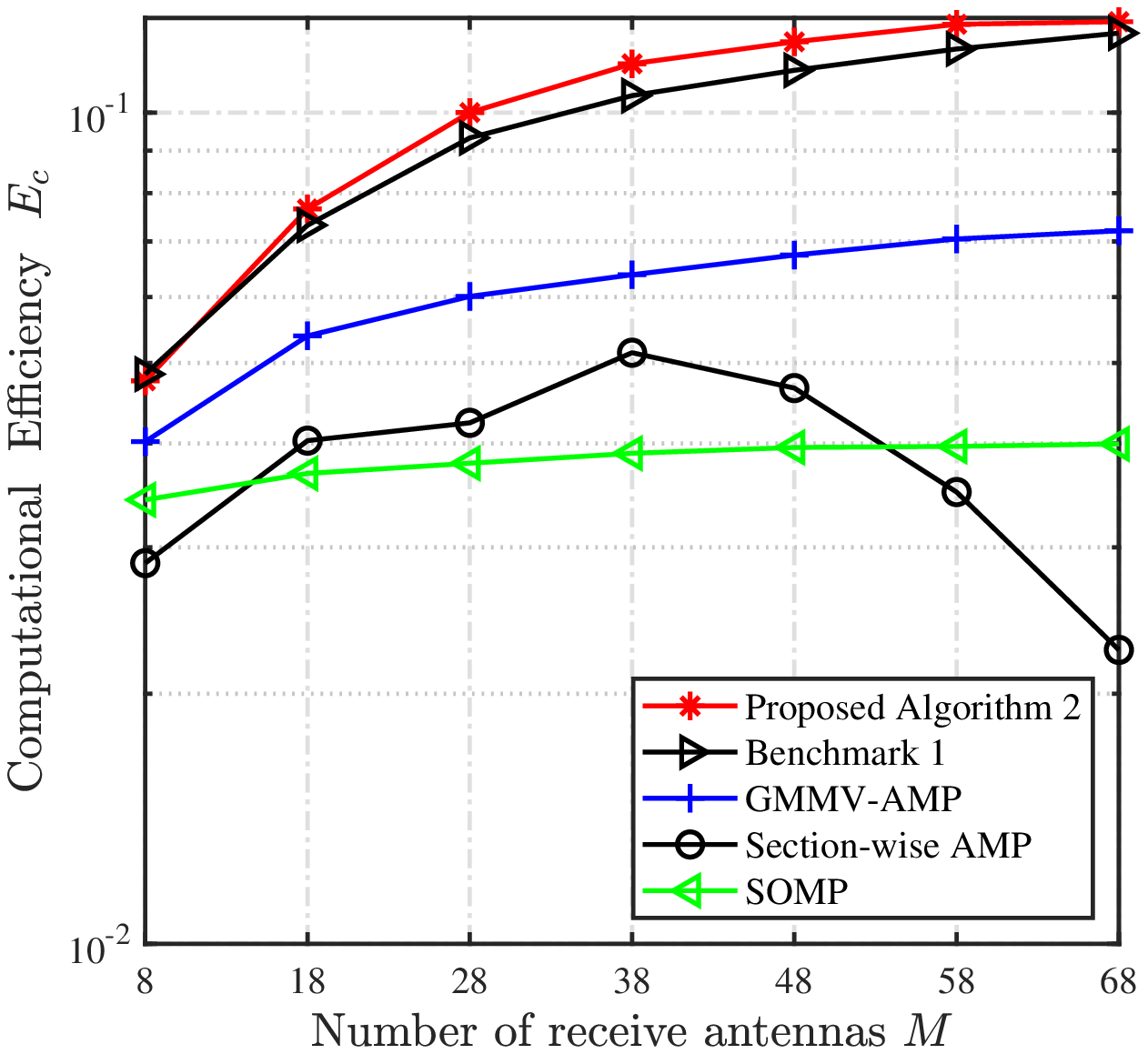}\\
    \end{minipage}%
}%
\centering
\setlength{\abovecaptionskip}{-1mm}
\captionsetup{font={footnotesize, color = {black}}, singlelinecheck = off, justification = raggedright,name={Fig.},labelsep=period}
\caption{Computational efficiency comparison of different algorithms: (a) $E_c$ versus the signature sequency length $L$; (b) $E_c$ versus the SNR; (c) $E_c$ versus the number of receive antennas. }
\label{fig:compEffi}
\vspace{-5mm}
\end{figure*}

\subsection{Computational Complexity}
{\color{black}Table I provides the computational complexity of the proposed STF-JABID and AE-JABID algorithms as well as other benchmark algorithms, where the order of complexity is calculated based on the number of complex-valued multiplications, denoted as $C_m$\cite{TWC-Qiao}. In addition, to verify the superiority of our proposed algorithms, we define the ``computational efficiency on BER$_{\rm total}$'' as: $E_c=\frac{-10{\rm log_{10}}({\rm BER}_{\rm total})}{10{\rm log}_{10}(C_m)}$. It can be seen that if BER$_{\rm total}$ or $C_m$ decreases, the defined $E_c$ will increase, which indicates a higher computational efficiency of the algorithm. A logarithmic scale is used to define $E_c$ in order to account for the wide range of $C_m$ and BER$_{\rm total}$.}

From Table I, it can be observed that the complexity of the greedy SOMP algorithm has a cubic relationship with the number of active devices $K_a$, due to the matrix inversion required for the least square estimation. {\color{black}In contrast, the proposed AE-JABID algorithm and other state-of-the-art AMP-based benchmarks have similar order of complexity, which increases linearly with $K$, $L$, $M$, $J$, and $\widetilde{N}$. {\color{black}While, the proposed AE-JABID algorithm significantly outperforms other AMP-based benchmarks in NMSE, ADER, and BER$_{\rm total}$ performance.} Hence, for massive IoT access scenarios with large $K_a$, the proposed AE-JABID algorithm is computationally efficient. Due to the joint signal processing of $J$ sub-frames and $\widetilde{N}$ subcarriers, the proposed STF-JABID algorithm has slightly higher complexity. Specifically, compared with the AE-JABID algorithm, the complexity of STF-JABID algorithm has an extra term that has quadratic relationship with $J$, $\widetilde{N}$, and $M$. Hence, the proposed STF-JABID algorithm significantly improves the DAD-EID performance in small-scale MIMO systems with a slightly higher computational complexity.}

{\color{black}As shown in Fig. \ref{fig:compEffi}, the performance of ``computational efficiency on BER$_{\rm total}$'' $E_c$ of different algorithms versus the signature sequence length $L$, the SNR, and the number of receive antennas $M$ is shown in Fig. \ref{fig:compEffi1}, Fig. \ref{fig:compEffi2}, and Fig. \ref{fig:compEffi3}, respectively. Note that $E_c=\frac{-10{\rm log_{10}}({\rm BER}_{\rm total})}{10{\rm log}_{10}(C_m)}$, where the complexity $C_m$ is calculated according to Table I and the number of iterations is set to the maximum $T_0=200$. The system parameters adopted in Fig. \ref{fig:compEffi1}, Fig. \ref{fig:compEffi2}, and Fig. \ref{fig:compEffi3} are the same as those in Fig. \ref{fig:Lsce1}, Fig. \ref{fig:SNRsce1}, and Fig. \ref{fig:AntennaSce2}, respectively. It can be observed that the proposed STF-JABID algorithm and the proposed AE-JABID algorithm have higher ``computational efficiency on BER$_{\rm total}$'', compared with the benchmark algorithms. Due to the space limitations, the ``computational efficiency on ADER'' performance is omitted. It reveals, however, the same computational efficiency superiority of the proposed algorithms.}

\vspace{-3mm}
\section{Conclusions}
This paper investigated the DAD-EID problem of grant-free NC-IM in UAV-based wide-area massive IoT access over both frequency-selective and time-selective channels with OFDM. For small-scale MIMO-based aerial BSs, to improve the DAD-EID performance, we proposed an efficient STF-JABID algorithm, by exploiting the space-time-frequency structured sparsity. Furthermore, for LS-MIMO-based aerial BSs, to further reduce the access latency of grant-free NC-IM, we proposed an AE-JABID algorithm by leveraging the virtual angular-domain structured sparsity. In addition, we developed a TFST strategy for achieving better JABID performance over doubly-selective fading channels. Compared with the state-of-the-art algorithms, the simulation results verified that the proposed STF-JABID algorithm can enhance the DAD-EID performance significantly and the proposed AE-JABID algorithm can reduce the access latency with improved DAD-EID performance. Also, simulation results demonstrated that the proposed TFST strategy is a promising yet straightforward solution for highly-mobile UAVs and/or IoT devices using NC-IM.

\vspace{-3mm}
\begin{appendices}
\section{Derivation of Message Update for STF-JABID Algorithm}
As shown in Fig. \ref{fig:FactorGraph}, we start with the message through the path $\{x_{k,i}^{m'}\} \to \{f_k^{m'}\} \to \{a_k\}$ in the $t$-th iteration of STF-JABID algorithm. Firstly, the message from variable node $\{x_{k,i}^{m'}\}$ to factor node $\{f_k^{m'}\}$ is
\begin{equation}\label{eq:ap1}
\begin{array}{l}
\mu_{x_{k,i}^{m'} \to f_k^{m'}}^t={\cal CN}(x_{k,i}^{m'};r_{k,i}^{m'},\varphi_{k,i}^{m'}).
\end{array}
\end{equation}

Then, the message from factor node $\{f_k^{m'}\}$ to variable node $\{a_k\}$ can be written as
\begin{align}\label{eq:ap2}
\vspace{-3mm}
\mu_{f_k^{m'} \!\!\!\to\! a_k}^t &\!\!\!\propto\!\!\! \int_{\backslash a_k} {\prod\nolimits_{i=1}^{I} {\mu_{x_{k,i}^{m'} \to f_k^{m'}}^t}f_k^{m'}(x_{k,1}^{m'},x_{k,2}^{m'},...,x_{k,I}^{m'})}\nonumber\\
& \!\!= \!\!(1-a_k)\prod\limits_{i=1}^{I}{\theta_{k,i}^{m'}}+\frac{a_k}{I}\sum\limits_{i=1}^{I}{\gamma_{k,i}^{m'}\prod\limits_{i'\in \backslash i}{\theta_{k,i'}^{m'}}},
\end{align}
where $\propto$ denotes the equality up to a constant scaling factor, $\backslash i$ denotes the set $\{x| x\in [I], x\ne i, i\in [I]\}$, and
\begin{align}\label{eq:ap22}
\vspace{-3mm}
\theta_{k,i}^{m'}&=\dfrac{1}{\pi \varphi_{k,i}^{m'}}{\rm exp}\left(-\dfrac{\left| {r_{k,i}^{m'}} \right|^2}{\varphi_{k,i}^{m'}}\right),\nonumber\\
~~\gamma_{k,i}^{m'}&=\dfrac{1}{\pi (\tau_0+\varphi_{k,i}^{m'})}{\rm exp}\left(-\dfrac{\left| {r_{k,i}^{m'}-\mu_0} \right|^2}{\varphi_{k,i}^{m'}+\tau_0}\right).
\end{align}

The backward message is calculated through the path $\{h_k\}\to \{a_k\} \to \{f_k^{m'}\} \to \{x_{k,i}^{m'}\}$. The message from factor node $\{h_k\}$ to variable node $\{a_k\}$ is
\begin{equation}\label{eq:ap3}
\begin{array}{l}
\mu_{h_k\to a_k}^t = (1-\lambda_k^{t-1})\delta(a_k)+\lambda_k^{t-1}\delta(a_k-1).
\end{array}
\end{equation}

Next, the message from the variable node $\{a_k\}$ to the factor node $\{f_k^{m'}\}$ is
\begin{align}\label{eq:ap4}
\mu_{a_k \to f_k^{m'}}^t &\propto \mu_{h_k\to a_k}^t \prod\limits_{\overline{m}\in \backslash {m'}}{\mu_{f_k^{\overline{m}} \to a_k}^t}\nonumber\\
&= (1-\lambda_k)\delta(a_k)\prod\limits_{\overline{m}\in \backslash {m'}}\prod\limits_{i=1}^{I}\theta_{k,i}^{\overline{m}}+\lambda_k\delta(a_k-1)\nonumber\\
&~~~~~~~~~~~\times \prod\limits_{\overline{m}\in \backslash m'}\left( {\dfrac{1}{I}\sum\limits_{i=1}^{I}\gamma_{k,i}^{\overline{m}}\prod\limits_{i'\in \backslash i}\theta_{k,i'}^{\overline{m}}} \right),
\end{align}
where $\backslash m'$ denotes the set $\{x| x\in [M'], x\ne m', m'\in [M']\}$.

Furthermore, the message from the factor node $\{f_k^{m'}\}$ to the variable node $\{x_{k,i}^{m'}\}$ is
\begin{equation}\label{eq:ap5}
\begin{array}{l}
\mu_{f_k^{m'} \to x_{k,i}^{m'}}^t \!\!\propto \!\!\int_{\backslash x_{k,i}^{m'}}{\prod\limits_{i'\in \backslash i} {\mu_{x_{k,i'}^{m'} \to f_k^{m'}}^t}\mu_{a_k \to f_k^{m'}}^t f_k^{m'}(x_{k,1}^{m'},...,x_{k,I}^{m'})}.
\end{array}
\end{equation}

Hence, the approximate posterior distribution of $x_{k,i}^{m'}$ is given by
\begin{align}\label{eq:ap6}
q(\!x_{k,i}^{m'}| {\bf{y}}_{m'}\!)&\!\!\propto\!\! \mu_{x_{k,i}^{m'} \to f_k^{m'}}^t \mu_{f_k^{m'} \to x_{k,i}^{m'}}^t\nonumber \\
&\!\!=\!\!(1\!-\!\pi_{k,i}^{m'})\delta(x_{k,i}^{m'})\!\!+\!\!\pi_{k,i}^{m'}{\cal CN}\!\!\left(\!\!x_{k,i}^{m'};\overline{\mu}_{k,i}^{m'},\overline{\tau}_{k,i}^{m'}\!\!\right),
\end{align}
where $\overline{\mu}_{k,i}^{m'}$, $\overline{\tau}_{k,i}^{m'}$, $\pi_{k,i}^{m'}$, and ${\cal L}_{k,i}^{m'}$ are indicated in (\ref{eq:post1})-(\ref{eq:post4}).

\section{Derivation of EM Update Rules for the Proposed Algorithms}
Here, we first derive the update rule of the activity indicator $\lambda_k$ in the proposed STF-JABID algorithm. We have the following relation
\begin{equation}\label{eq:ap7}
\begin{split}
{\rm ln}p({\bf X}\!,\!{\bf Y}\!)\!=\!{\rm ln}p({\bf Y}|{\bf X})\!+\!{\rm ln}f({\bf X})\!=\!\!\!\sum\limits_{m'=1}^{M'}\!\!{\rm ln}\!f(\!{\bf x}_k^{m'}\!)\!\!+\!{\rm Const},
\end{split}
\end{equation}
where ${\rm Const}$ is independent of $\lambda_k$, and $f({\bf x}_k^{m'})$ is the prior probability defined in (\ref{eq:prior}).

Then, we have
\begin{align}\label{eq:ap8}
{\dfrac{\partial Q\left({{\bm \theta},{\bm \theta}^t}\right)}{\partial \lambda_k}}&=\sum\limits_{m'=1}^{M'}\dfrac{\partial}{\partial \lambda_k}{\int}p\left({\bf x}_k^{m'}|{\bf y}_{m'} \right){\rm ln}p\left( {\bf x}_k^{m'}\right)d{\bf x}_k^{m'} \nonumber \\
&\approx\sum\limits_{m'=1}^{M'}\dfrac{\partial}{\partial \lambda_k}{\int}\prod\limits_{i=1}^{I}q\left( x_{k,i}^{m'}|{\bf{y}}_{m'}\right){\rm ln}p\left( {\bf x}_k^{m'}\right)d{\bf x}_k^{m'},
\end{align}
where $q\left( x_{k,i}^{m'}|{\bf{y}}_{m'}\right)$ is defined in (\ref{eq:post}).

Since the Dirac function is discontinuous, we adopt the following approximation: ${\cal CN}(x;x_0,\epsilon)\to \delta(x-x_0)$, when $\epsilon\to 0^{+}$. Under this assumption, by exchanging the order of the integral and the derivative \cite{AMPmeng}, we have
\begin{align}\label{eq:ap9}
{\dfrac{\partial }{\partial \lambda_k}}\!{\rm ln}\!p\!\left(\! {\bf x}_k^{m'}\!\right)&\!\!\!=\!\!\dfrac{1}{p\left( {\bf x}_k^{m'}\right)}{\dfrac{\partial }{\partial \lambda_k}}\Bigg[ (1\!-\!\lambda_k)\sum\limits_{i=1}^{I}\delta\left( x_{k,i}^{m'}\right)\nonumber\\
&\!\!\!+\!\!\lambda_k\sum\limits_{i=1}^{I}{\cal CN}\left(x_{k,i}^{m'};\mu_0,\tau_0 \right)\prod\limits_{g\neq i} \delta\left( x_{k,g}^{m'}\right)\Bigg],
\end{align}
where ${\frac{\partial }{\partial \lambda_k}}{\rm ln}p\left( {\bf x}_k^{m'}\right)=\frac{-1}{1-\lambda_k}$ if ${\bf x}_k^{m'}={\bf 0}$, $\epsilon\to 0^{+}$; ${\frac{\partial }{\partial \lambda_k}}{\rm ln}p\left( {\bf x}_k^{m'}\right)=\frac{1}{\lambda_k}$ if ${\bf x}_k^{m'}\neq{\bf 0}$, $\epsilon\to 0^{+}$. Denote $\mathbb{B}_{\epsilon\to 0^{+}}$ as a set that ${\bf x}_k^{m'}={\bf 0}$ and $\epsilon\to 0^{+}$, while $\overline{\mathbb{B}}_{\epsilon\to 0^{+}}$ is denoted as a set that ${\bf x}_k^{m'}\neq{\bf 0}$ and $\epsilon\to 0^{+}$.

By substituting (\ref{eq:ap9}) and (\ref{eq:post}) into (\ref{eq:ap8}), we have
\begin{align}\label{eq:ap10}
{\dfrac{\partial Q\left({{\bm \theta},{\bm \theta}^t}\right)}{\partial \lambda_k}}&=\sum\limits_{m'=1}^{M'}\Bigg(\dfrac{-1}{1-\lambda_k}{\int}_{\mathbb{B}_{\epsilon\to 0^{+}}}p\left({\bf x}_k^{m'}|{\bf y}_{m'} \right)d{\bf x}_k^{m'} \nonumber\\
&~~~~~~~+\dfrac{1}{\lambda_k}{\int}_{\overline{\mathbb{B}}_{\epsilon\to 0^{+}}}p\left({\bf x}_k^{m'}|{\bf y}_{m'} \right)d{\bf x}_k^{m'}\Bigg)\nonumber\\
&\approx\sum\nolimits_{m'=1}^{M'}\Bigg(\dfrac{-1}{1-\lambda_k}\prod\nolimits_{i=1}^{I} \left(1-\pi_{k,i}^{m'}\right) \nonumber\\
&~~~~~~~+\dfrac{1}{\lambda_k}\sum\nolimits_{i=1}^{M} \pi_{k,i}^{m'}\prod_{j\neq i}\left(1-\pi_{k,j}^{m'} \right)\Bigg).
\end{align}

By setting (\ref{eq:ap10}) to zero, we obtain the update rule for $\lambda_k$ in (\ref{eq:EMACT}), $\forall k\in[K]$. Similarly, the update rule (\ref{eq:EMACT2}) for the AE-JABID algorithm, the update rules of the prior means and variances for both algorithms can also be derived, but are omitted for brevity.


\section{Proof of Remark 1}
Given a frequency-domain channel response vector ${\bf h}(\theta)=h{\bf a}(\theta)\in\mathbb{C}^{M\times 1}$, where ${\bf a}(\theta)$ is the steering vector denoted in (\ref{eq:SVec}) and $\theta$ is the AoA. Define the channel gain in the spatial domain as $h^2$.
According to \cite{VAD_Channel}, the matrix that transforms the spatial domain channel to the virtual angular domain is a unitary DFT matrix with $d=\lambda/2$, denoted as ${\bf A}_R=[{\bf a}(\widetilde{\theta}_1),{\bf a}(\widetilde{\theta}_2),...,{\bf a}(\widetilde{\theta}_{M})]/\sqrt{M}\in \mathbb{C}^{M\times M}$, where $\widetilde{\theta}_m$ is the $m$-th virtual angle, denoted as $\widetilde{\theta}_m=\frac{m-\overline{M}}{M}$, $m\in[M]$, and $\overline{M}=\frac{M+1}{2}$. If the AoA of channel ${\bf h}$ is on the $m$-th ($m\in[M]$) virtual angular grid, we have the virtual angular-domain channel as $\overline{\bf h}=({\bf h}(\widetilde{\theta}_m))^T{{\bf A}_R}^*\in\mathbb{C}^{M\times 1}$, where only the $m$-th element of $\overline{\bf h}$ equals $h\sqrt{M}$ and others equal zeros. In addition, the unitary DFT matrix does not change the noise power. Hence, the maximum channel gain in the virtual angular domain, i.e., the channel gain at the $m$-th virtual angle is $Mh^2$, which is $M$ times larger than that of the spatial domain. This completes the proof of Remark 1.

\section{List of Abbreviations}
\begin{center}
\small
\begin{tabular}{ll}
ADER&Activity detection error rate\\
AMP&Approximate message passing\\
AoA& Angle of arrival\\
BER$_{\rm total}$&Total bit error rate\\
CE& Channel estimation\\
CS&Compressed sensing\\
CSI& Channel state information\\
DAD&Device activity detection\\
DD&Data detection\\
EID&Embedded information detection\\
EM&Expectation maximization\\
GFRA&Grant-free random access\\
GMMV-AMP&Generalized multiple measurement vector AMP\\
ICI&Inter-sub-carrier interference\\
IM&Index modulation\\
JABID& Joint activity and blind information detection\\
AE-JABID&Angular-domain enhanced JABID\\
JDAD-CE&Joint DAD and CE\\
JDAD-DD&Joint DAD and DD\\
LoS (NLoS) & (Non)-Line-of-Sight\\

%

LS-MIMO&Large-scale MIMO\\
MIMO&Multiple-input multiple-output\\
MMSE& Minimum mean square error\\
MMV&Multiple measurement vector\\
NC-IM&Non-coherent index-modulation\\
NMSE&Normalized mean square error\\
OFDM&Orthogonal frequency division multiplexing\\
RIS&Reconfigurable intelligent surface\\ 
SM&Spatial modulation\\
SNR&Signal-to-noise ratio\\
SOMP&Simultaneous orthogonal matching pursuit\\
STF-JABID &Space-time-frequency JABID\\ 
TFST &Time-frequency spread transmission\\
UAV&Unmanned aerial vehicle\\
ULA&Uniform linear array\\
\end{tabular}
\end{center} 
\end{appendices}


\end{document}